\documentclass[twocolumn]{aastex631}
\usepackage{array}
\usepackage{subfigure}
\usepackage{natbib}
\usepackage{enumitem}
\usepackage{booktabs}

\usepackage{graphicx}	% Including figure files
\usepackage{amsmath}	% Advanced maths commands
\usepackage{amssymb}	% Extra maths symbols
\usepackage{bm}		% Bold maths symbols, including upright Greek
\usepackage{color}

\usepackage{siunitx}
\usepackage{anyfontsize}
\usepackage{xspace}
\usepackage{threeparttable}
\usepackage{tablefootnote}

\makeatletter
\let\TPT@hookin\@gobble
\let\TPT@hookarg\@gobble
\makeatother

\def\hi{\textsc{H\,i}}
\def\oh{\textsc{OH}}
\newcommand{\kms}{\,km\,s$^{-1}$} % kilometres per second
\newcommand{\cm}{\ensuremath{\,{\rm cm}}}

\renewcommand{\kms}{\ensuremath{\,{\rm km\,s^{-1}}}} % kilometres per second

\newcommand{\K}{\ensuremath{\, {\rm K}}}

\newcommand{\mJy}{\ensuremath{\,{\rm mJy}}}

\shorttitle{FAST \oh\ absorption search}
\shortauthors{Hu et al.}
\begin{document}

\title[FAST \oh\ absorption search]
{The FAST HI 21-cm Absorption Blind Survey. III. \\ OH Absorption Search in the 21-cm Absorber Sample and Continued HI Absorption Search}

\correspondingauthor{Wenkai Hu}
\author[0000-0002-3108-5591]{Wenkai Hu}
\email{wkhu@nao.cas.cn}
\affiliation{State Key Laboratory of Radio Astronomy and Technology, National Astronomical Observatories, Chinese Academy of Sciences, Beijing 100101, China}
\affiliation{Department of Physics and Astronomy, University of the Western Cape, Robert Sobukhwe Road, Bellville, 7535, South Africa}

\author[0000-0003-0631-568X]{Yougang Wang}
\affiliation{State Key Laboratory of Radio Astronomy and Technology, National Astronomical Observatories, Chinese Academy of Sciences, Beijing 100101, China}
\affiliation{School of Astronomy and Space Science, University of Chinese Academy of Sciences, Beijing 100049, China}
\affiliation{Key Laboratory of Cosmology and Astrophysics (Liaoning) \& College of Sciences, Northeastern University, Shenyang 110819, China}

\author[0000-0003-2511-2060]{Jeremy Darling}
\affiliation{Center for Astrophysics and Space Astronomy, Department of Astrophysical and Planetary Sciences, University of Colorado, 389 UCB, Boulder, CO 80309-0389, USA}

\author[0009-0005-9546-4573]{Zheng Zheng}
\affiliation{National Astronomical Observatories, Chinese Academy of Sciences, Beijing 100101, China}
\affiliation{Research Center for Intelligent Computing Platforms, Zhejiang Laboratory, Hangzhou 311100, China}

\author[0000-0003-0436-4680]{James R. Allison}
\affiliation{First Light Fusion Ltd., Unit 10 Oxford Pioneer Park, Yarnton, OX5 1QU, UK}

\author[0000-0002-1136-2555]{Elaine M. Sadler}
\affiliation{ATNF, CSIRO Space and Astronomy, PO Box 76, Epping, NSW 1710, Australia}
\affiliation{Sydney Institute for Astronomy, School of Physics A28, University of Sydney, NSW 2006, Australia}
\affiliation{ARC Centre of Excellence for All Sky Astrophysics in 3 Dimensions (ASTRO 3D), Australia}

\author[0009-0006-2521-025X]{Wenxiu Yang}
\affiliation{State Key Laboratory of Radio Astronomy and Technology, National Astronomical Observatories, Chinese Academy of Sciences, Beijing 100101, China}
\affiliation{School of Astronomy and Space Science, University of Chinese Academy of Sciences, Beijing 100049, China}

\author[0000-0003-1962-2013]{Yichao Li}
\affiliation{Key Laboratory of Cosmology and Astrophysics (Liaoning) \& College of Sciences, Northeastern University, Shenyang 110819, China}

\author[0000-0003-3224-4125]{Yidong Xu}
\affiliation{State Key Laboratory of Radio Astronomy and Technology, National Astronomical Observatories, Chinese Academy of Sciences, Beijing 100101, China}

\author[0000-0002-9937-2351]{Jie Wang}
\affiliation{National Astronomical Observatories, Chinese Academy of Sciences, Beijing 100101, China}
\affiliation{School of Astronomy and Space Science, University of Chinese Academy of Sciences, Beijing 100049, China}

\author[0000-0002-6174-8640]{Fengquan Wu}
\affiliation{State Key Laboratory of Radio Astronomy and Technology, National Astronomical Observatories, Chinese Academy of Sciences, Beijing 100101, China}

\author[0000-0003-3010-7661]{Di Li}
\affiliation{Department of Astronomy, Tsinghua University, 30 Shuangqing Road, Beijing 100084, People’s Republic of China}
\affiliation{State Key Laboratory of Radio Astronomy and Technology, National Astronomical Observatories, Chinese Academy of Sciences, Beijing 100101, China}

\author[0000-0001-6083-956X]{Ming Zhu}
\affiliation{State Key Laboratory of Radio Astronomy and Technology, National Astronomical Observatories, Chinese Academy of Sciences, Beijing 100101, China}
\affiliation{Guizhou Radio Astronomical Observatory, Guizhou University, Guiyang 550000, China}

\author[0000-0001-6475-8863]{Xuelei Chen}
\affiliation{State Key Laboratory of Radio Astronomy and Technology, National Astronomical Observatories, Chinese Academy of Sciences, Beijing 100101, China}
\affiliation{School of Astronomy and Space Science, University of Chinese Academy of Sciences, Beijing 100049, China}
\affiliation{Key Laboratory of Cosmology and Astrophysics (Liaoning) \& College of Sciences, Northeastern University, Shenyang 110819, China}

\begin{abstract}

We present the first blind search for \oh\ 18-cm absorption with the Five-hundred-meter Aperture Spherical Telescope (FAST), conducted alongside the \hi\ 21-cm absorption search. Our previous FAST blind \hi\ absorption search identified 34 systems. In this work, we extend the search using 2024 and part of the 2025 CRAFTS and FASHI data (394.4 hr and 1622.1 deg$^{2}$) together with FATHOMER observations, yielding four known and three new \hi\ absorbers, for a total of 41 \hi\ absorption systems. We search for \oh\ absorption in 19 \hi\ absorption systems whose \oh\ redshifted frequencies fall within the FAST band. The known \oh\ absorber towards PKS 1413\allowbreak+135 was re-detected, making our survey the first blind survey to detect \oh\ absorption. No new \oh\ absorbers were identified. We examine the relationship between $N_{\rm{OH}}$ and $N_{\hi}$, applying survival analysis to account for upper limits. The analysis does not provide statistically significant evidence for either an $N_{\rm{OH}}$-$N_{\hi}$ correlation or redshift evolution of $N_{\rm{OH}}$/$N_{\hi}$. Finally, spectral stacking sets 3$\sigma$ \oh\ column density upper limits of 4.93, 1.64, and 1.72 $T_{\rm{ex}}$/$c_{\rm{f,\oh}}\times$10$^{12}$cm$^{-2}$K$^{-1}$ for associated, intervening, and combined samples, corresponding to [\rm{OH}]/[\hi] ratios of $<$1.66$\times$10$^{-8}$, $<$1.42$\times$10$^{-8}$, and $<$0.90$\times$10$^{-8}$, assuming $T_{\rm{ex}}$=10K for \oh\ and $T_{\rm{s}}$=100K for \hi. These results place the strongest constraints to date on the \oh\ content in radio-selected \hi\ absorbers and establish a blind-survey benchmark for future studies of molecular gas in \hi-selected systems. They also demonstrate that known \hi\ 21-cm absorbers provide an effective parent sample for systematic \oh\ absorption searches, paving the way for future larger surveys.

%Intervening absorbers show modest correlations, while associated systems show none. A moderate positive trend between redshift and the $N_{\rm{OH}}$/$N_{\hi}$ ratio is found for both intervening and associated absorbers, consistent with higher gas fractions in galaxies at earlier cosmic times. Limited by the sample size, these trends and correlations are constrained with relatively low statistical precision.

\end{abstract}

%% You will be asked to select these concepts during the submission process


\section{Introduction} \label{sec:intro}

The hydroxyl radical (OH) is a fundamental tracer of the physical and chemical conditions in the interstellar and circumgalactic medium. In most galaxies, \oh\ resides in both diffuse and dense molecular gas clouds; however, its spontaneous emission is typically weak. While stimulated \oh\ emission can be observed in both the main OH lines (e.g., in OH megamaser galaxies; \citealt{2002AJ....124..100D}) and the satellite OH lines \citep{2004ApJ...612...58D,2004PhRvL..93e1302K,2005PhRvL..95z1301K}, such cases are exceptionally rare, typically associated with extreme starburst, AGN activity or specific pumping conditions (e.g., shock waves) in the molecular gas. In normal galaxies where \oh\ is not radiatively or collisionally excited, population inversion does not occur, preventing masing. Consequently, \oh\ absorption against background radio continuum sources remains the only viable method for detecting \oh\ in non-masing galaxies at cosmological distances, providing a crucial probe of molecular gas in diverse astrophysical environments.

The importance of \oh\ absorption stems from its ability to trace atomic-to-molecular transitions, determine gas kinematics, reveal the physical conditions within the Galaxy and high-redshift galaxies, and provide insight into how cold gas regulates star formation. Unlike carbon monoxide (CO), which is commonly used to trace molecular gas but requires cold, dense conditions to emit efficiently, \oh\ absorption can be detected even in diffuse gas. Moreover, CO is more abundant and often optically thick, while \oh\ is much less abundant and typically optically thin. This makes OH a powerful tool for studying molecular reservoirs in a broader range of astrophysical conditions, providing further constraints on `dark' molecular gas fractions \citep{2018ApJ...860L..22G}.

In addition, cospatial \hi\ and \oh\ absorption lines provide a powerful means to constrain the cosmic evolution of fundamental physical constants, including the electron–proton mass ratio $\mu$, the fine structure constant $\alpha$, and the proton g-factor $g_{\rm{p}} $\citep{2005PhRvL..95z1301K,2018ApJ...860L..22G}. Variations in physical constants over cosmic time would result in spectral lines deviating from their present-day measured values. By comparing the frequencies of different transition lines from the same object, one can experimentally test Grand Unified Theories, which have significant implications for both modern physics and cosmology. While such tests can be conducted by comparing the \hi\ 21-cm line to ionized metal lines in the optical/ultraviolet bands \citep{1976PhRvL..37..179W,PhysRevLett.95.041301,2010ApJ...712L.148K,2012MNRAS.425..556R}, the use of \hi–\oh\ line comparisons or \rm{OH}–\oh\ line comparisons \citep{2003PhRvL..91x1302C,2003PhRvL..91a1301D,2004ApJ...612...58D,2018PhRvL.120f1302K} is more robust. Since \oh\ and \hi\ lines originate along the same sight line within the beam, mitigating potential line-of-sight systematics that could otherwise mimic variations in fundamental constants. 

However, only six \oh\ absorbers at cosmological distances have been detected to date. This limited number is due to a combination of factors, including the insufficient sensitivity of current radio telescopes, radio frequency interference (RFI), potential selection biases in existing surveys, and the intrinsic low abundance of \textsc{OH}. Among these detections, four are intervening absorbers: B0218+357 \citep{2003MNRAS.345L...7K}, G0248+430 \citep{2018ApJ...860L..22G}, PKS 1830-211 \citep{1999A&A...343L..79C}, and PMN J0134-0931 \citep{2005PhRvL..95z1301K}. The remaining two are associated absorbers: B3 1504+377 \citep{2002A&A...381L..73K} and PKS 1413+135 \citep{2004ApJ...612...58D}. Previous studies have shown that all known \oh\ absorbers have also been detected in \hi\ 21-cm absorption. Motivated by this correlation and aiming to expand the sample of \oh\ absorbers, we used FAST to perform an unbiased \oh\ search in the spectra of radio sources where \hi\ 21-cm absorption was previously detected in FAST surveys.

In \citet{2023A&A...675A..40H,2025ApJS..277...25H}, we conducted a blind search for \hi\ absorption using data from the Commensal Radio Astronomy FasT
Survey (CRAFTS; \citealt{2018IMMag..19..112L}) and FAST All Sky \hi\ Survey (FASHI; \citealt{2024SCPMA..6719511Z}) surveys, covering the redshift from 0 to 0.35 and completed between 2020 and 2023, covering 1325.6 hours and a sky area of 6072.0 deg$^{2}$. This search resulted in the detection of 34 \hi\ absorbers, including 14 previously identified systems and 20 newly discovered ones. These comprised 14 associated systems, 11 intervening systems, and 9 systems with undetermined classifications. In this paper, we extend the search to include additional CRAFTS and FASHI data (394.4 hours and 1622.1 deg$^{2}$) obtained in 2024, as well as data from the FAst neuTral HydrOgen intensity Mapping ExpeRiment (FATHOMER; \citealt{2023ApJ...954..139L}) survey. This continued effort led to the detection of 7 more \hi\ absorption systems, consisting of four associated systems (NVSS J080601\allowbreak+190611, NVSS J090150\allowbreak+030422, NVSS J150034\allowbreak+364844 and NVSS J234106\allowbreak+001833), one intervening system (NVSS J024516\allowbreak+240535), and two systems with undetermined classifications (NVSS J010015\allowbreak+201710 and NVSS J053538\allowbreak+643141). In total, our catalog now includes 41 \hi\ absorption systems: 18 associated systems, 12 intervening systems, and 11 systems with undetermined classifications. Using this expanded \hi\ absorption catalog, we perform a systematic search for \oh\ absorption.

This paper is structured as follows: Section~\ref{sec:Data} outlines the survey data and follow-up observations utilized in this study. Section~\ref{sec:Data_Analysis} details the data processing procedures, candidate selection methodology, and the approach used to measure absorption. The physical properties of the newly confirmed \hi\ absorption systems are presented in Section~\ref{sec:Absorption_Systems}. Section~\ref{sec:OH_absorption} focuses on the search for \oh\ absorption. The \oh\ abundance is analyzed in Section~\ref{sec:OH_abundance}. A discussion is provided in Section~\ref{sec:Discussion}, and a summary of the work is given in Section~\ref{sec:Summary}. Throughout this paper, we adopt the cosmological parameters H$_{0}$=70 \kms Mpc$^{-1}$, $\Omega_{\rm m}$=0.3 and $\Omega_{\Lambda}$=0.7. The velocity reference frame used in the paper and for all of the spectra is the heliocentric frame. The value of the speed of light employed throughout is $c$=299792.458\kms.

\section{Radio Data}
\label{sec:Data}
\subsection{Radio Survey Data}

The radio data used for the \hi\ absorption search in this work were sourced from three surveys: CRAFTS, FASHI, and FATHOMER. CRAFTS is a multi-purpose drift-scan survey designed to observe galactic and extragalactic \hi\ emissions\citep{2025ApJS..279...32Y}, measure continuum signals, and search for new pulsars and fast radio bursts (FRBs).
FASHI is a comprehensive observational program aiming to conduct a detailed survey of \hi\ gas in the nearby universe. Both surveys are planned to cover approximately 22,000 square degrees, spanning declinations from $-14^\circ$ to $66^\circ$. FATHOMER is a drift-scan survey focused on studying cosmological large-scale structures using the \hi\ intensity mapping technique. Each year, from 2019 through 2024 (excluding 2023), approximately 30 hours were allocated for FATHOMER observations. All three surveys are conducted in drift-scan mode, utilizing the FAST L-band Array of 19 feed horns (FLAN; \citealt{8105012}), which operates in the frequency range of 1050 MHz to 1450 MHz (corresponding to $z$=0-0.35 for \hi\ 21-cm line and $z$=0.15-0.59 for \oh\ 1667 line.).

\begin{figure*}[hbt!]
    \centering
    \includegraphics[width=0.9\textwidth]{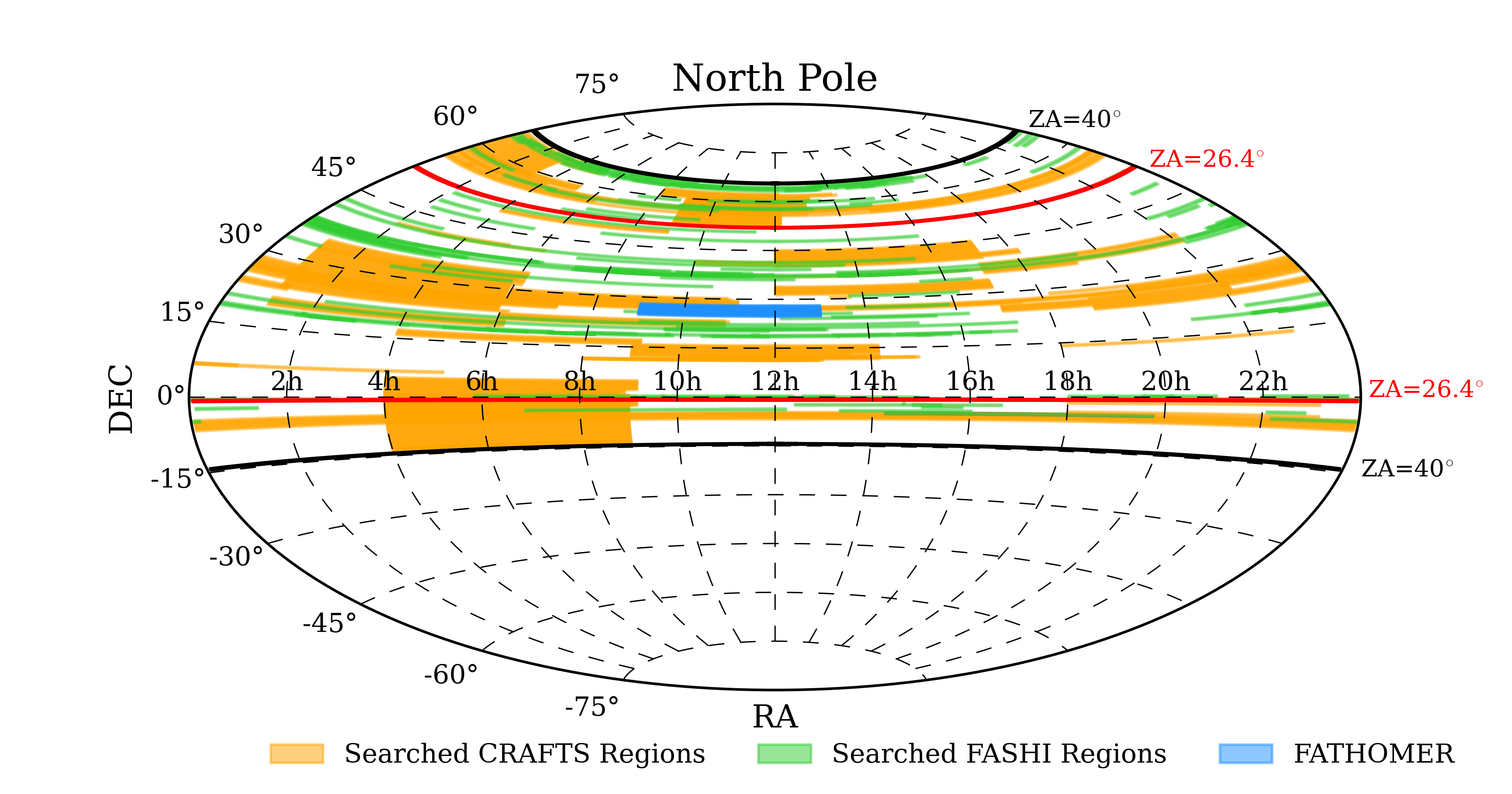}
    \caption{Up-to-date (2025-10-25) searched CRAFTS, FASHI, and FATHOMER sky coverage in Equatorial coordinates. Search areas in CRAFTS, FASHI, and FATHOMER regions are highlighted by orange, green, and blue labels, respectively. Zenith angles of \ang{40} (maximum zenith angle for FAST) and \ang{26.4} (zenith angle within which FAST has full gain) are shown as black circles and red circles, respectively.}
    \label{surveyed_sky}
\end{figure*}

As of October 25, 2024, our search has been conducted in part of the CRAFTS, FASHI, and FATHOMER regions, as shown in Figure~\ref{surveyed_sky}. \footnote{Further information about the finished CRAFTS scans can be found in \url{http://groups.bao.ac.cn/ism/CRAFTS/CRAFTS/}.} 

\subsection{Candidate \hi\ detections and follow-up observations}

In a blind search conducted using 2024 and part of 2025 CRAFTS and FASHI data, as well as data from the FATHOMER survey, a total of seven absorptions were detected. These include four previously known \hi\ absorption systems (NVSS J080601\allowbreak+190611, NVSS J090150\allowbreak+030422, NVSS J150034\allowbreak+364844, and NVSS J234106\allowbreak+001833) and three newly discovered ones (NVSS J010015\allowbreak+201710, NVSS J024516\allowbreak+240535, and NVSS J053538\allowbreak+643141). All absorption systems were verified using data from several neighboring beams with high signal-to-noise ratios. To confirm the signals of the three newly discovered candidates and to enhance the detection of \hi\ absorption features in the known systems, follow-up observations were conducted with FAST in December, 2024 and March, 2026, utilizing the FLAN.

To mitigate bandpass fluctuations and eliminate sky signals, the follow-up observations employed an ON-OFF tracking mode and were conducted at night. For each target, two cycles were obtained, with each cycle consisting of a on-source scan followed by an off-source scan. To optimize the utilization of observational time and the 19-beam receiver of FAST, pointing positions are selected such that Beam 12 is aligned with the target source when Beam 01 is positioned off-source, and vice versa. This alternating beam strategy ensures efficient sky coverage while minimizing systematic uncertainties. For the \oh\ absorption search, data from both beams will be combined to enhance sensitivity and improve the reliability of detections. The integration time for each source searched for \oh\ absorption is presented in Table~\ref{OH_absorption_table}. The integration times vary because they were limited by the observation time awarded in different follow-up observation campaigns. The raw follow-up data have a frequency resolution of 7.63 kHz, corresponding to velocity resolutions of 1.6 \kms at redshift $z$=0 and 2.2 \kms at $z$=0.35.

\section{Analysis}
\label{sec:Data_Analysis}
\subsection{Searching Algorithms}
\label{sec:algorithms}

The search technique is described in detail in \citet{2023A&A...675A..40H,2025ApJS..277...25H}. In brief, we first cross-matched the NRAO VLA Sky Survey (NVSS; \citealt{1998AJ....115.1693C}) catalog with our radio dataset and selected sources with flux densities above 12 mJy. This threshold ensures detectability of \hi\ absorption with peak S/N $\gtrsim$5.5 using FAST, assuming a channel width of $\Delta\nu$=15.26 kHz and an integration time of 12 s per beam. For each selected source, spectra were generated by rebinning the raw data centered on the source into a time resolution of approximately $\sim$ 12$/\cos\delta_{\rm{dec}}$ seconds (the transit time during drift scan, where $\delta_{\rm{dec}}$ is the declination of the source) and a frequency resolution of $\Delta\nu$=15.26 kHz. After removing the baseline (estimated using a low-pass filter), \hi\ absorbers are identified through a blind search using a matched-filtering approach, where flux spectra are cross-correlated with Gaussian templates \citep{2007AJ....133.2087S}. The search is conducted for \hi\ absorption signals in both XX and YY polarizations for each beam. Candidates are selected if they have a combined velocity-integrated signal-to-noise ratio (S/N) greater than 5.5 and are detected at nearly the same frequency ($\delta \nu$$<$0.04 MHz) with an individual S/N exceeding 3.5 in both XX and YY polarizations. Final candidates are confirmed using transit information recorded by the 19 beams of FAST.

\subsection{Absorption Characterization}
\label{sec:theory}

Data from the follow-up observations were utilized to derive the physical properties of the absorption. The spectra of \hi\ absorption lines were calibrated using a built-in noise diode, which was activated for 1 second every 8 seconds. To remove the baseline, a polynomial function was fitted to the absorption-free regions of the spectrum. For each \hi\ absorption system, we modeled the follow-up spectra using Gaussian functions, varying the number of components, and determined the best-fitting model by calculating the Bayesian Information Criterion (BIC; \citealt{1978AnSta...6..461S,24ce203a-855a-3aa9-952f-976d23b28943}). The model with the lowest BIC was selected for the final \hi\ absorption profile fitting and the estimation of physical parameters.

The \hi\ column density of the absorbing gas can be determined by integrating the observed absorption over velocity, following the relation given by \cite{2018A&ARv..26....4M}: $N_{\hi}[\mathrm{cm^{-2}}]$=1.82$\times 10^{18} T_{\mathrm{s}}[\mathrm{K}]c^{-1}_{\rm f,\hi}\int \tau_{\rm \hi}(V)dV[\mathrm{km\,s^{-1}}]$, where $\tau(V)$ is the optical depth, $c_{\rm f,\hi}$ is covering factor for \hi\ absorption, and $T_{\rm s}$ is the spin temperature of the \hi\ gas. Similarly, for \oh\ absorption, the column density can be expressed as: $N_{\rm{OH}}[\mathrm{cm^{-2}}]$=$X\times T_{\mathrm{ex}}[\mathrm{K}] c^{-1}_{\rm f,OH} \int \tau_{\rm OH}(V)dV[\mathrm{km\,s^{-1}}]$, where $T_{\rm ex}$ denotes the excitation temperature of the corresponding \oh\ transition, and $c_{\rm f,OH}$ is covering factor for \oh\ absorption. The constant $X$ varies depending on the specific transition: for the \oh\ 1667 MHz transition, $X$=2.38$\times$10$^{14}$ \citep{2008MNRAS.391..765C}, while for the \oh\ 1612 MHz transition, $X$=2.14$\times$10$^{15}$ \citep{2019ApJS..245....3G}. 
  
Under the assumption that $T_{\rm s} \ll c_{\rm f}T_{\rm c}$, the optical depth $\tau(V)$ can be expressed as: $\tau(V)$=$-\mathrm{ln}(1+\Delta T(V)/(c_{\rm f}T_{\rm c}))$. $T_{\rm c}$ is the brightness temperature of the background source, which, in this work, is deduced from the line-free parts of the spectrum. $\Delta T(V)$ is the difference between the signal of absorbing gas and that of the background source. Throughout this paper we take the assumption that $T_{\rm s} \ll c_{\rm f}T_{\rm c}$, $c_{\rm f,OH}$=1 and $c_{\rm f,\hi}$=1 \citep{2017A&A...604A..43M} for the calculation of $\tau$ and $N_{\hi}$. The new absorption will be classified as ``associated absorption'' if the velocity difference between the absorption and the background is within 1500 $\kms$.

\subsection{Infrared and Optical Counterparts}

To determine the type of detected \hi\ absorption (associated or intervening) and to study their physical properties, we cross-match our detected \hi\ absorption systems with the catalogs from the Wide-field Infrared Survey Explorer (WISE; \citealt{2010AJ....140.1868W}) and the Sloan Digital Sky Survey (SDSS; \citealt{2000AJ....120.1579Y}). The WISE and SDSS counterparts of the \hi\ absorption systems are identified using the NASA/IPAC Extragalactic Database (NED; \citealt{1991ASSL..171...89H}). When available, the WISE magnitudes and continuum map for each \hi\ absorption system are retrieved from the NASA/IPAC Infrared Science Archive\footnote{\url{https://irsa.ipac.caltech.edu/frontpage/} (IRSA), while the SDSS magnitudes and images are retrieved from the SDSS SkyServer website\footnote{\url{https://skyserver.sdss.org/dr16/en/home.aspx}}. These WISE and SDSS data products are presented below.} For NVSS J053538\allowbreak+643141 that lies outside the SDSS footprint, we extract its optical information from the Panoramic Survey Telescope and Rapid Response System (Pan-STARRS; \citealt{2016arXiv161205560C}).

\section{Verified \hi\ Absorption Systems}
\label{sec:Absorption_Systems}

During follow-up observations in 2024 and 2026, seven \hi\ absorption systems were confirmed, including four previously identified absorbers and three newly discovered ones. For each system, we present the absorption spectrum and its Gaussian fit. Additionally, we provide an image centered on the radio source, created using radio data at the S-band from the Karl G. Jansky Very Large Array Sky Survey (VLASS; \citealt{2020PASP..132c5001L}), complemented by infrared maps from WISE and optical maps from SDSS or Pan-STARRS. The observed properties of the radio source corresponding to each absorption are summarized in Table~\ref{radiosource_table}. The infrared and optical magnitudes of the WISE and SDSS counterparts are listed in Tables~\ref{WISE_counterpart_table} and \ref{SDSS_counterpart_table}, respectively. Table~\ref{HI_absorption_table} and Table~\ref{HI_absorption_gaucomp_table} present the measurements of \hi\ absorption signals. For each source, the total integrated flux density, $\int S_{\hi}dv$, and the total integrated \hi\ optical depth, $\int \tau dv$, were obtained by direct velocity integration over the observed absorption profile and the corresponding optical depth profile, respectively.

\begin{table*}
    %\fontsize{8}{10}\selectfont
	\centering
	\caption{Some observed parameters of the associated/background radio sources. The positional values and the redshifts for the radio sources are obtained from NED. Except for NVSS J080601\allowbreak+190611, for which we adopt the continuum flux density measured from WSRT observations, the continuum flux densities of the other sources are measured from the line-free regions adjacent to the redshifted \hi\ 21-cm frequency in our calibrated spectra. The redshift of the \hi\ absorption line is determined from the Gaussian fit to the absorption profile. The associated, intervening, and unknown types are labeled with stars, diamonds, and crosses, respectively.}
	\label{radiosource_table}
	\begin{tabular}{|c|c|c|c|c|c|c|}
	\hline
        Radio Source & Source Type & Ra(J2000) & Dec(J2000) & Redshift & Redshift$_{\hi}$ & $S_{\nu}$ (mJy)\\
	  \hline
        NVSS J010015+201710$^{\diamond}$ &  QSO & 01h 00m 14.92s & +20d 17m 10.23s & - & 0.265594 & 171.6 $\pm$ 2.9 \\
        \hline
        NVSS J024516+240535$^{\diamond}$ & QSO & 02h 45m 16.86s & +24d 05m 35.17s & 2.245 & 0.141344 & 507.4 $\pm$ 8.1 \\
        \hline
        NVSS J053538+643141$^{\otimes}$ &  Radio Galaxy & 05h 35m 39.05s & +64d 31m 41.03s & - & 0.080328 & 57.8 $\pm$ 1.9 \\
        \hline
        NVSS J080601+190611$^{\star}$ &  Radio Galaxy & 08h 06m 01.54s & +19d 06m 14.62s & 0.097882 & 0.098045 & 142.0 \\
        \hline
        NVSS J090150+030422$^{\star}$ &  Radio Galaxy & 09h 01m 50.98s & +03d 04m 22.70s & 0.287544 & 0.287150 & 396.4 $\pm$ 6.6 \\
        \hline
        NVSS J150034+364844$^{\star}$ & Radio Galaxy & 15h 00m 34.56s & +36d 48m 44.80s & 0.066 & 0.06618 & 65.4 $\pm$ 1.4 \\
        \hline
        NVSS J234106+001833$^{\star}$ & QSO & 23h 41m 06.91s & +00d 18m 33.34s & 0.276720 & 0.276988 & 414.8 $\pm$ 6.7\\
   
        \hline
        \end{tabular}
\end{table*}

\begin{table*}
    %\fontsize{8}{10}\selectfont
	\centering
	\caption{WISE magnitudes for the WISE counterparts of background radio sources (WISE magnitudes for foreground sources of intervening systems, if available, are presented in the context).}
	\label{WISE_counterpart_table}
	\begin{tabular}{|c|c|c|c|c|c|c|}
	\hline
        Radio Source & WISE Counterpart & W1 & W2 & W3 & W4 & WISE color-color classification \\
	  \hline
        NVSS J010015+201710$^{\diamond}$ &  WISEA J010014.92+201710.2 & 15.540 & 14.747 & 11.577 & $<$8.798 & QSOs \\
        \hline
        NVSS J024516+240535$^{\diamond}$ & WISEA J024516.83+240535.0 & 15.614 & 14.541 & 11.408 & 8.857 & QSO \\
        \hline
        NVSS J053538+643141$^{\otimes}$ &  WISEA J053539.05+643141.0 & 14.442 & 13.300 & 9.880 & 6.780 & Seyferts \\
        \hline
        NVSS J080601+190611$^{\star}$ &  WISEA J080601.54+190614.6 & 12.403 & 12.158 & 10.080 & 7.846 & Spirals \\
        \hline
        NVSS J090150+030422$^{\star}$ &  WISEA J090150.97+030422.7 & 14.441 & 13.553 & 10.310 & 7.638 & Seyferts \\
        \hline
        NVSS J150034+364844$^{\star}$ & WISEA J150034.59+364845.1 & 12.739 & 12.634 & 9.884 & 8.063 & Spirals \\
        \hline
        NVSS J234106+001833$^{\star}$ & WISEA J234106.92+001833.1 & 13.160 & 12.374 & 9.771 & 7.497 & QSOs \\
        \hline
        \end{tabular}
\end{table*}

\begin{table*}
    %\fontsize{8}{10}\selectfont
	\centering
	\caption{SDSS optical magnitudes for the optical counterparts of background sources (optical magnitudes for foreground sources of intervening systems, if available, are presented in the context).}
	\label{SDSS_counterpart_table}
	\begin{tabular}{|c|c|c|c|c|c|c|c|}
	\hline
        Radio Source & Optical Counterpart & $u$ & $g$ & $r$ & $i$ & $z$ & $g-r$\\
	  \hline
        
        NVSS J010015+201710$^{\diamond}$ &  SDSS J010015.03+201709.9 & 21.53 & 20.76 & 20.60 & 20.49 & 20.17 & 0.16\\
        \hline
        NVSS J024516+240535$^{\diamond}$ & Pan-STARRS 136910413201582279 & - & 20.12 & 19.93 & 19.57 & 19.35 & 0.19\\
        \hline
        NVSS J053538+643141$^{\otimes}$ &  Pan-STARRS 185430839123424385 & - & 19.89 & 19.47 & 18.71 & 18.71 & 0.42\\
        \hline
        NVSS J080601+190611$^{\star}$ &  SDSS J080601.51+190614.7 & 18.57 & 16.58 & 15.61 & 15.14 & 14.78 & 0.97\\
        \hline
        NVSS J090150+030422$^{\star}$ &  SDSS J090150.96+030423.0 & 20.13 & 19.06 & 18.19 & 17.84 & 17.44 & 0.87\\
        \hline
        NVSS J150034+364844$^{\star}$ & SDSS J150034.56+364845.1 & 18.11 & 16.37 & 15.55 & 15.12 & 14.77 & 0.82\\
        \hline 
        NVSS J234106+001833$^{\star}$ &  SDSS J234106.89+001833.3 & 19.69 & 18.34 & 17.30 & 16.85 & 16.35 & 1.04\\
        \hline
        \end{tabular}
\end{table*}

\subsection{Previously Known \hi\ Absorbers}

\subsubsection{NVSS J080601+190611}

NVSS J080601\allowbreak+190611 is an extended radio source. It has been little studied individually before. The WISE counterpart of NVSS J080601\allowbreak+190611 is WISEA J080601.54\allowbreak+190614.6, it has W1-W2=0.245, W2-W3=2.078, located in the spiral galaxy region of the WISE color-color diagram \citep{2010AJ....140.1868W}. The W1-W2 color of WISEA J080601.54\allowbreak+190614.6 is $<$0.8, suggesting no AGN presence \citep{2012ApJ...753...30S}. Its associated \hi\ absorption was first detected by \citet{2015A&A...575A..44G}, who presented an analysis of the \hi\ 21 cm absorption in a sample of 101 flux-selected radio AGN (S$_{\rm1.4GHz}$$>$50 mJy) observed with the Westerbork Synthesis Radio Telescope (WSRT). They reported the parameter of absorption as peak optical depth $\tau_{\rm peak} \sim 0.099$ and $N_{\hi}/(T_{\rm{s}}/c_{f,\hi}) \sim 26.9\times$10$^{18}$cm$^{-2}$K$^{-1}$. 

\begin{figure*}[hbt!]
    \centering
    \includegraphics[width=0.25\textwidth]{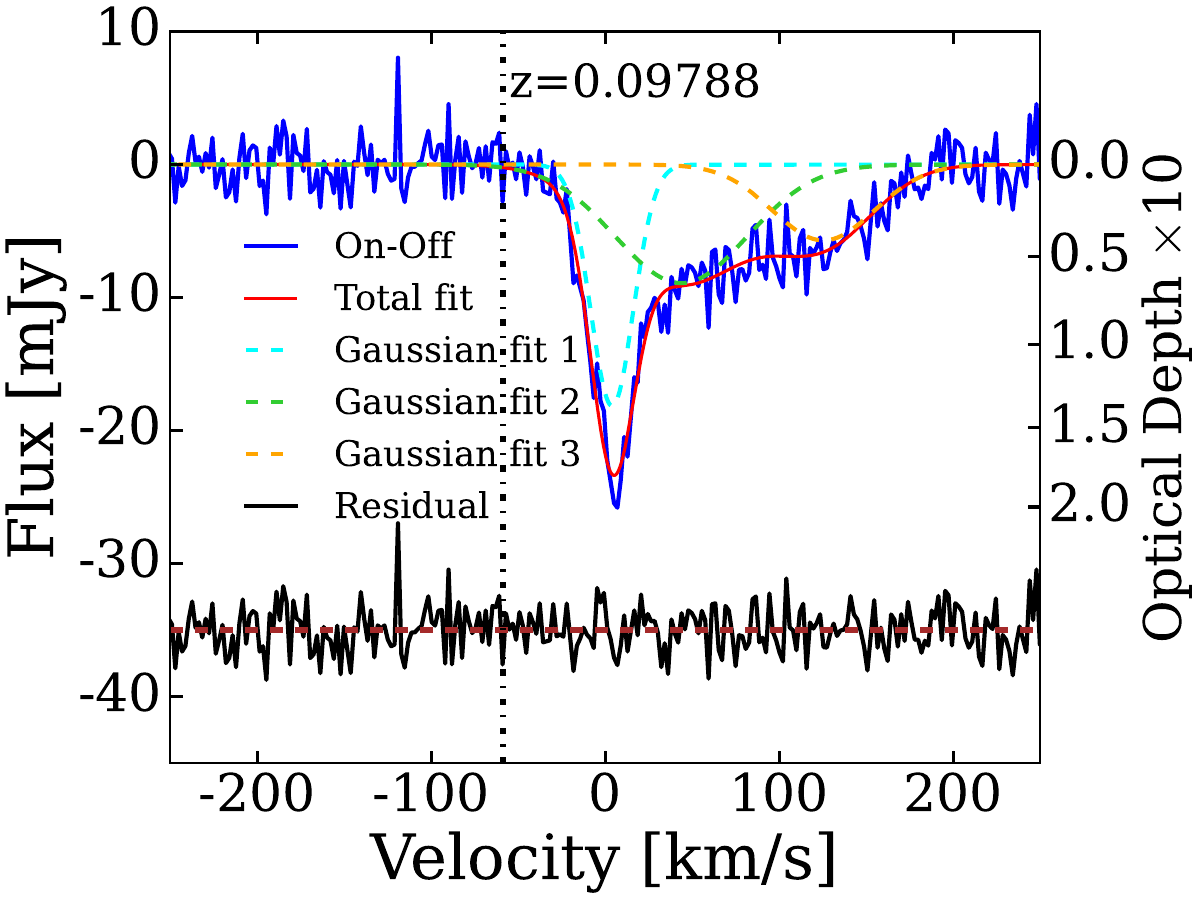}
    \includegraphics[width=0.25\textwidth]{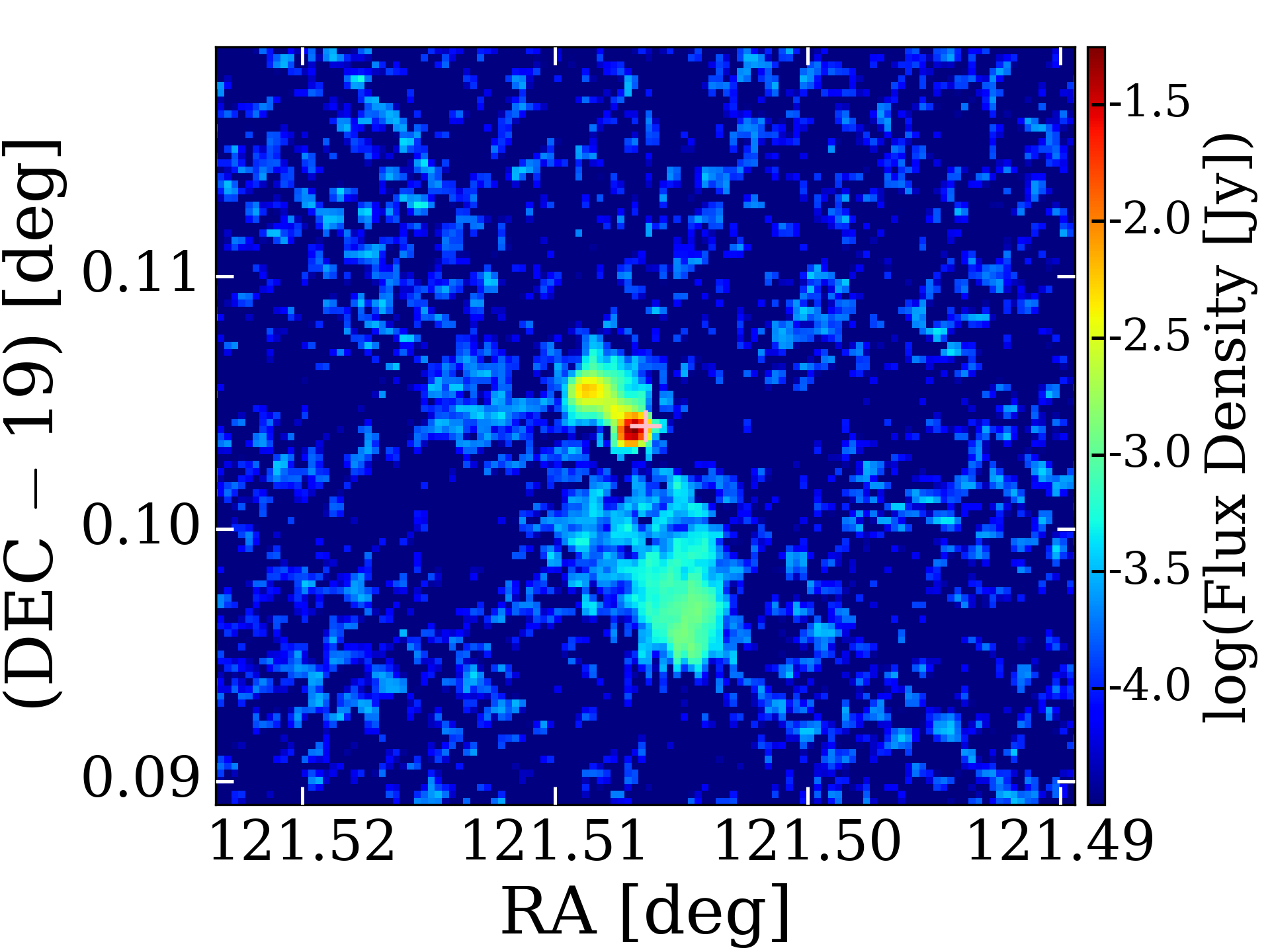}
    \includegraphics[width=0.25\textwidth]{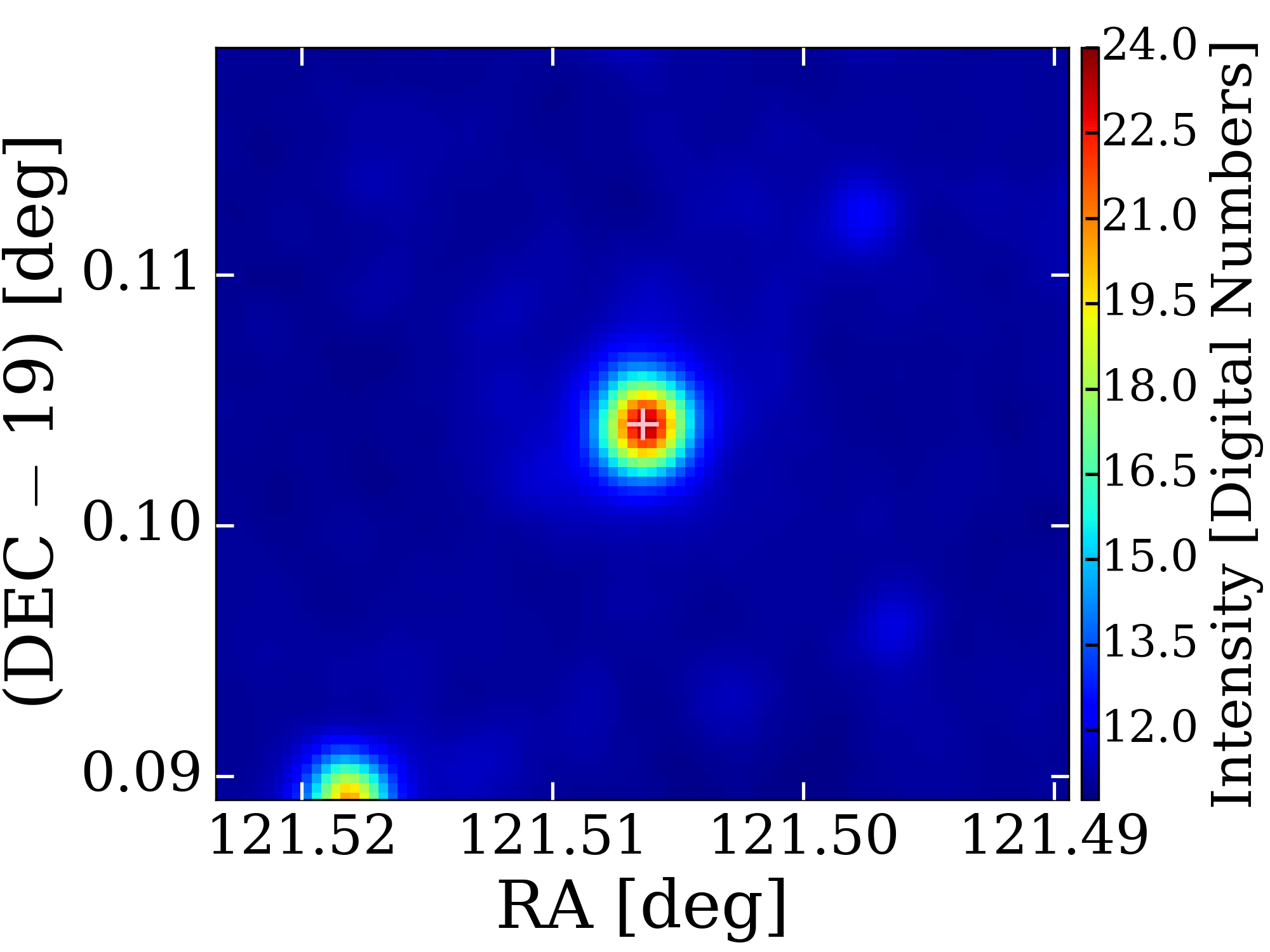}
    \includegraphics[width=0.18\textwidth]{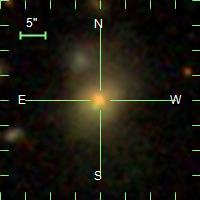}
    \caption{Left: \hi\ absorption feature of NVSS J080601\allowbreak+190611. The blue solid line shows the absorption spectrum, and the red solid line shows the fit with a three-component Gaussian model. The fitting residual is shown as a black-solid line at the bottom. The optical depth value for \hi\ absorption is shown on the right scale. Middle left: radio map from VLASS centered at NVSS J080601\allowbreak+190611. Middle right: W2 band infrared map of WISEA J080601.54\allowbreak+190614.6 (shown as a purple cross) from WISE. Right: SDSS optical map of the optical counterpart of NVSS J080601\allowbreak+190611.}
    \label{NVSS_J080601+190611_fit}
\end{figure*}

We re-detected the absorption profile of NVSS J080601\allowbreak+190611 through blind observations. The \hi\ absorption along with its VLASS radio, WISE infrared, and SDSS optical images are shown in Figure~\ref{NVSS_J080601+190611_fit}. The absorption spectrum is well-modeled using a three-component Gaussian function. Our measurements reveal a flux density depth of $S_{\hi,\rm peak}$=$-23.37 \mJy$, a FWHM of $36.84 \kms$, a peak optical depth of $\tau_{\rm peak}$=$0.18$ and a column density of $N_{\hi}$=$0.24 T_{\rm{s}}\times 10^{20}\cm^{-2}\K^{-1}$. The relatively broad component suggests the presence of unsettled gas in this galaxy. The VLASS image of NVSS J080601\allowbreak+190611 shows two extended areas of radio emission located to the north and south of the core. These features are absent in the infrared and optical maps, indicating that the emission is dominated by non-thermal plasma radiation, likely associated with AGN-driven jets or lobes.

Our independent blind detection of the \hi\ absorption line yielded a consistent line profile that aligns with previous observations. The FAST flux measurement may be overestimated due to its 3-arcminute beam, potentially including contributions from other bright radio components in the vicinity of the target source (as shown in the VLASS map in Figure~\ref{NVSS_J080601+190611_fit}). To mitigate the effects of source confusion within the large FAST beam, we adopt the continuum flux density measured with the WSRT by \citet{2015A&A...575A..44G} for the optical depth and column density calculations.

%However, the continuum flux we derive differs from that reported by \citet{2015A&A...575A..44G}. We measured the flux measured in the line-free region near the absorption, which amounts to 223.50 mJy, while \citet{2015A&A...575A..44G} used a continuum flux of 142 mJy measured by WSRT. The FAST flux measurement may be overestimated due to its 3-arcminute beam, potentially including contributions from other bright radio components in the vicinity of the target source (as shown in the VLASS map in Figure~\ref{NVSS_J080601+190611_fit}). To mitigate the effects of source confusion within the large FAST beam and to estimate the intrinsic continuum flux density at the redshifted \hi\ absorption frequency, we adopted a power-law radio spectrum: $S(\nu)\propto\nu^{a}$, where $S(\nu)$ is the flux density, $\nu$ is the observing frequency, and $a$ is the spectral index. Using the NVSS measurement of $S_{\mathrm{NVSS}}=176.0\pm0.45$ mJy at 1.4 GHz and the VLASS measurement of $S_{\mathrm{VLASS}}=66.17\pm0.54$ mJy at 3 GHz, we derived a spectral index of $a=-1.28\pm0.01$. Extrapolating this spectrum to the redshifted \hi\ absorption frequency gives a continuum flux density of $194.8\pm1.1$ mJy. We therefore use this value in the subsequent calculations of the \hi\ optical depth and column density for \hi\ absorption in NVSS J080601\allowbreak+190611.

\subsubsection{NVSS J090150+030422}
\label{sec:absHI_NVSS_J090150+030422}

The \hi\ absorption associated with NVSS J090150\allowbreak+030422 was first detected by \citet{2016AJ....151...74Y}, who also presented high-resolution radio continuum images and detailed optical and near-infrared (NIR) properties of the source. Using the line ratios of its moderately strong narrow emission, the optical and near-infrared (NIR) spectral energy distribution (SED), the VLBA map, and its gigahertz-peaked spectrum (GPS) SED, \citet{2016AJ....151...74Y} classified NVSS J090150\allowbreak+030422 as an Sc galaxy, likely a LINER or a weak Seyfert ionization source, and possibly a very small Compact Symmetric Object (CSO). Additionally, its optical and NIR images indicate the possibility of an interacting system, with diffuse structures extending northeast of the main galaxy. The WISE counterpart of NVSS J090150\allowbreak+030422 is WISEA J090150.97\allowbreak+030422.7, it has W1-W2=0.888, W2-W3=3.243, located in the spiral galaxy region of the WISE color-color diagram. The W1-W2 color of WISEA J090150.97\allowbreak+030422.7 is $>$0.8, suggesting AGN presence.

\begin{figure*}[hbt!]
    \centering
    \includegraphics[width=0.2\textwidth]{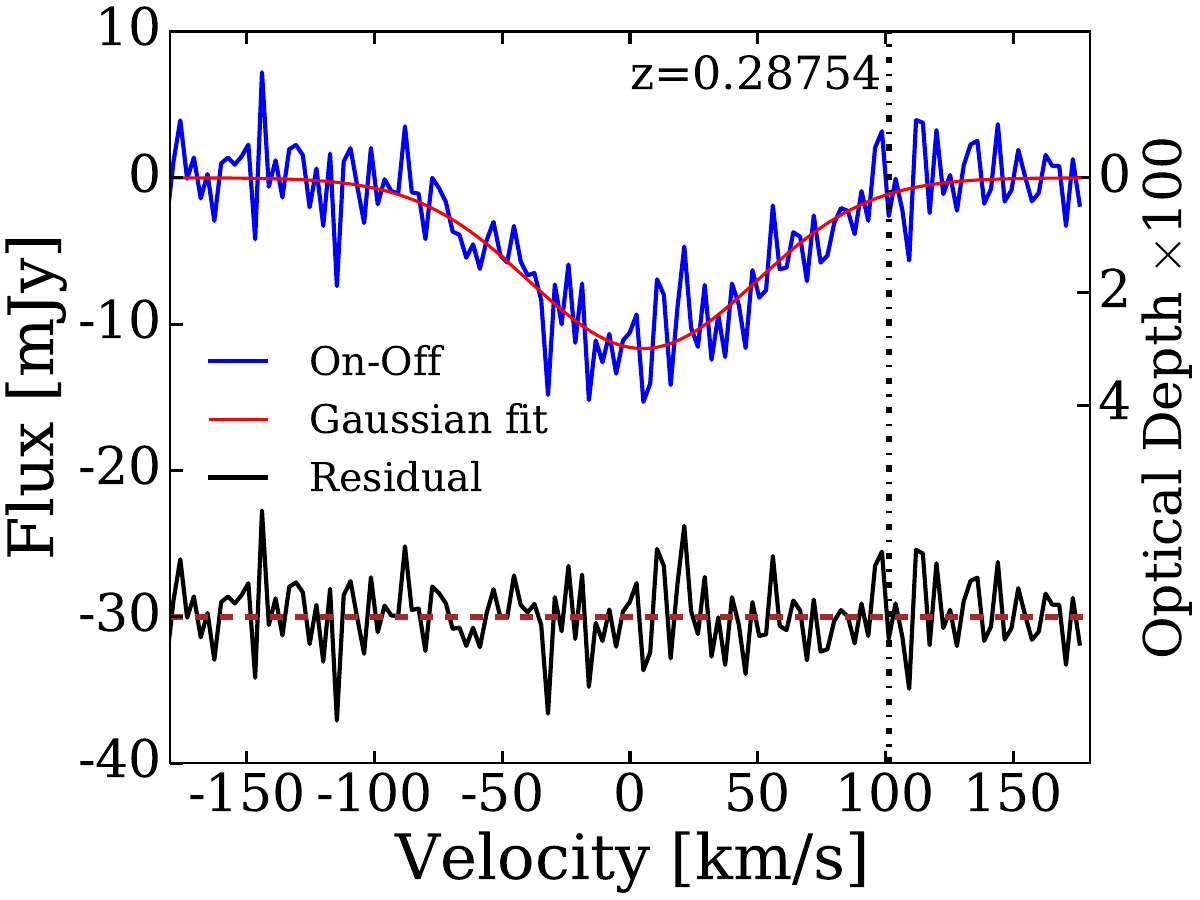}
    \includegraphics[width=0.2\textwidth]{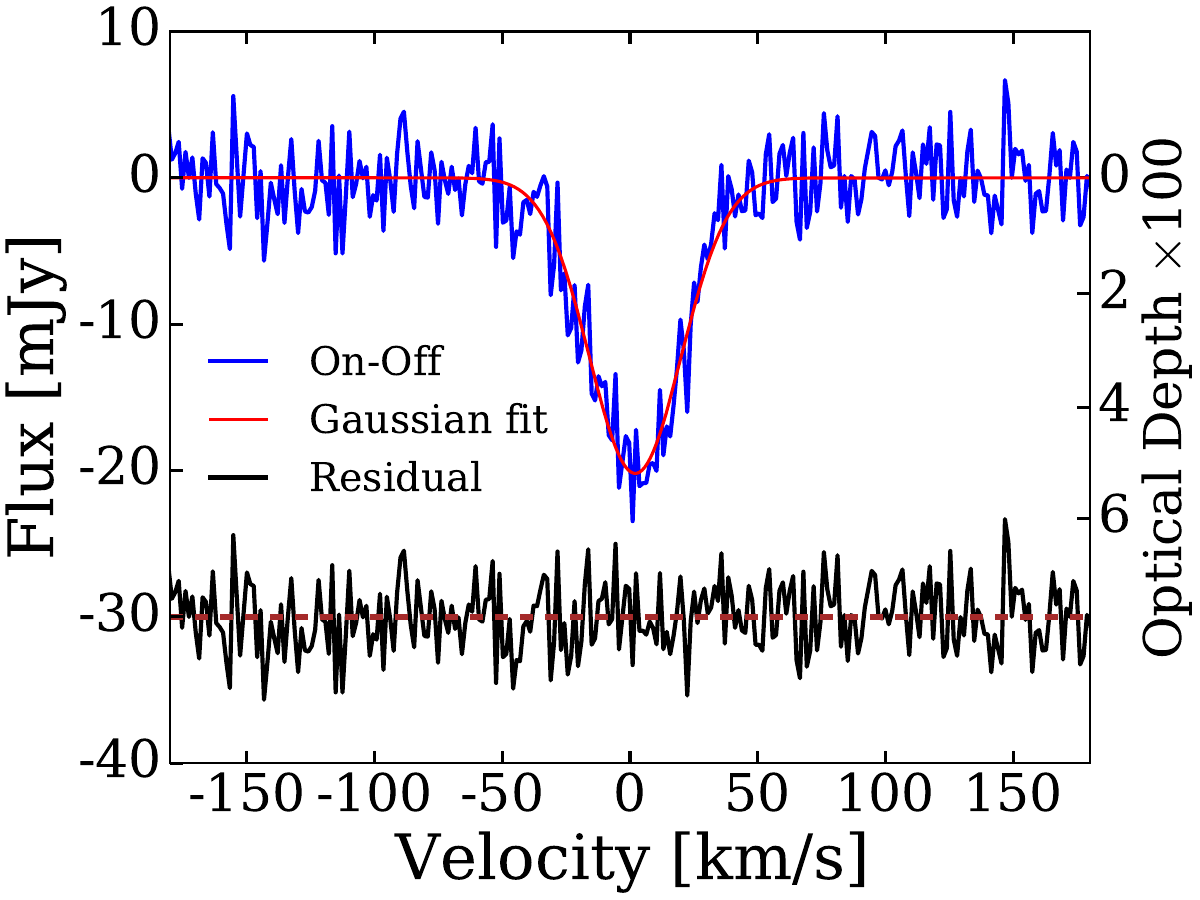}
    \includegraphics[width=0.2\textwidth]{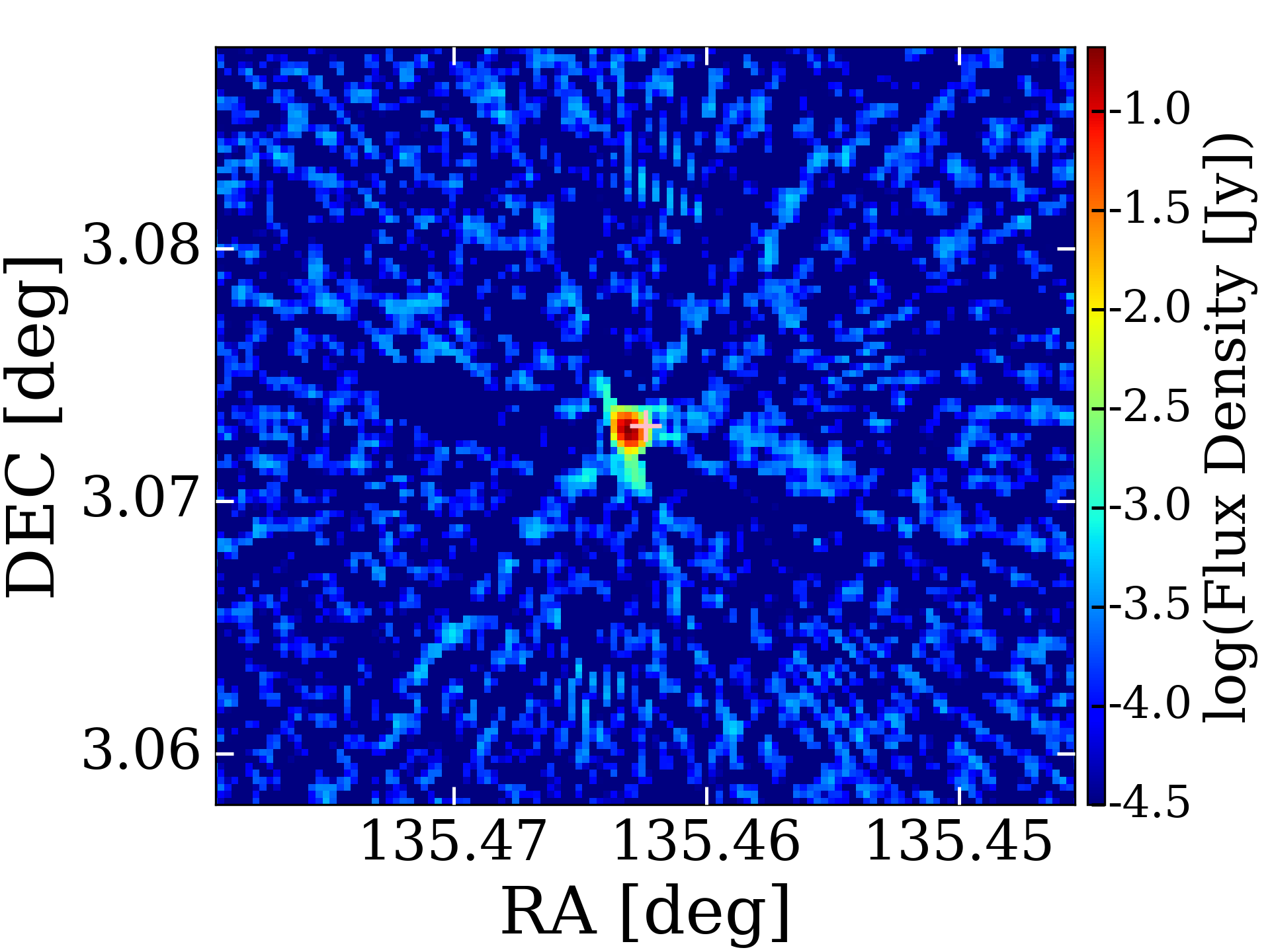}
    \includegraphics[width=0.2\textwidth]{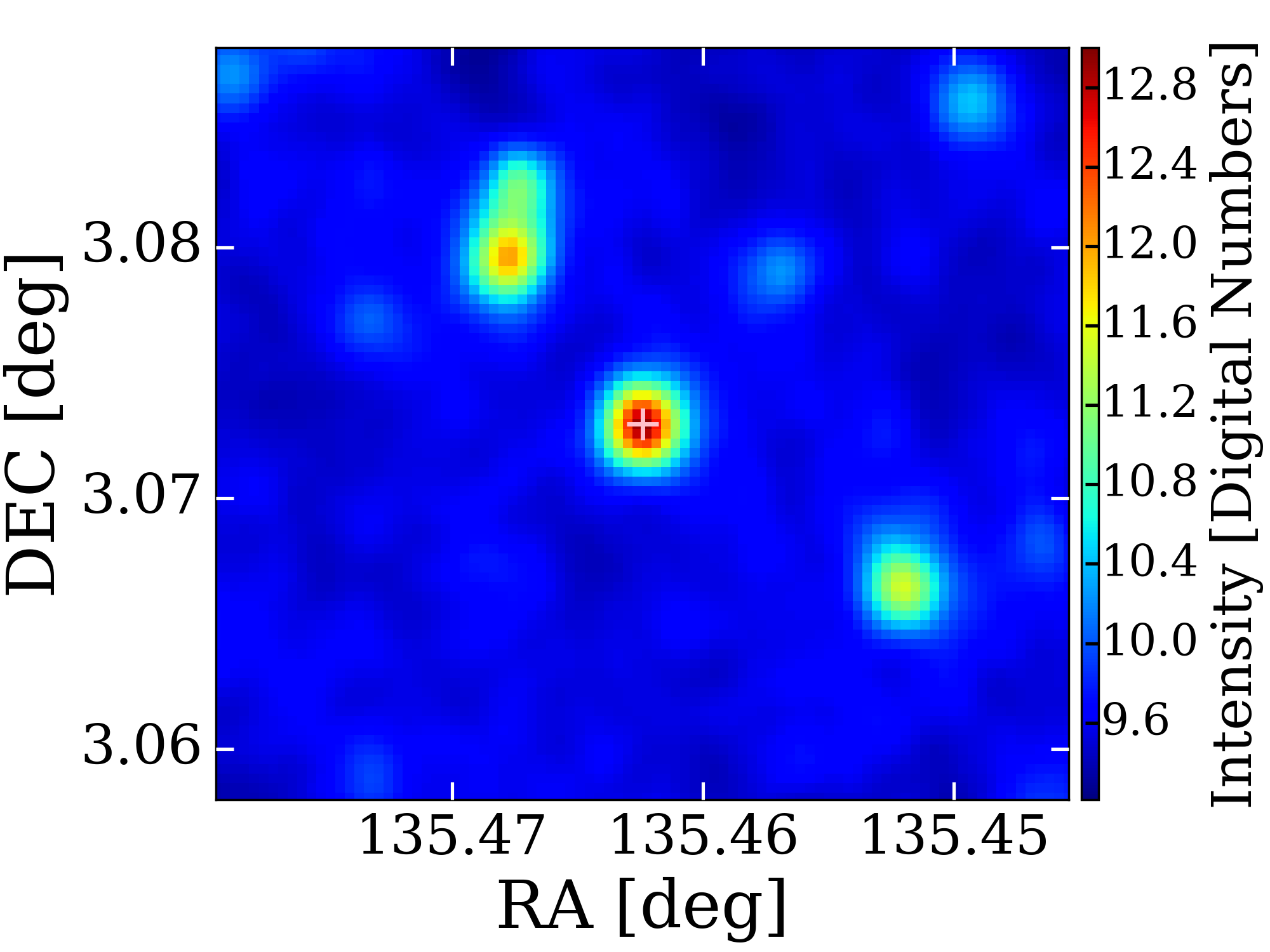}
    \includegraphics[width=0.144\textwidth]{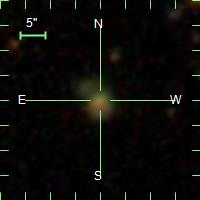}
    \caption{Left and Middle left: same as Figure~\ref{NVSS_J080601+190611_fit}, but for NVSS J090150\allowbreak+030422 and the HVC. Middle : radio map from VLASS centered at NVSS J090150\allowbreak+030422. Middle right: W2 band infrared map of WISEA J090150.97\allowbreak+030422.7 (shown as a purple cross) from WISE. Right: SDSS optical map of the optical counterpart of NVSS J090150\allowbreak+030422.}
    \label{NVSS_J090150+030422_fit}
\end{figure*}

Two \hi\ absorption lines were detected: one with a broad profile at the optical redshift of NVSS J090150\allowbreak+030422, and the other redshifted by approximately $\sim$400\kms. The broad absorption is attributed to disk gas, while the narrow component is likely an infalling high-velocity cloud (HVC). Our independent detection of the \hi\ absorption line, presented in Figure~\ref{NVSS_J090150+030422_fit}, produced a consistent line profile that agrees with previous observations. The absorption spectrum is accurately modeled with a single-component Gaussian function.

\subsubsection{NVSS J150034+364844}

\begin{figure*}[hbt!]
    \centering
    \includegraphics[width=0.25\textwidth]{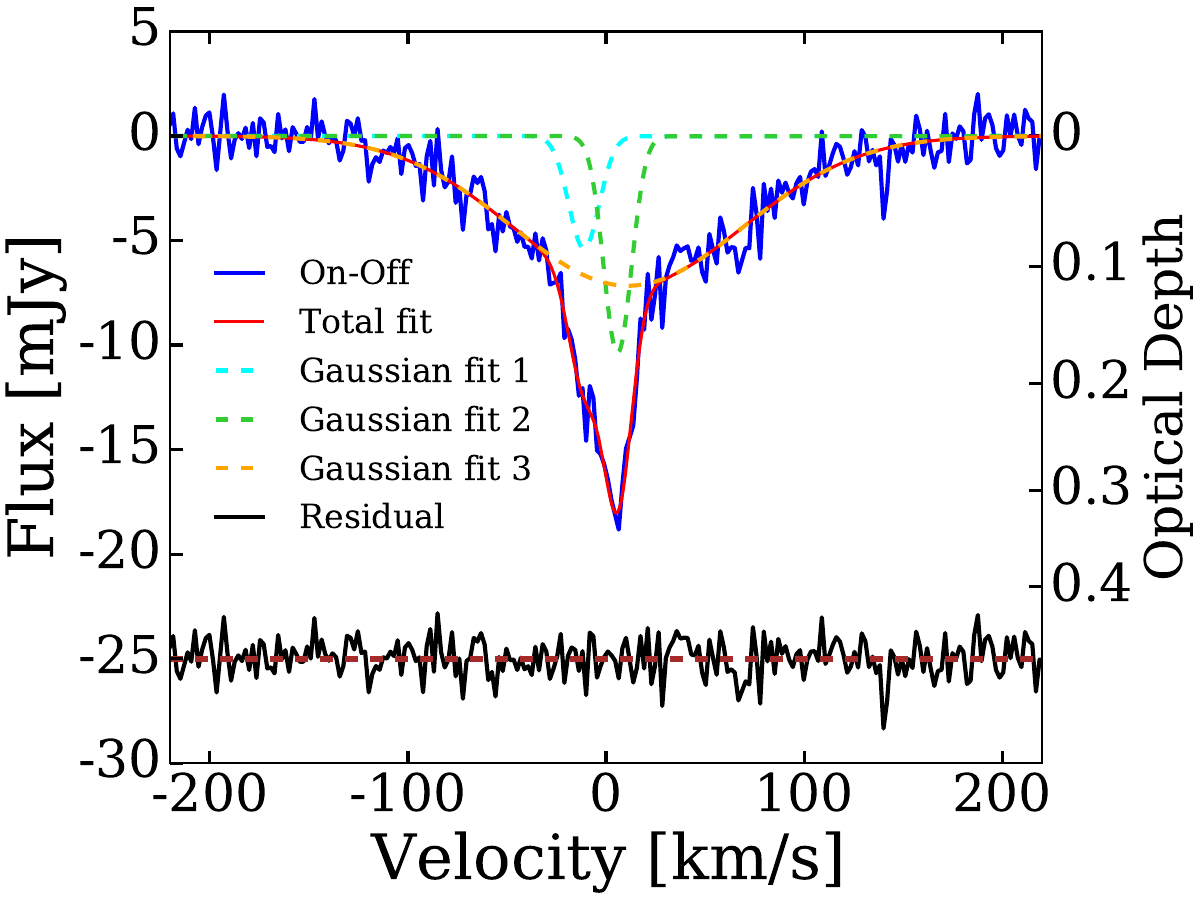}
    \includegraphics[width=0.25\textwidth]{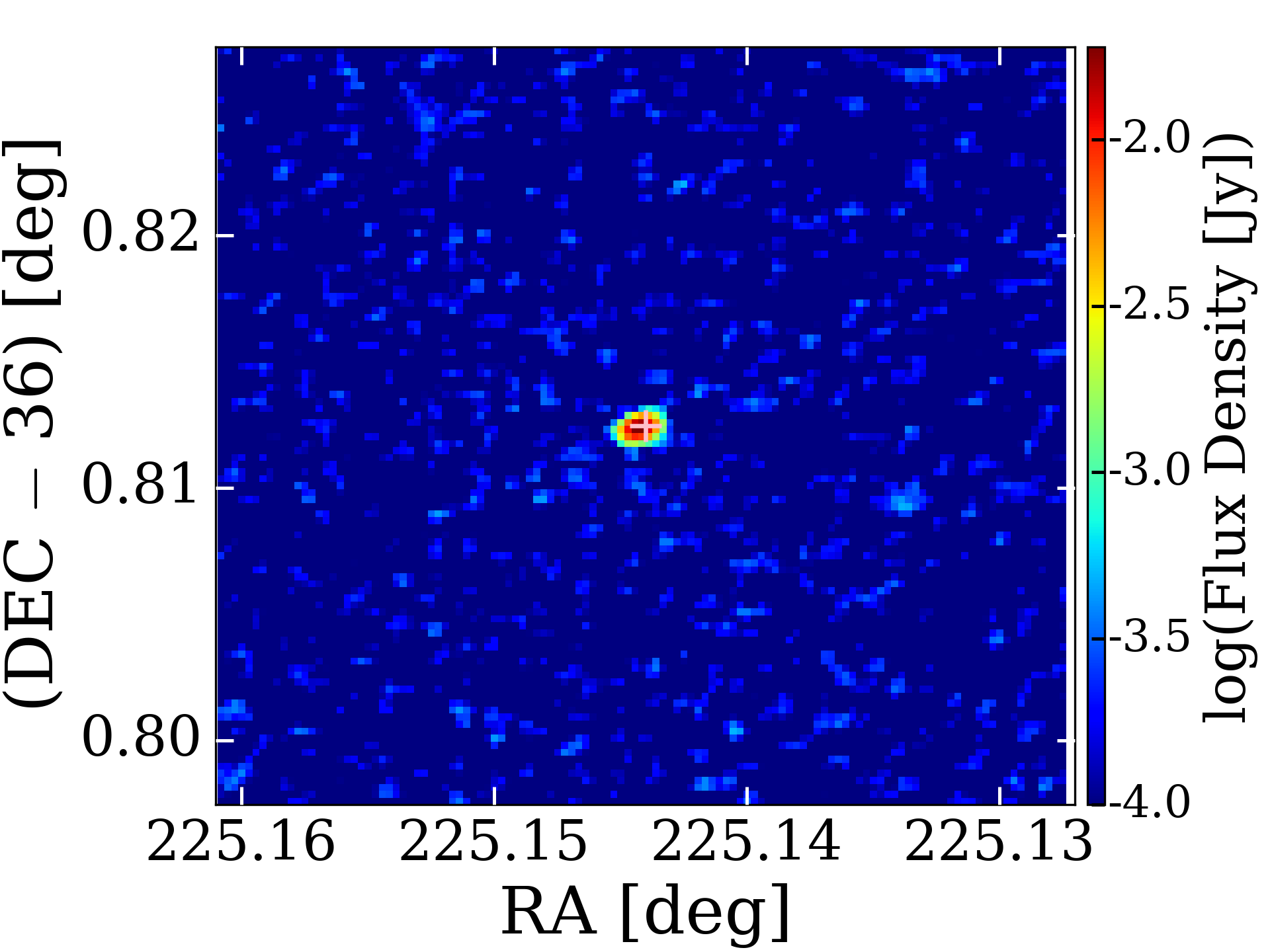}
    \includegraphics[width=0.25\textwidth]{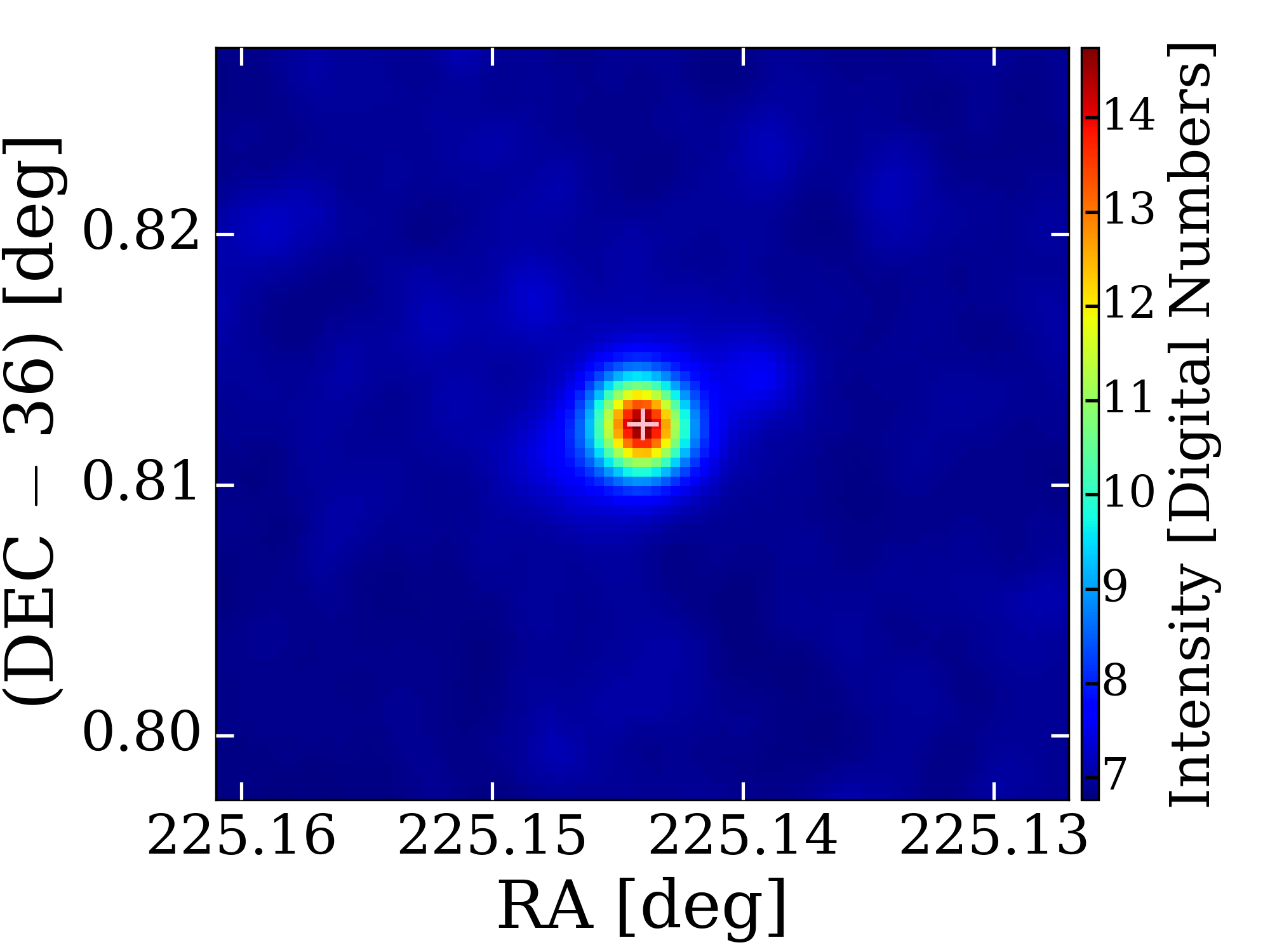}
    \includegraphics[width=0.18\textwidth]{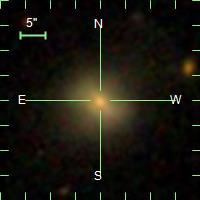}
    \caption{Left: same as Figure~\ref{NVSS_J080601+190611_fit}, but for NVSS J150034\allowbreak+364844. Middle left: radio map from VLASS centered at NVSS J150034\allowbreak+364844. Middle right: W2 band infrared map of WISEA J150034.59\allowbreak+364845.1 (shown as a purple cross) from WISE. Right: SDSS optical map of the optical counterpart of NVSS J150034\allowbreak+364844.}
    \label{NVSS_J150034+364844_fit}
\end{figure*}

NVSS J150034\allowbreak+364844 (optically identified as SDSS J150034.56\allowbreak+364845.1) is a radio galaxy at redshift $z=0.066$ that has received relatively little exploration in the literature. Its \hi 21 cm absorption was firstly detected by
\citet{2015A&A...575A..44G} and verified by \citet{2025ApJS..276....6Z} and this work. Figure~\ref{NVSS_J150034+364844_fit} presents the \hi\ 21-cm absorption detected in this source, together with the VLASS radio continuum image, the WISE infrared image, and the SDSS optical image. The SDSS optical image reveals a bright, compact galaxy with a prominent stellar core and a diffuse, slightly extended envelope. The galaxy appears relatively smooth and roundish, with no obvious signs of major morphological disturbance or strong tidal features at this depth. The SDSS $u,g,r,i,$ and $z$-band magnitudes are 18.11, 16.37, 15.55, 15.12, and 14.77, respectively. The \hi\ 21-cm absorption profile was well modeled with a three-component Gaussian fit. The spectrum displays relatively broad absorption wings on both sides of the line center, which may be indicative of gas inflows/outflows or a component with large velocity dispersion.

\subsubsection{NVSS J234106+001833}

NVSS J234106\allowbreak+001833 is a radio galaxy situated within the galaxy cluster WHL J234106.9+001833 \citep{2012ApJS..199...34W}. It has received limited individual attention in the literature. Absorption associated with this source was detected by \citet{2021A&A...654A..94M}, who conducted a \hi\ 21 cm absorption study of 26 radio-loud AGN using the Karl G. Jansky Very Large Array. The SDSS spectrum of NVSS J234106\allowbreak+001833 reveals significantly broad emission lines, suggesting that it resides in an environment with disturbed kinematics. Based on its position on the `Baldwin, Phillips $\&$ Terlevich' (BPT) diagrams \citep{1981PASP...93....5B}, \citet{2021A&A...654A..94M} identified it as a low-ionization source, located in the composite region between LINERs and star-forming galaxies. Furthermore, its WISE magnitudes place it near the edge of the QSO region in the WISE color-color diagram, indicating its potential association with quasar-like properties. 

\begin{figure*}[hbt!]
    \centering
    \includegraphics[width=0.25\textwidth]{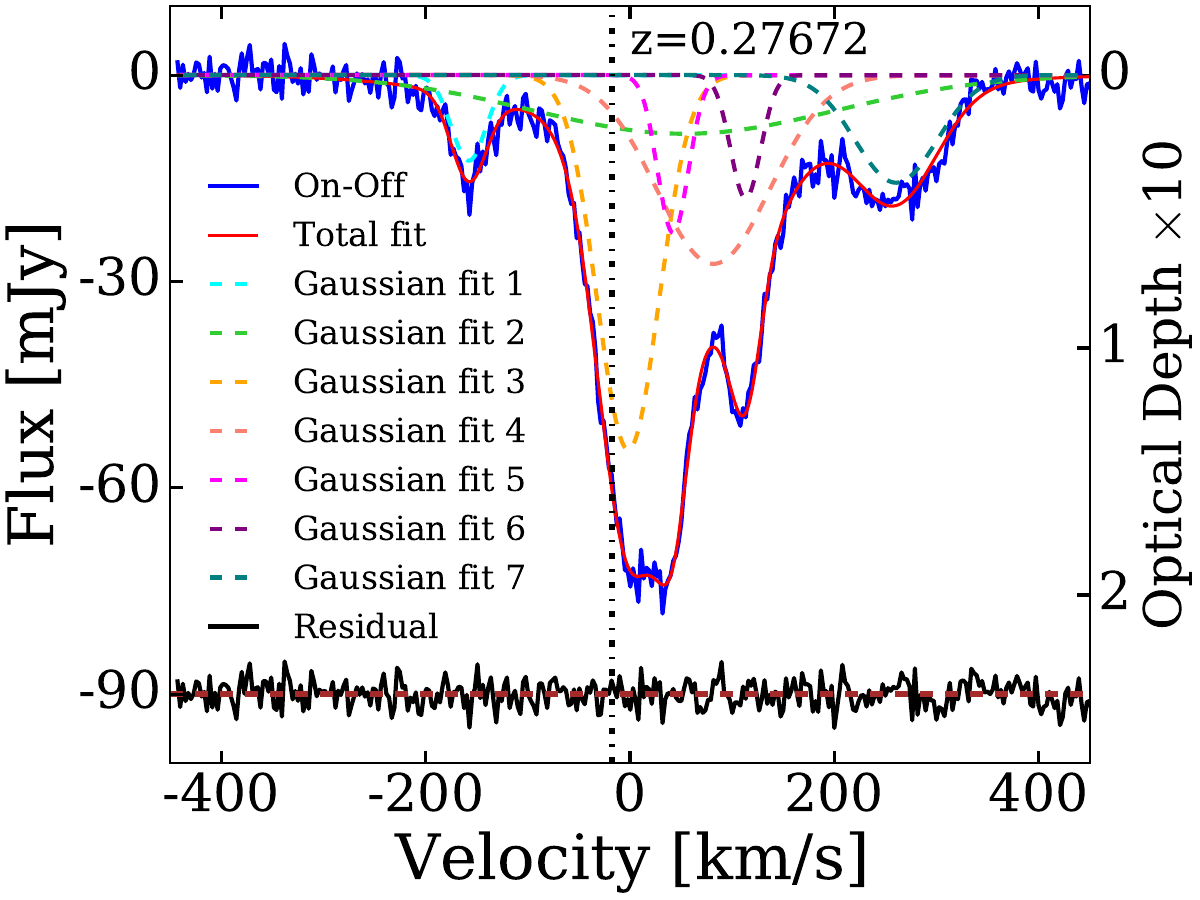}
    \includegraphics[width=0.25\textwidth]{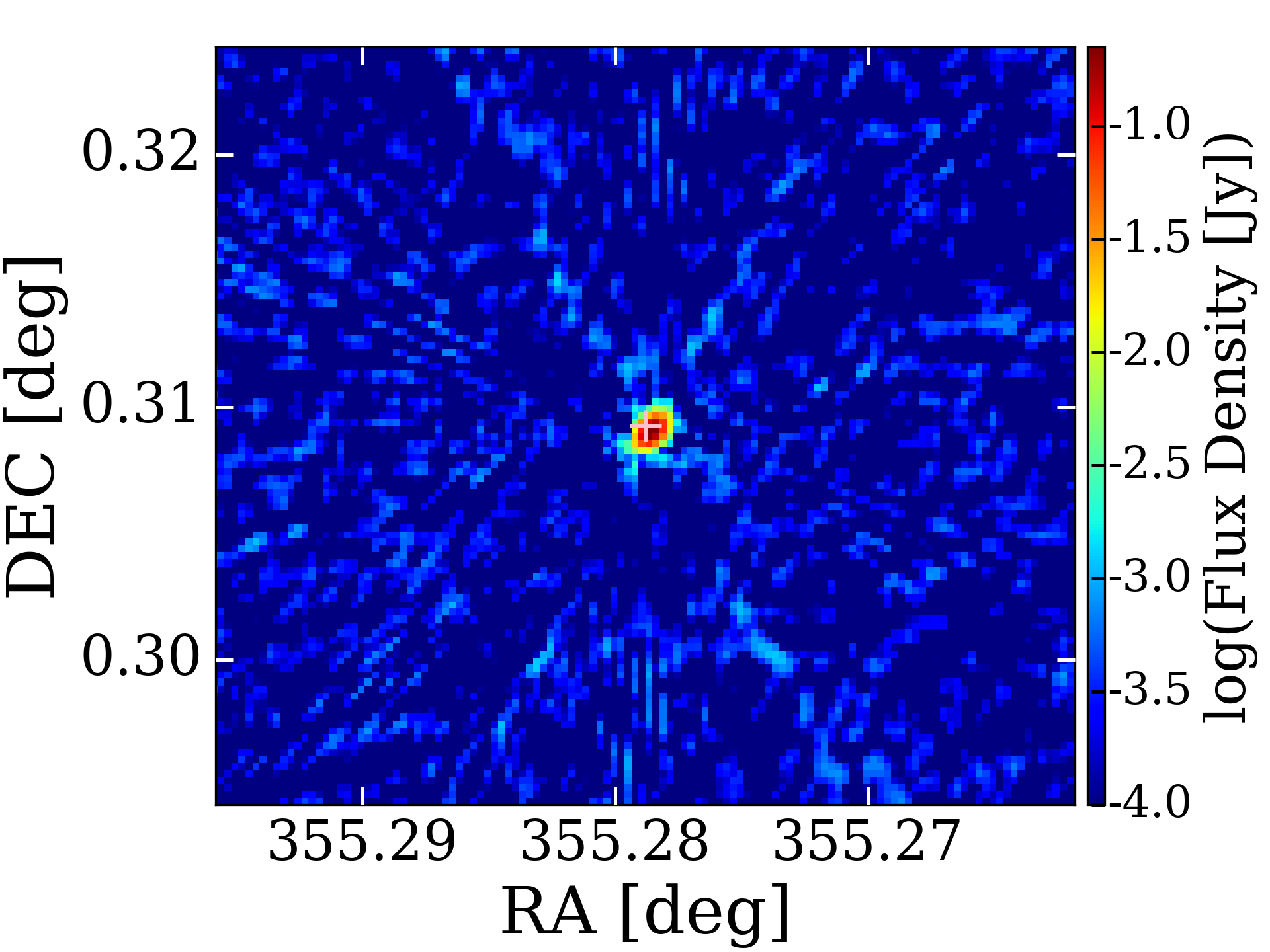}
    \includegraphics[width=0.25\textwidth]{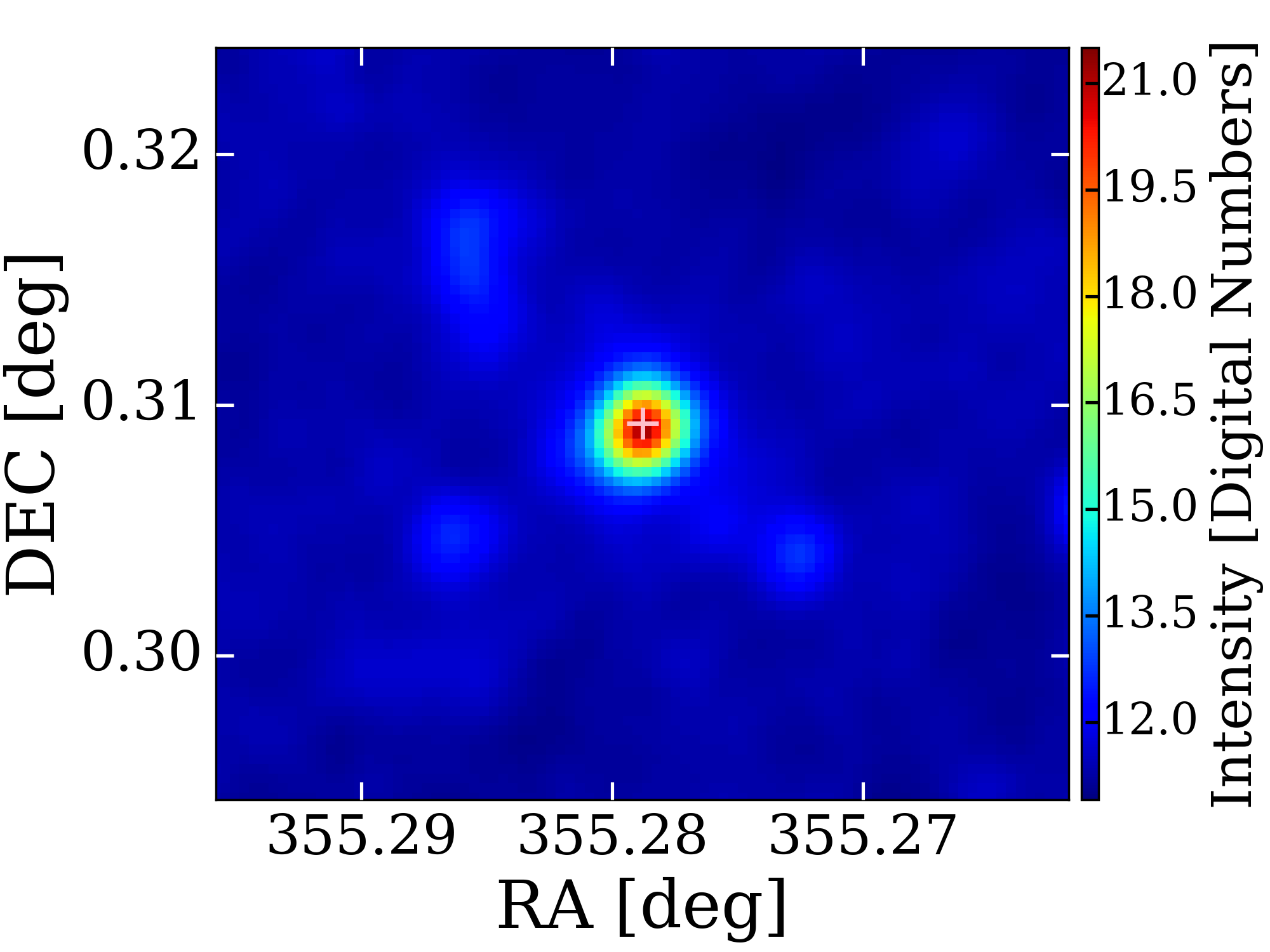}
    \includegraphics[width=0.18\textwidth]{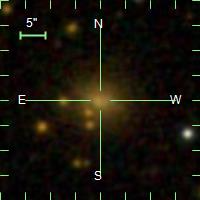}
    \caption{Left: same as Figure~\ref{NVSS_J080601+190611_fit}, but for NVSS J234106\allowbreak+001833. Middle left: radio map from VLASS centered at NVSS J234106\allowbreak+001833. Middle right: W2 band infrared map of WISEA J234106.92\allowbreak+001833.1 (shown as a purple cross) from WISE. Right: SDSS optical map of the optical counterpart of NVSS J234106\allowbreak+001833.}
    \label{NVSS_J234106+001833_fit}
\end{figure*}

Shown in Figure~\ref{NVSS_J234106+001833_fit} are the \hi\ absorption along with VLASS radio, WISE infrared, and SDSS optical images, all centered on NVSS J234106\allowbreak+001833. The \hi\ 21-cm absorption observed in NVSS J234106\allowbreak+001833 exhibits a complex structure, consisting of a prominent deep absorption component, indicative of a gas disk, along with shallow redshifted and blueshifted wings that point to infalling and outflowing gas. This complexity suggests the possible presence of unsettled gas, potentially driven by interactions with other galaxies within the cluster, or the existence of gas clouds within a disc.

%The WISE counterpart of NVSS J234106\allowbreak+001833 is WISEA J234106.92+001833.1, it has W1-W2=0.786, W2-W3=2.603, located in the QSOs region of the WISE color-color diagram.

\subsection{New \hi\ Absorbers}

\subsubsection{NVSS J010015+201710}

\begin{figure*}[hbt!]
    \centering
    \includegraphics[width=0.25\textwidth]{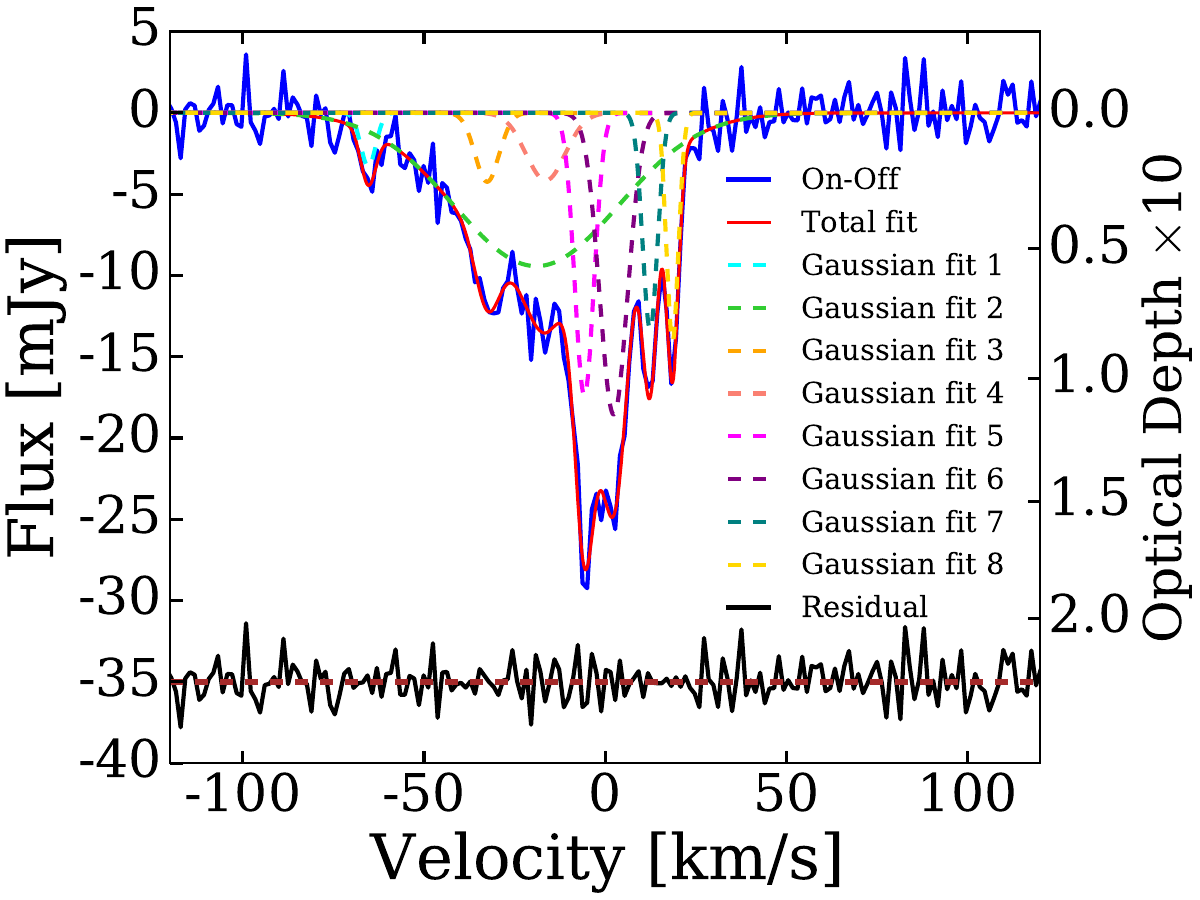}
    \includegraphics[width=0.25\textwidth]{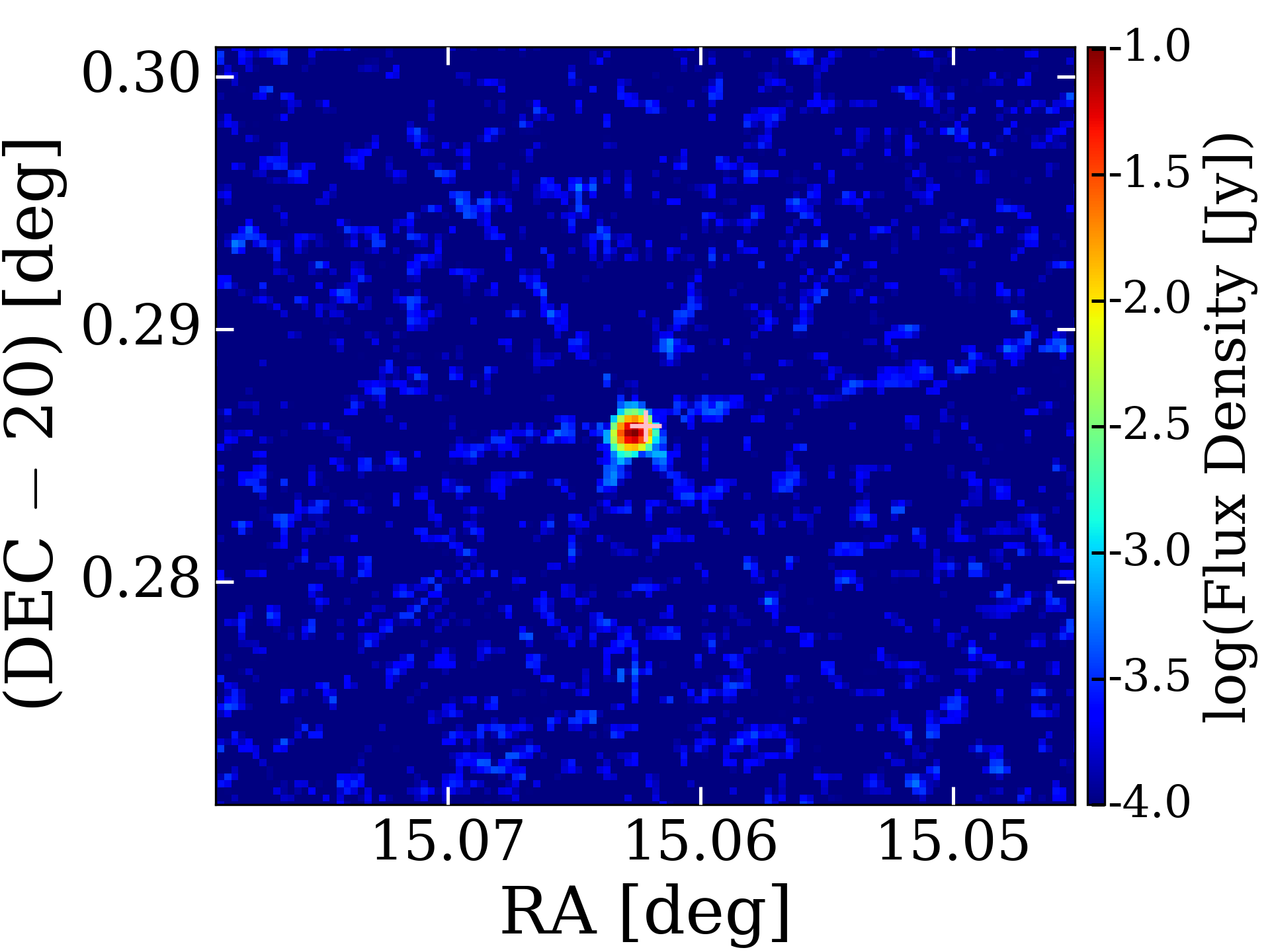}
    \includegraphics[width=0.25\textwidth]{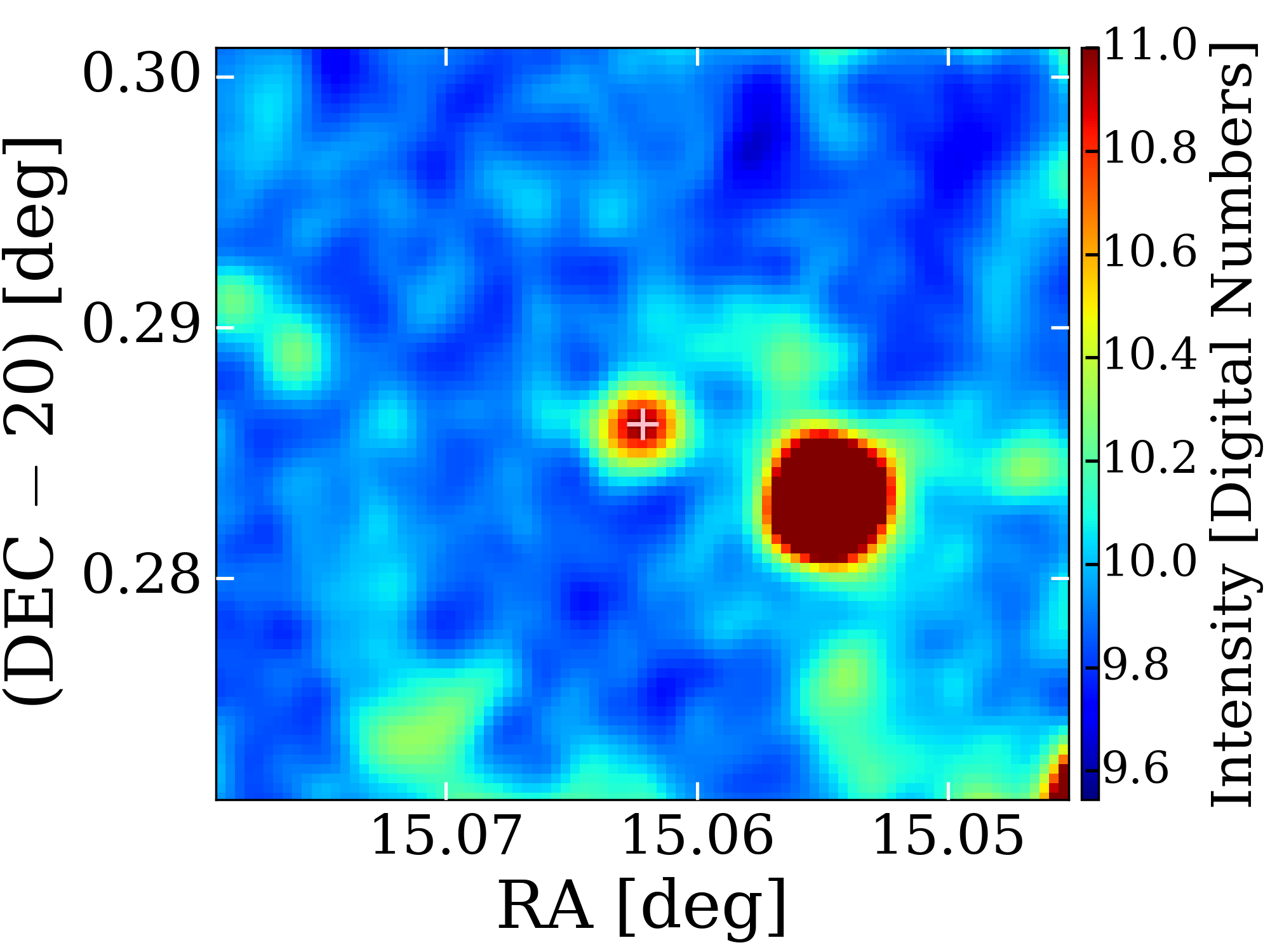}
    \includegraphics[width=0.18\textwidth]{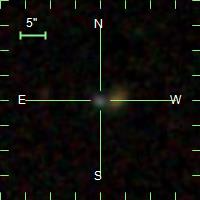}
    \caption{Left: same as Figure~\ref{NVSS_J080601+190611_fit}, but for NVSS J010015\allowbreak+201710. Middle left: radio map from VLASS centered at NVSS J010015\allowbreak+201710. Middle right: W2 band infrared map of WISEA J010014.92\allowbreak+201710.2 (shown as a purple cross) from WISE. Right: optical map obtained from the SDSS, showing the optical counterpart of NVSS J010015\allowbreak+201710.}
    \label{NVSS_J010015+201710_fit}
\end{figure*}

NVSS J010015\allowbreak+201710 is a QSO that has been relatively unexplored in the literature. Figure~\ref{NVSS_J010015+201710_fit} presents the \hi\ absorption toward NVSS J010015\allowbreak+201710, along with the VLASS radio image, WISE infrared image, and SDSS optical image of the same source. The optical image reveals a galaxy located to the west of NVSS J010015\allowbreak+201710. Although no SDSS spectrum is available for this galaxy, its photometric redshift is determined to be 0.209$\pm$0.0645, consistent with the redshift of the detected \hi\ absorption ($z$=0.265) within the margin of error. This suggests that the galaxy is likely the foreground object responsible for the absorption. The SDSS $u,g,r,i$ and $z$-band magnitudes for this galaxy are 22.16, 20.63, 19.64, 19.24, and 19.00, respectively. With a $g-r$ color of 0.99, it is likely classified as a red galaxy.

The \hi\ 21-cm absorption towards NVSS J010015\allowbreak+201710 exhibits a complex profile, which we modeled using an eight-component Gaussian function. The spectrum reveals relatively broad absorption on the blueward side of the line, potentially indicative of gas outflows. Additionally, the peaks on the redward side suggest the presence of infalling gas.

%The WISE counterpart of NVSS J010015+201710 is WISEA J010014.92+201710.2, it has W1-W2=0.793, W2-W3=3.17, located in the QSOs region of the WISE color-color diagram.

\subsubsection{NVSS J024516+240535}

\begin{figure*}[hbt!]
    \centering
    \includegraphics[width=0.25\textwidth]{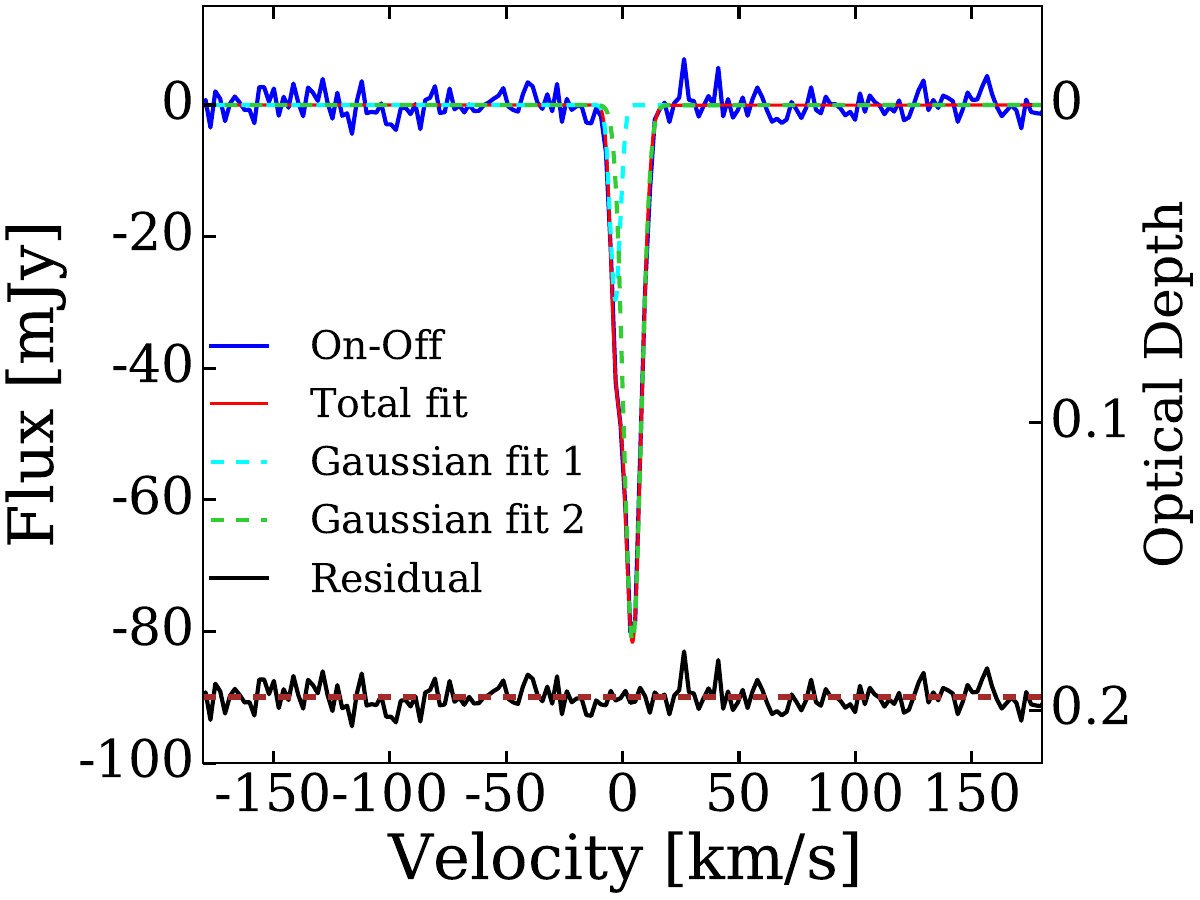}
    \includegraphics[width=0.25\textwidth]{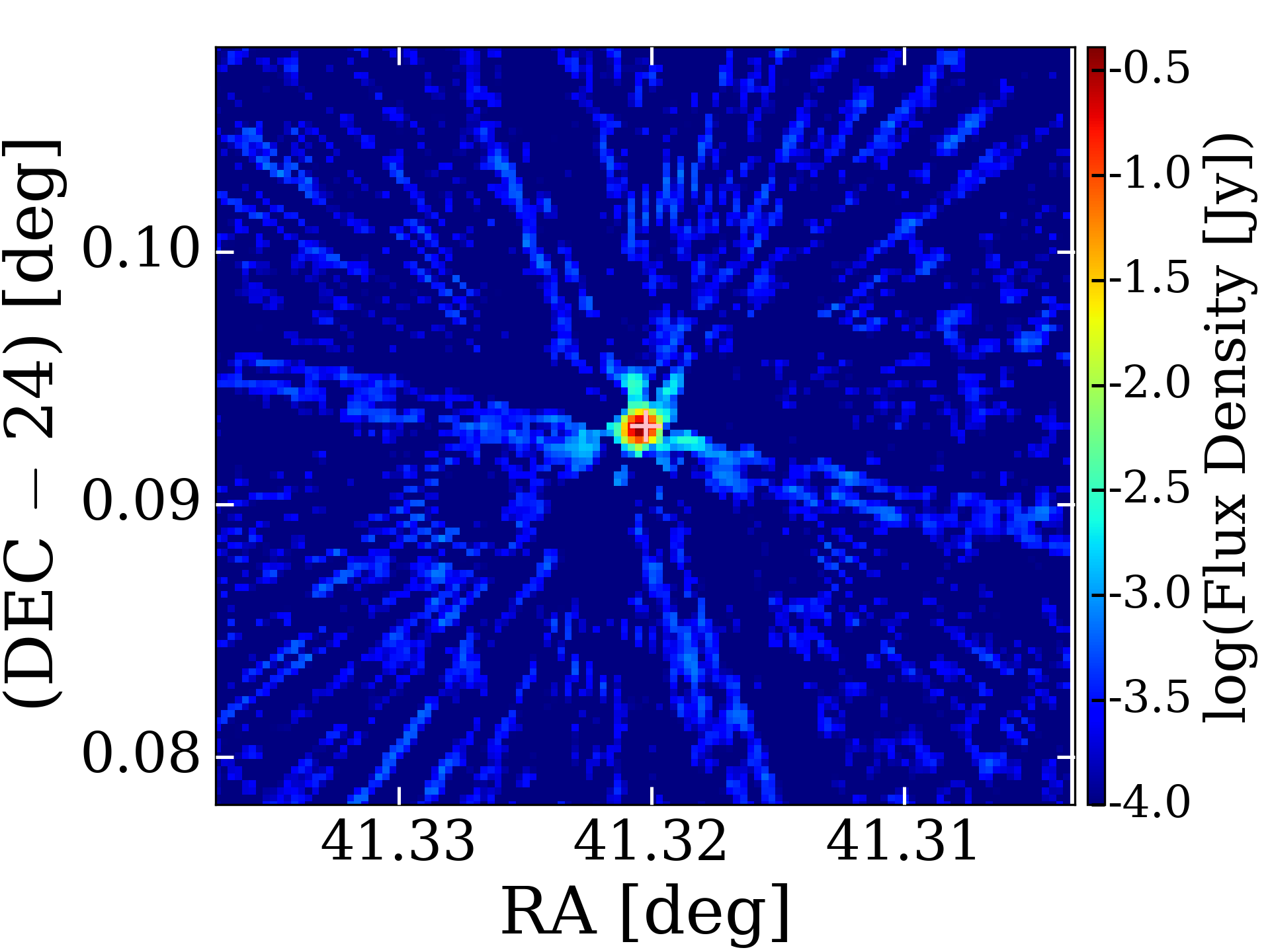}
    \includegraphics[width=0.25\textwidth]{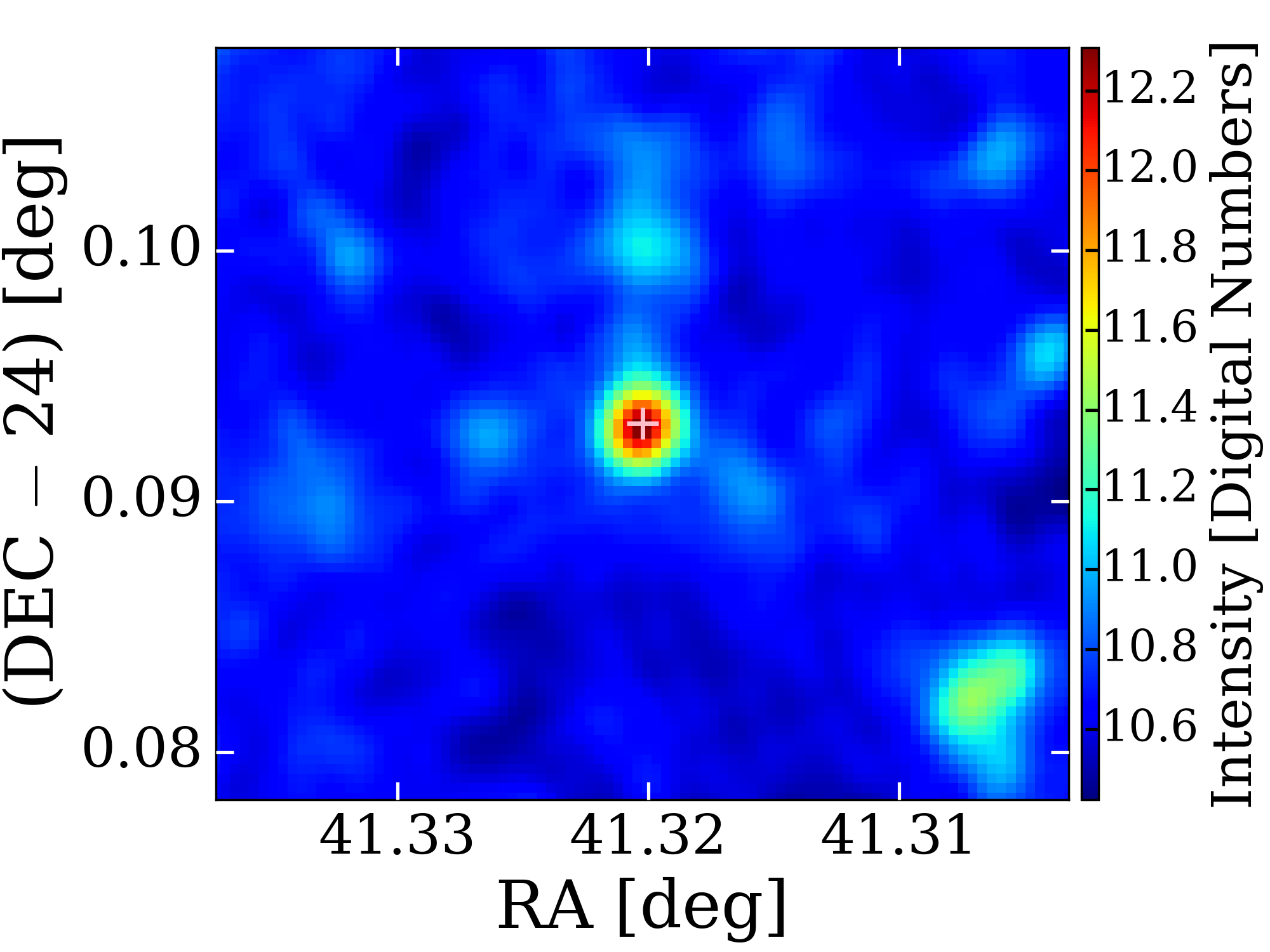}
    \includegraphics[width=0.23\textwidth]{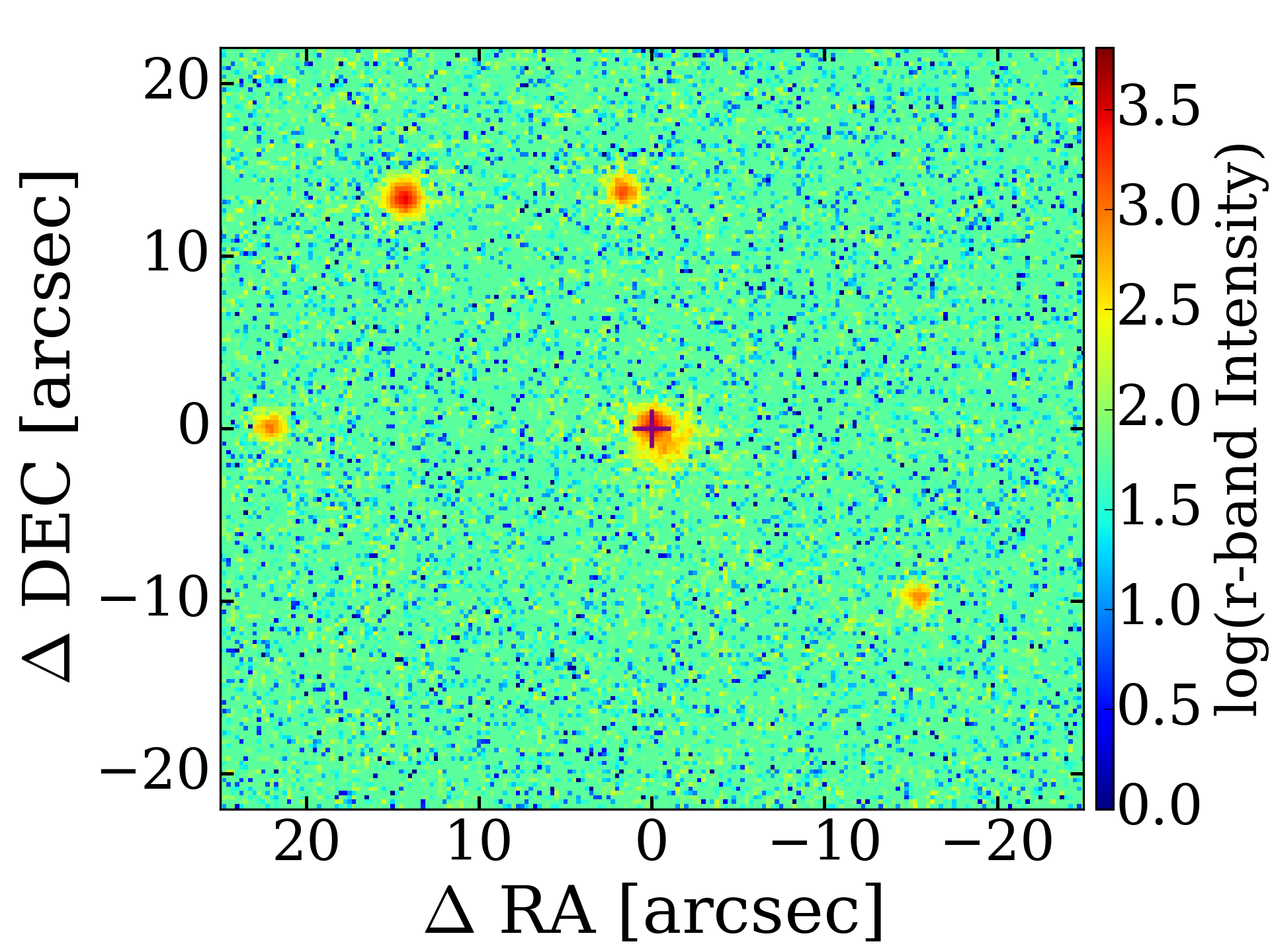}
    \caption{Left: same as Figure~\ref{NVSS_J080601+190611_fit}, but for NVSS J024516\allowbreak+240535. Middle left: radio map from VLASS centered at NVSS J024516\allowbreak+240535. Middle right: W2 band infrared map of WISEA J024516.83\allowbreak+240535.0 (shown as a purple cross) from WISE. Right: Pan-STARRS r-band optical map centered at NVSS J024516\allowbreak+240535.}
    \label{NVSS_J024516+240535_fit}
\end{figure*}

NVSS J024516+240535 is a flat-spectrum radio quasar (FSRQ) at a redshift of $z=2.245$ and is classified as a blazar. Figure~\ref{NVSS_J024516+240535_fit} shows the \hi\ absorption spectrum toward NVSS J024516\allowbreak+240535, along with its radio image from VLASS, infrared image from WISE, and optical r-band image from Pan-STARRS. The Pan-STARRS image reveals no galaxies within 20 arcsec of NVSS J024516\allowbreak+240535 that could serve as potential foreground absorbers, implying that the foreground object is either optically faint, low surface brightness, or located at a larger impact parameter. The intervening \hi\ 21-cm absorption, detected at $z=0.141$, exhibits a narrow, symmetric, and relatively simple profile, which is well described by a two-component Gaussian model. Its narrow and simple profile suggests the presence of cold neutral gas.

\subsubsection{NVSS J053538+643141}

\begin{figure*}[hbt!]
    \centering
    \includegraphics[width=0.25\textwidth]{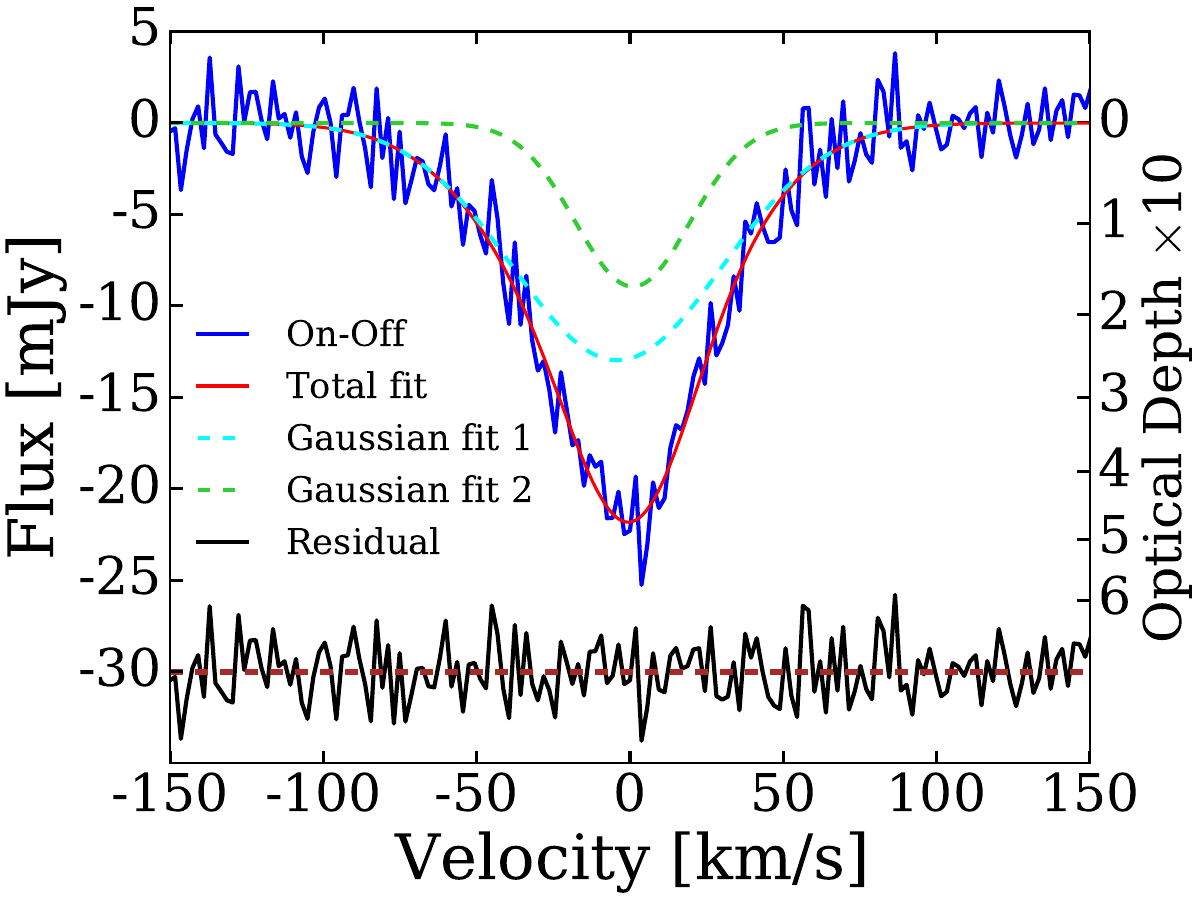}
    \includegraphics[width=0.25\textwidth]{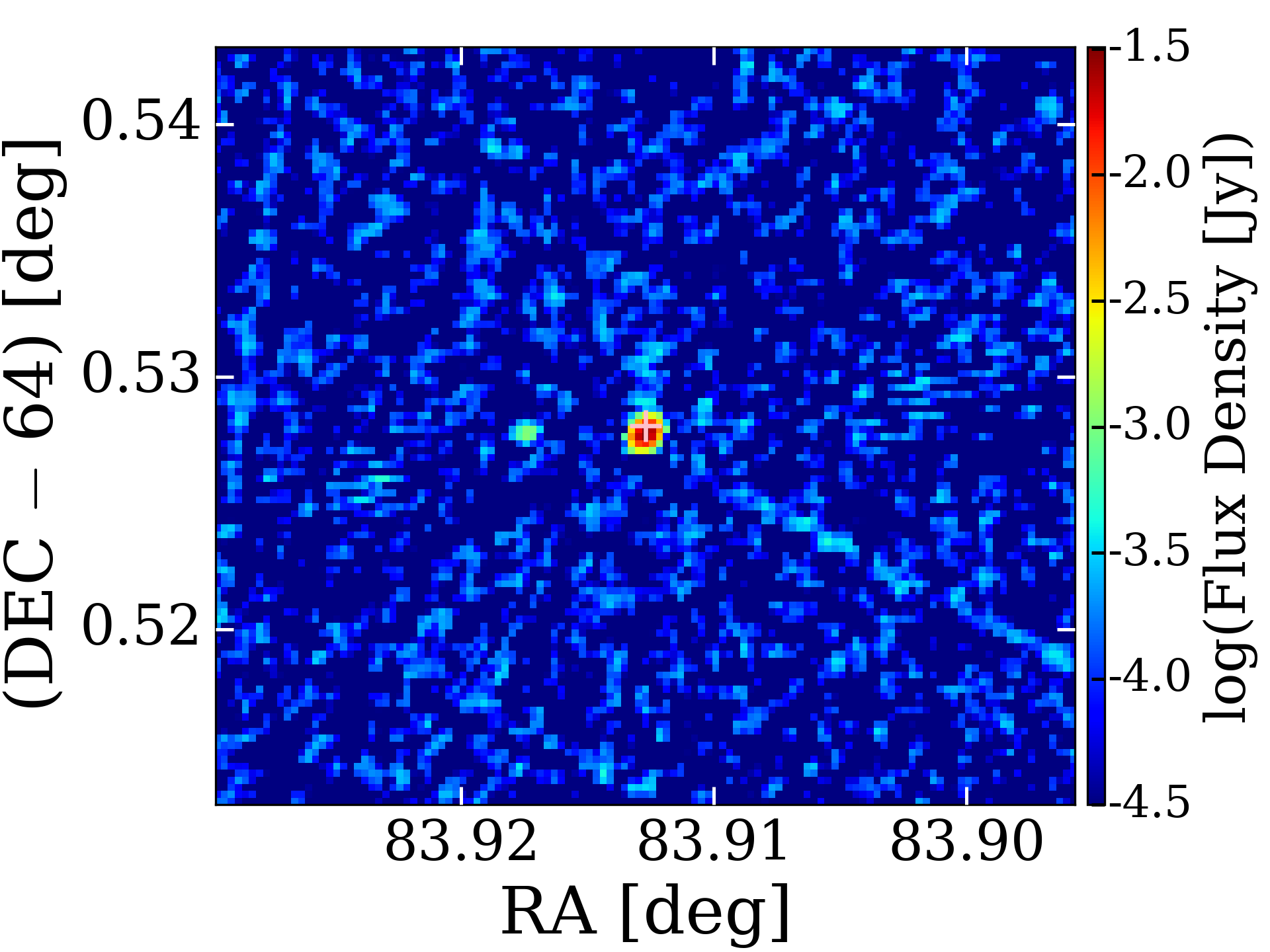}
    \includegraphics[width=0.25\textwidth]{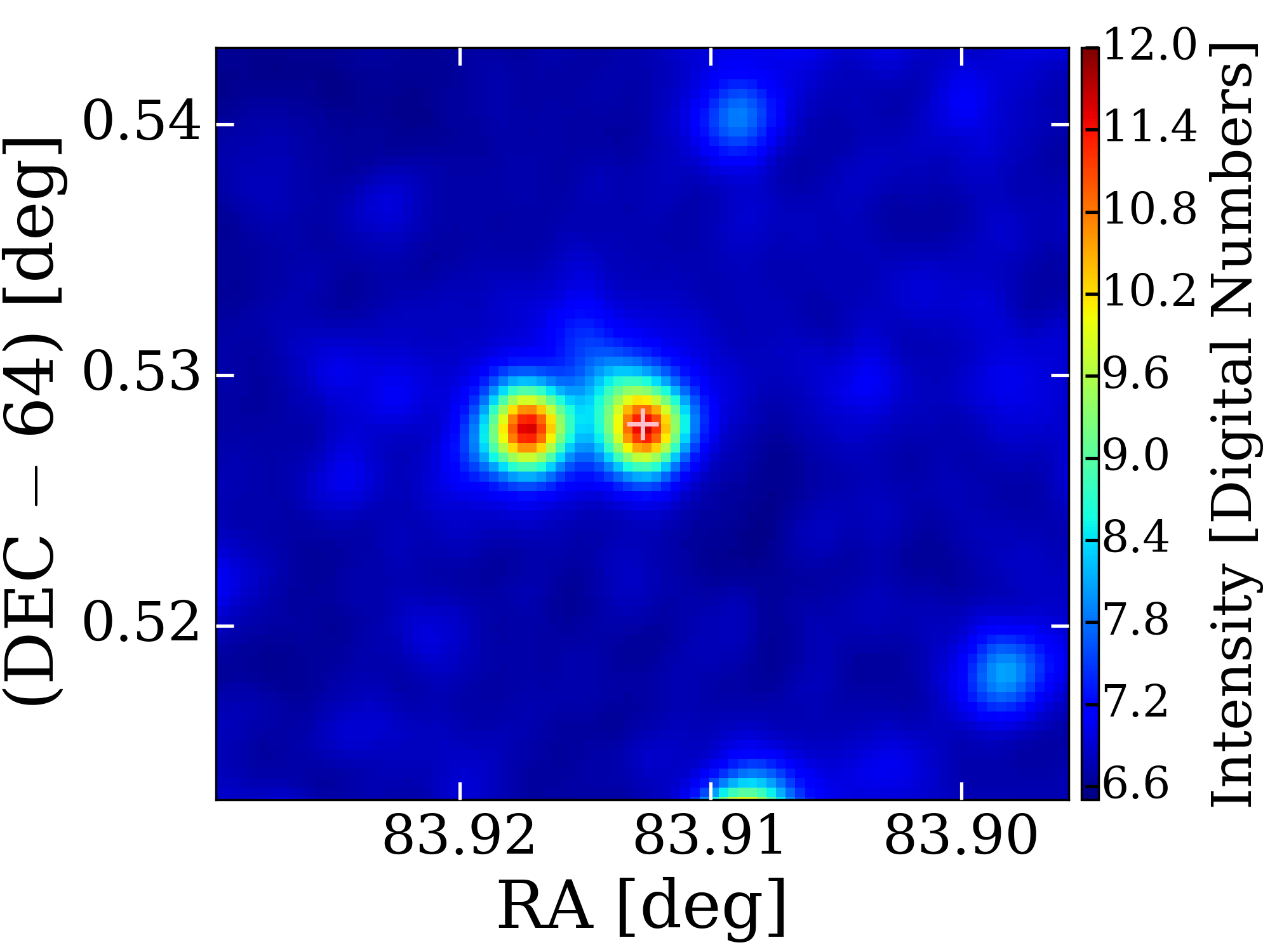}
    \includegraphics[width=0.23\textwidth]{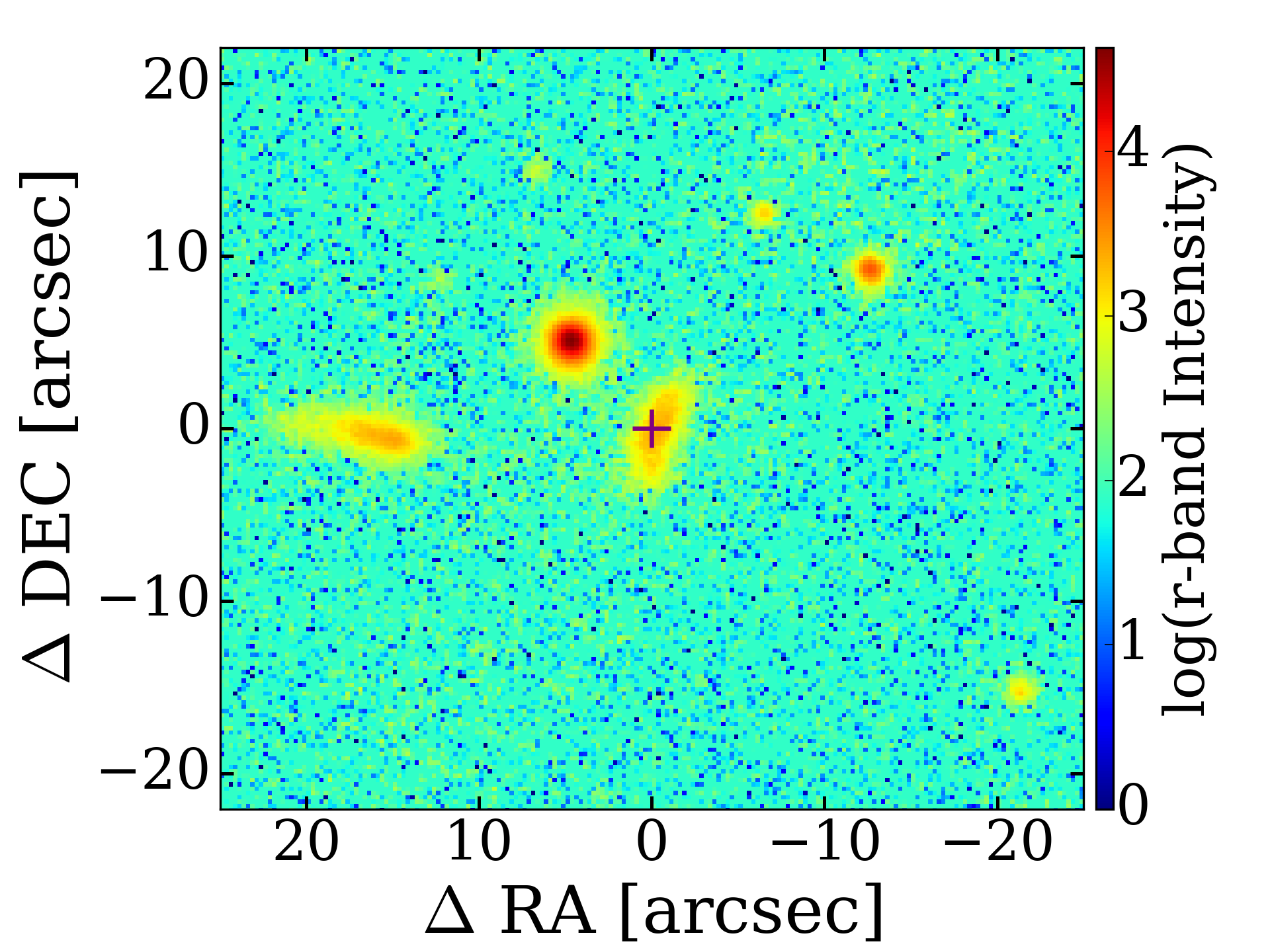}
    \caption{Left: same as Figure~\ref{NVSS_J080601+190611_fit}, but for NVSS J053538\allowbreak+643141. Middle left: radio map from VLASS centered at NVSS J053538\allowbreak+643141. Middle right: W2 band infrared map of WISEA J053539.05\allowbreak+643141.0 (shown as a purple cross) from WISE. Right: Pan-STARRS r-band optical map centered at NVSS J053538\allowbreak+643141.}
    \label{NVSS_J053538+643141_fit}
\end{figure*}

NVSS J053538\allowbreak+643141 is a radio source situated within a galaxy pair system, with limited exploration in the current literature. The WISE counterpart of NVSS J053538\allowbreak+643141 is WISEA J053539.05\allowbreak+643141.0, it has W1-W2=1.142, W2-W3=3.42, locating in the Seyferts region of the WISE color-color diagram. The W1-W2 color of WISEA J053539.05\allowbreak+643141.0 is larger than 0.8, suggesting AGN presence. The absence of precise redshift information for NVSS J053538\allowbreak+643141 introduces uncertainty about whether it is the foreground object. The \hi\ absorption profile can be well represented by a two-component Gaussian model, with its symmetric shape suggesting the presence of a potential gas disk. The \hi\ absorption, VLASS radio image, WISE infrared image, and Pan-STARRS optical image centered at NVSS J053538\allowbreak+643141 are presented in Figure~\ref{NVSS_J053538+643141_fit}.

\begin{table*}
    \fontsize{6.8}{8}\selectfont
	\centering
	\caption{Basic observed parameters of the \hi\ absorptions detected in our 2024 and 2025 FAST \hi\ absorption survey. The associated, intervening, and unknown types are labeled with stars, diamonds, and crosses, respectively.}
	\label{HI_absorption_table}
	\begin{tabular}{|c|c|c|c|c|c|c|c|}
	\hline
	Radio Source & cz$_{\rm peak}$ & FWHM & $S_{\hi, \rm peak}$ & $\int S_{\hi}dv$ & $\tau_{\rm peak}\times10^{2}$ & $\int\tau dv$ & $N_{\hi}/T_{\rm{s}}$\\
	& (\kms) & (\kms) & (mJy) & (mJy\kms) &  & (\kms) & (10$^{18}$cm$^{-2}$K$^{-1}$)\\
        \hline
        
        NVSS J010015+201710$^{\diamond}$ & 79623.16$\pm$1.56 & 17.98$\pm$1.46 & -28.11$\pm$2.23 & -1066.44$\pm$18.62 & 17.90$\pm$1.53 & 6.52$\pm$0.04 & 11.87$\pm$0.06\\
        
        \hline
        
        NVSS J053538+643141$^{\otimes}$ & 24081.77$\pm$2.30 & 61.90$\pm$9.72 & -21.82$\pm$5.20 & -1529.66$\pm$25.38 & 47.37$\pm$14.51 & 30.67$\pm$1.38 & 55.82$\pm$2.69\\

        \hline
        
        NVSS J080601+190611$^{\star}$ & 29393.07$\pm$5.98 & 36.84$\pm$9.08 & -23.37$\pm$0.88 & -1822.47$\pm$34.52 & 17.98$\pm$0.72 & 13.40$\pm$0.08 & 24.40$\pm$0.16\\

        \hline

        NVSS J090150+030422$^{\star}$ & 86085.49$\pm$2.39 & 104.55$\pm$5.62 & -11.66$\pm$0.57 & -1298.20$\pm$47.66 & 3.00$\pm$0.15 & 3.32$\pm$0.25 & 6.04$\pm$0.20\\
        
        HVC$^{\star}$ & 86525.42$\pm$0.58 & 42.27$\pm$1.36 & -20.17$\pm$0.64 & -907.65$\pm$14.21 & 5.18$\pm$0.15 & 2.31$\pm$0.10 & 4.21$\pm$0.18\\

        \hline
        
        NVSS J234106+001833$^{\star}$ & 83038.90$\pm$12.07 & 165.66$\pm$20.00 & -73.75$\pm$5.12 & -14613.41$\pm$235.19 & 19.58$\pm$1.47 & 37.28$\pm$0.13 & 67.85$\pm$0.23\\
        
        \hline
        
        NVSS J024516+240535$^{\diamond}$ & 42373.77$\pm$0.17 & 11.74$\pm$0.36 & -81.65$\pm$2.04 & -919.25$\pm$14.98 & 17.54$\pm$0.38 & 1.92$\pm$0.01 & 3.50$\pm$0.02\\

        \hline
        
        NVSS J150034+364844$^{\star}$ & 19841.56$\pm$1.55 & 38.74$\pm$3.54 & -18.03$\pm$0.48 & -1333.84$\pm$29.94 & 32.26$\pm$0.97 & 21.93$\pm$0.26 & 39.90$\pm$0.47\\

        \hline
        \end{tabular}
\end{table*}

\section{\oh\ absorption}
\label{sec:OH_absorption}

In this section, we present the search for \oh\ absorption within our \hi\ 21-cm absorption catalog. Here, we focus on the radio ground-state transitions of the \oh\ molecule, which occur at wavelengths around 18 cm (frequencies $\sim$ 1.6–1.7 GHz). The four primary radio transitions are at 1612, 1665, 1667, and 1720 MHz. Among these, the 1665 MHz and 1667 MHz lines are known as the main lines and are the most commonly observed in \rm{OH}-related studies, while the 1612 MHz and 1720 MHz lines are satellite lines. The main lines correspond to transitions characterized by $\Delta$F=0, signifying no change in the total angular momentum quantum number $F$, whereas the satellite lines are associated with $\Delta$F=$\pm$1.

For each calibrated spectrum with detected \hi\ 21-cm absorption, we analyzed the regions corresponding to the redshifted \oh\ lines, including both the main and satellite lines. Since the L-band receiver of FAST covers a frequency range of 1.0–1.5 GHz, 22 \hi\ 21-cm absorptions have their potential redshifted \oh\ emissions outside this usable frequency range. The remaining 18 absorbers—including the HVC associated with NVSS J090150\allowbreak+030422—have redshifted \oh\ transitions within the L-band coverage and were selected for the \oh\ absorption search.

\begin{figure}[hbt!]
    \centering
    \includegraphics[width=0.45\textwidth]{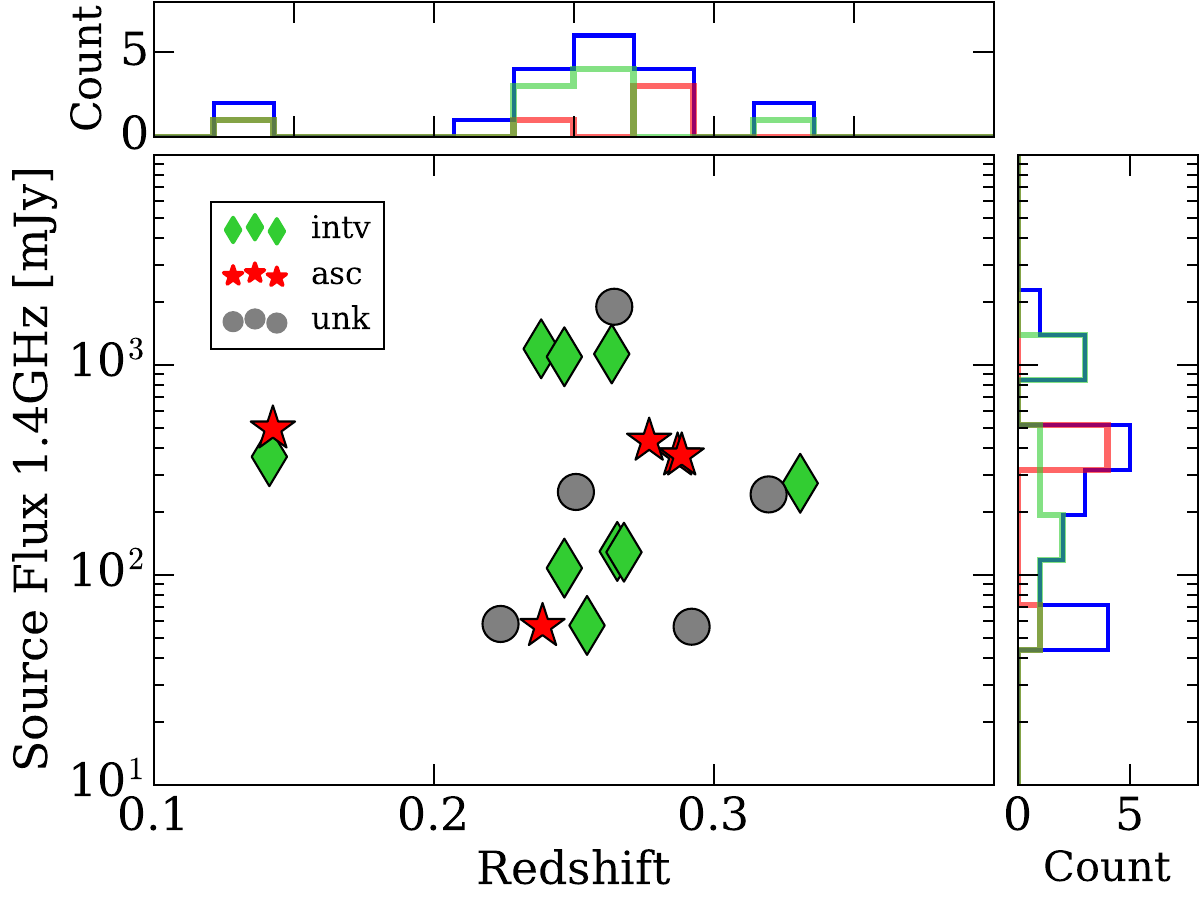}
    \caption{Redshift versus 1.4-GHz flux density of radio sources for the systems searched for \oh\ absorption. The flux densities were obtained from the NVSS data available through NED.}
    \label{redshift_source_flux}
\end{figure}

In Figure~\ref{redshift_source_flux} we plot the redshift versus 1.4-GHz flux density of the radio sources for the systems searched for \oh\ absorption, along with their corresponding histograms for \hi\ absorption systems across all considered samples. For the \hi\ absorption detected in this work, the scarcity of \hi\ absorption systems found between redshifts 0.15 and 0.2 is attributed to interference caused by RFI in the frequency range from 1150 MHz to 1250 MHz. 

We analyzed the spectral range corresponding to the redshifted \oh\ 18 cm transitions and successfully re-detected \oh\ 1612 MHz absorption and \oh\ 1720 MHz emission in NVSS J141558\allowbreak+132024. However, no \oh\ absorption or emission was identified in the remaining cases.

\subsection{Redetecting \oh\ Absorption in PKS 1413+135}

\begin{figure}[hbt]
    \centering
    \includegraphics[width=0.48\textwidth]{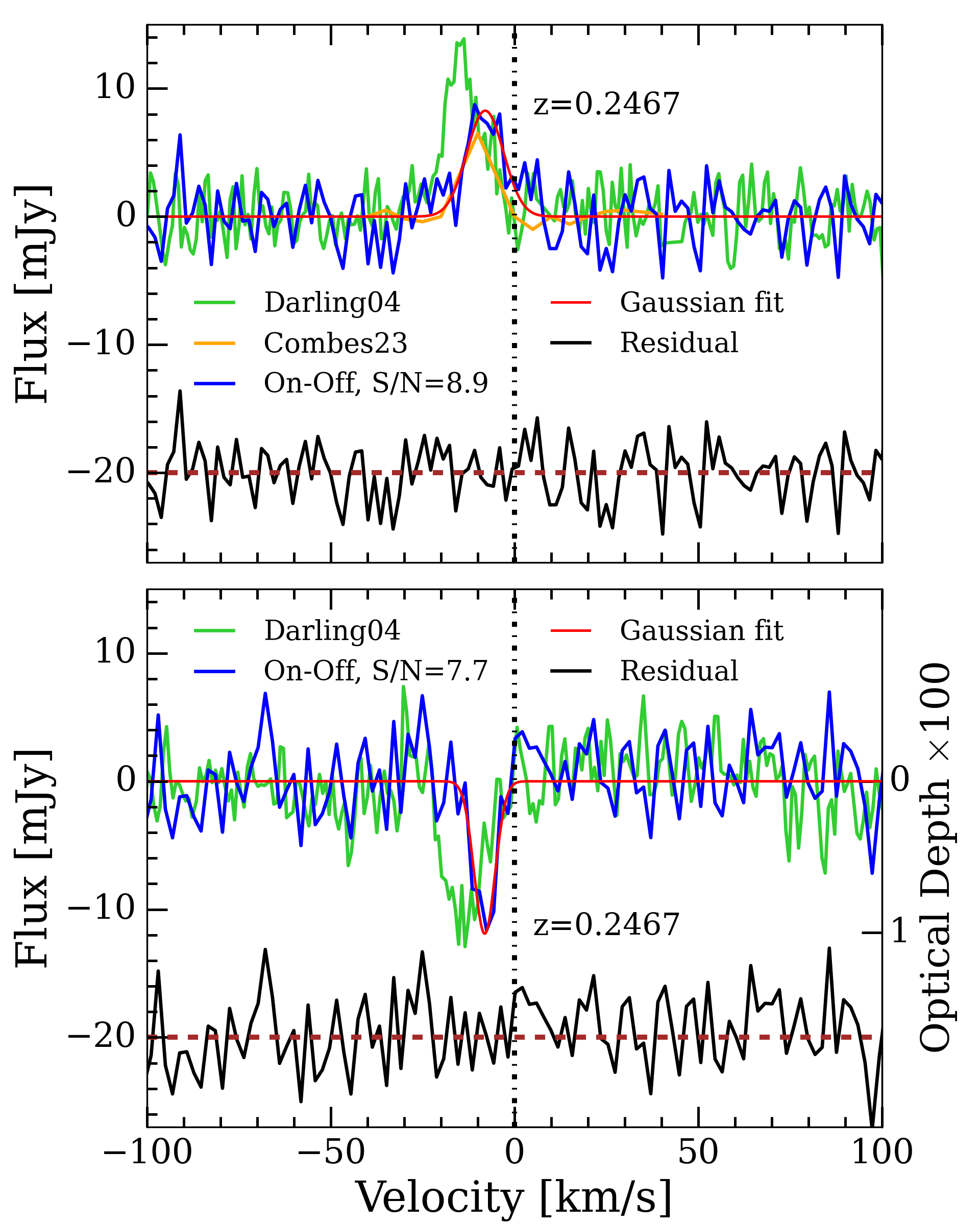}
    \caption{Similar to Figure~\ref{NVSS_J080601+190611_fit}, but showing the \oh\ 1612 MHz absorption line (bottom) for PKS 1413\allowbreak+135, compared with the GBT spectrum from \citet{2004ApJ...612...58D} (green), and the \oh\ 1720 MHz emission line (top) for the same source, compared with the GBT spectrum from \citet{2004ApJ...612...58D} (green) and the MeerKAT spectrum from \citet{2023A&A...671A..43C} (orange).}
    \label{NVSS_J141558+132024_OHabs_emi}
\end{figure}

NVSS J141558\allowbreak+132024 (PKS 1413\allowbreak+135) is a radio galaxy that has been classified as both a disk galaxy and a BL Lac object. It has been extensively studied in previous literature. The details of PKS 1413\allowbreak+135, including information about its associated \hi\ absorption, were summarized in \citet{2025ApJS..277...25H}. The \oh\ 1667 MHz absorption line in  PKS 1413\allowbreak+135 was first detected by \citet{2002A&A...381L..73K} using the Giant Metrewave Radio Telescope (GMRT), with 5.5 hours of on-source integration. They gave the $N_{\rm{OH}}$ = 1.16$\times$10$^{15}$(T$_{\rm{ex}}$/10)(0.044/$c_{\rm f,OH}$)cm$^{-2}$. Limited by our integration time, we don't detect its \oh\ 1667 MHz absorption signal. The \oh\ 1612 MHz absorption and 1720 MHz emission lines in PKS 1413\allowbreak+135 were first independently detected by \citet{2004ApJ...612...58D} using the NRAO Green Bank Telescope (GBT), with 90 minutes of on-source integration in tracking mode, and \citet{2004PhRvL..93e1302K} using the Westerbork Synthesis Radio Telescope (WSRT). We successfully re-detected the 1612 MHz and 1720 MHz \oh\ lines in the FAST follow-up observations, with the corresponding spectra presented in Figure~\ref{NVSS_J141558+132024_OHabs_emi}. For the \oh\ 1720 MHz emission line, we measured a peak flux of 8.27$\pm$1.10 mJy, an FWHM of 11.78$\pm$2.17 \kms, and a peak optical depth of $\tau_{\rm peak} \sim -0.007$. The \oh\ 1612 MHz absorption line exhibited a peak flux of -11.90$\pm$1.48 mJy, an FWHM of 6.68$\pm$1.83 \kms and a peak optical depth of $\tau_{\rm peak} \sim$ 0.010, with a corresponding \oh\ column density estimate of $N_{\rm{OH}} \sim 0.15T_{\rm{s}}$10$^{15}$cm$^{-2}$K$^{-1}$. These measurements are consistent with previous findings shown by \citet{2023A&A...671A..43C}.

\subsection{Non-detections}

Apart from the previously known \oh\ absorption and emission in PKS 1413\allowbreak+135, no additional \oh\ transition detections were observed. The 19 spectra (including two objects with detected \hi\ absorption towards NVSS J090150\allowbreak+030422) centered on the redshifted \oh\ 1667 MHz line are shown in Figure~\ref{OH_abs_1667_1} and Figure~\ref{OH_abs_1667_2} as examples. For the non-detections, we assume the lines are optically thin, and the 3$\sigma$ upper limit on the optical depth can be approximated by: $\tau_{3\rm{\sigma}} \sim 3\rm{\sigma}/S$, where $\rm{\sigma}$ is the root-mean-square (RMS) noise level, and 
$S$ is the continuum flux density. The 3$\sigma$ upper limits on the integrated optical depths is then given by $\int^{\Delta v/2}_{-\Delta v/2}\tau_{3\rm{\sigma}}dv$, where $\Delta v$ is the velocity range over which the integration is performed. Under the assumptions that the noise is Gaussian and uncorrelated between spectral channels, and that the RMS is uniform across the velocity range, the integration can be calculated as:
\begin{eqnarray}
    \int^{\Delta v/2}_{-\Delta v/2}\tau_{3\rm{\sigma}}dv=\tau_{3\rm{\sigma}}\delta v\sqrt{\Delta v/\delta v},
    \label{optical_path_int}
\end{eqnarray}
where $\delta v$ denotes the velocity resolution at which the RMS noise, $\rm{\sigma}$, is evaluated. We adopt $\Delta v=30\kms$, consistent with the fact that the majority of known OH absorbers exhibit FWHM values of $\lesssim 30\kms$. The 3$\sigma$ upper limits on the integrated optical depths for the 1612, 1665, 1667, and 1720 MHz lines for each radio source searched for \oh\ absorption are summarized in Table~\ref{OH_absorption_table}.

\begin{figure*}[hbt!]
    \centering
    \includegraphics[width=0.32\textwidth]{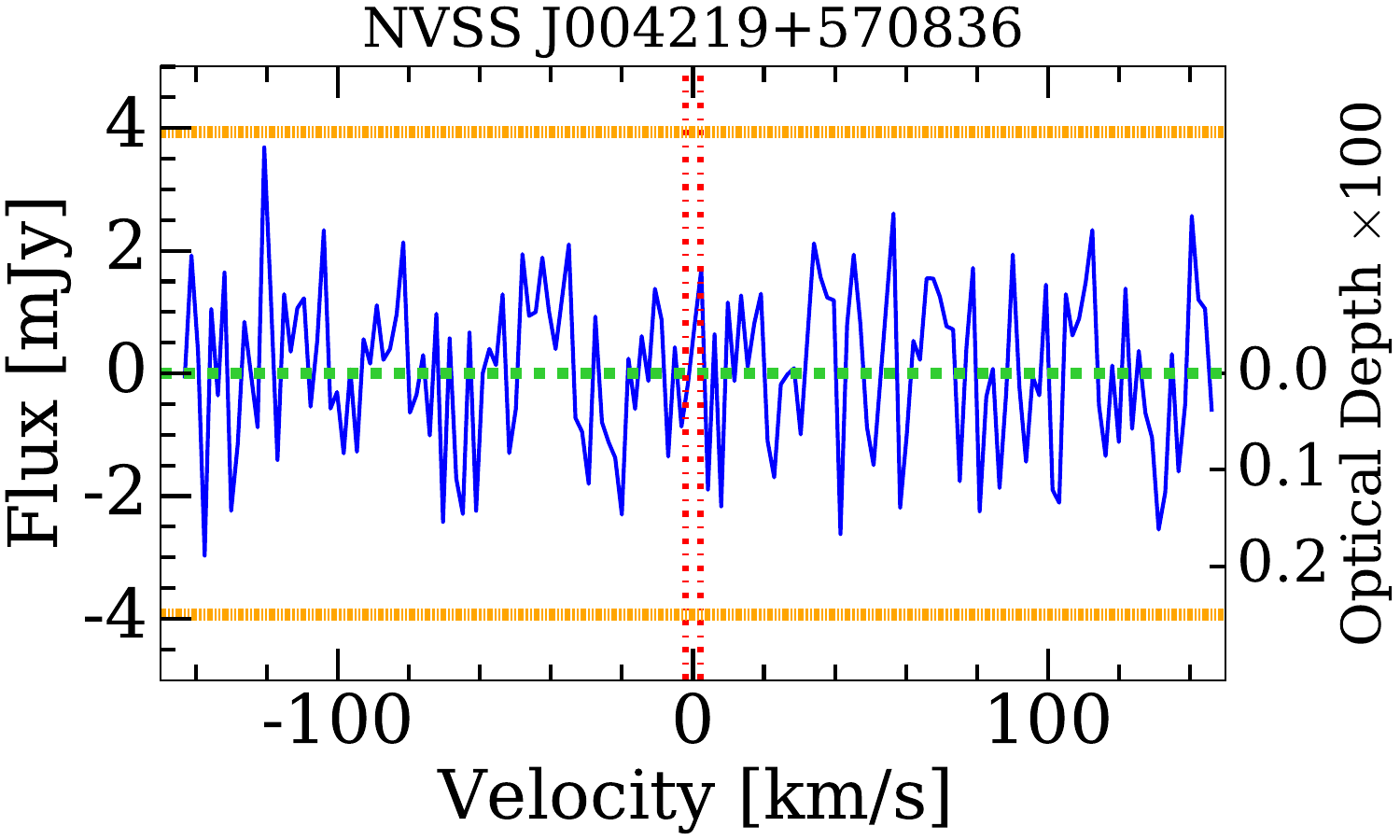}
    \includegraphics[width=0.32\textwidth]{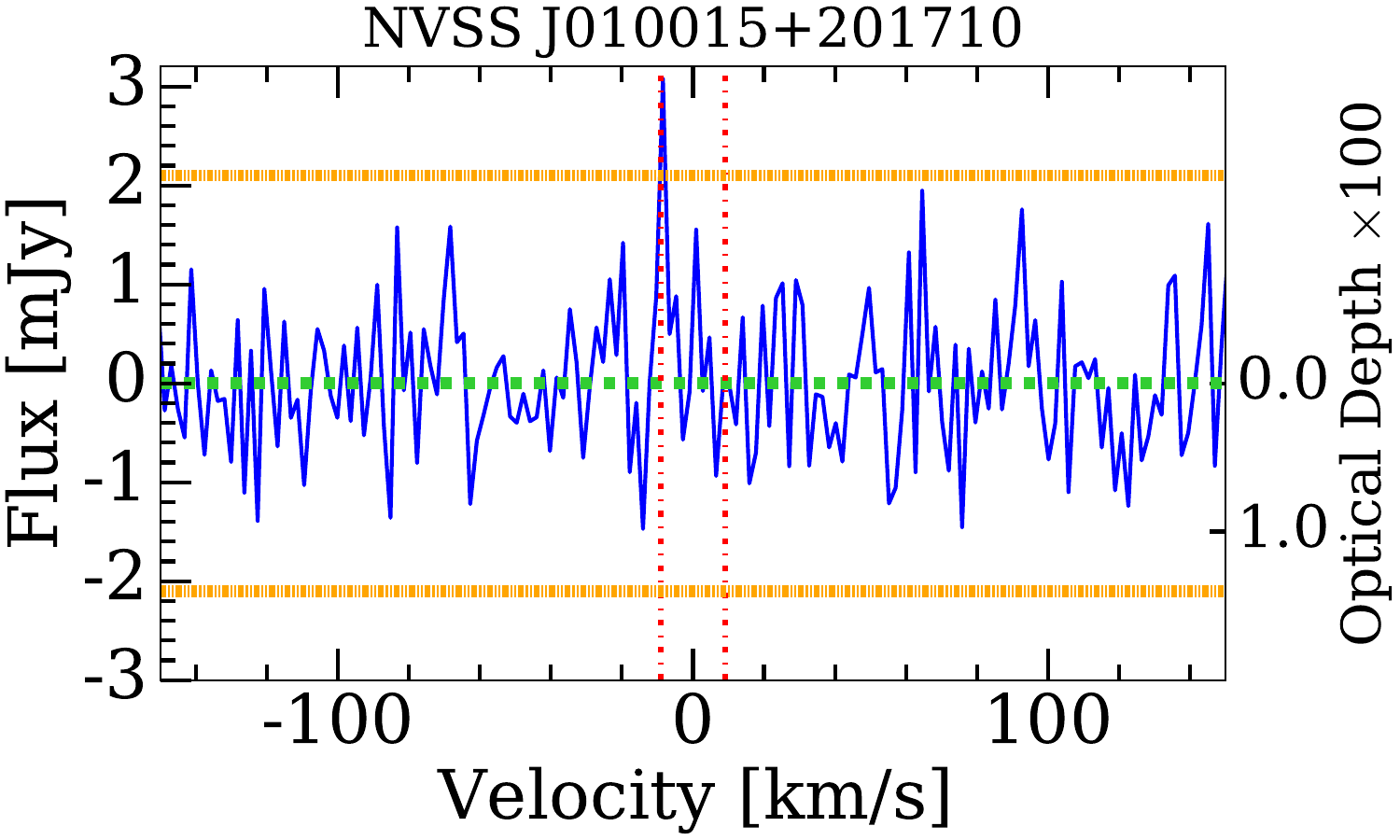}
    \includegraphics[width=0.32\textwidth]{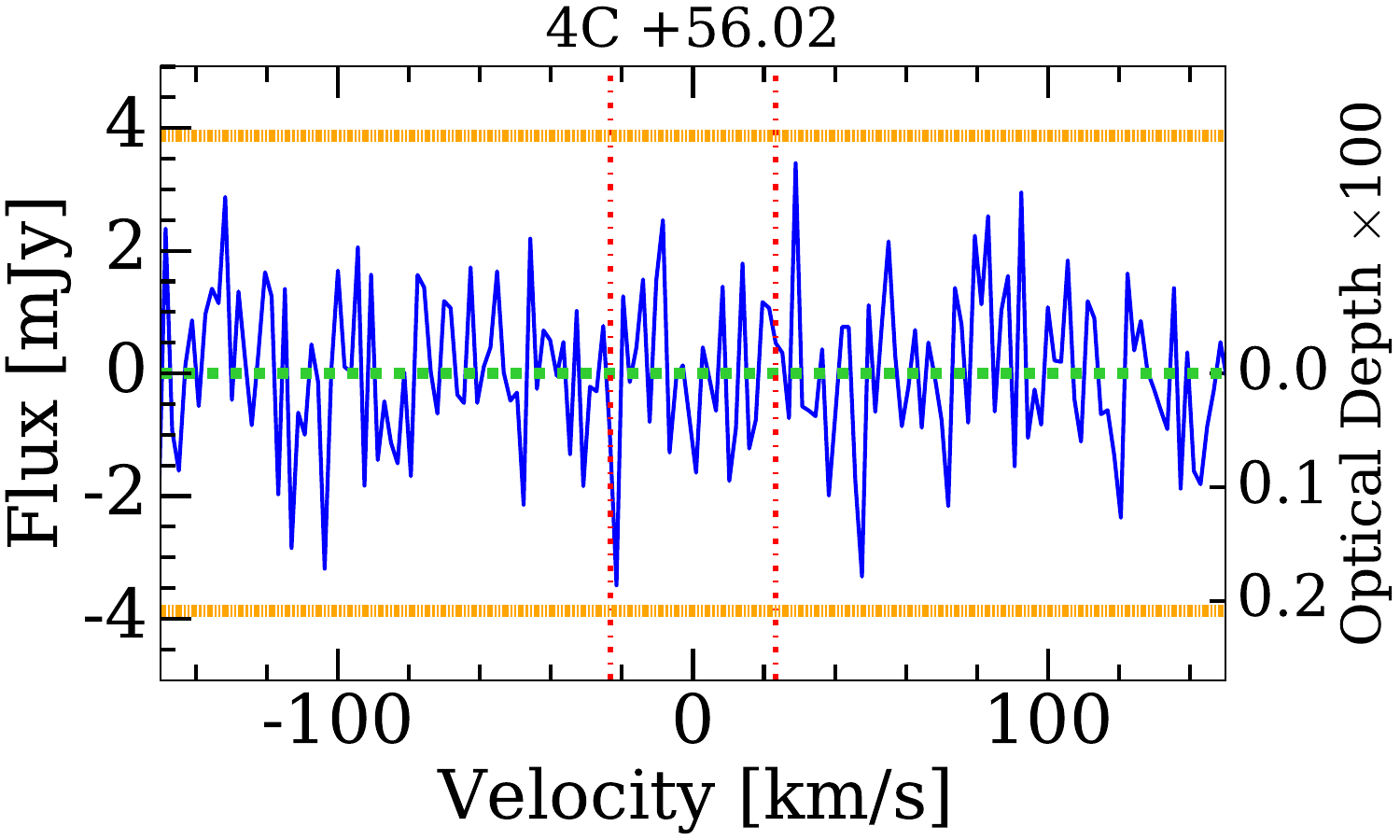}
    \includegraphics[width=0.32\textwidth]{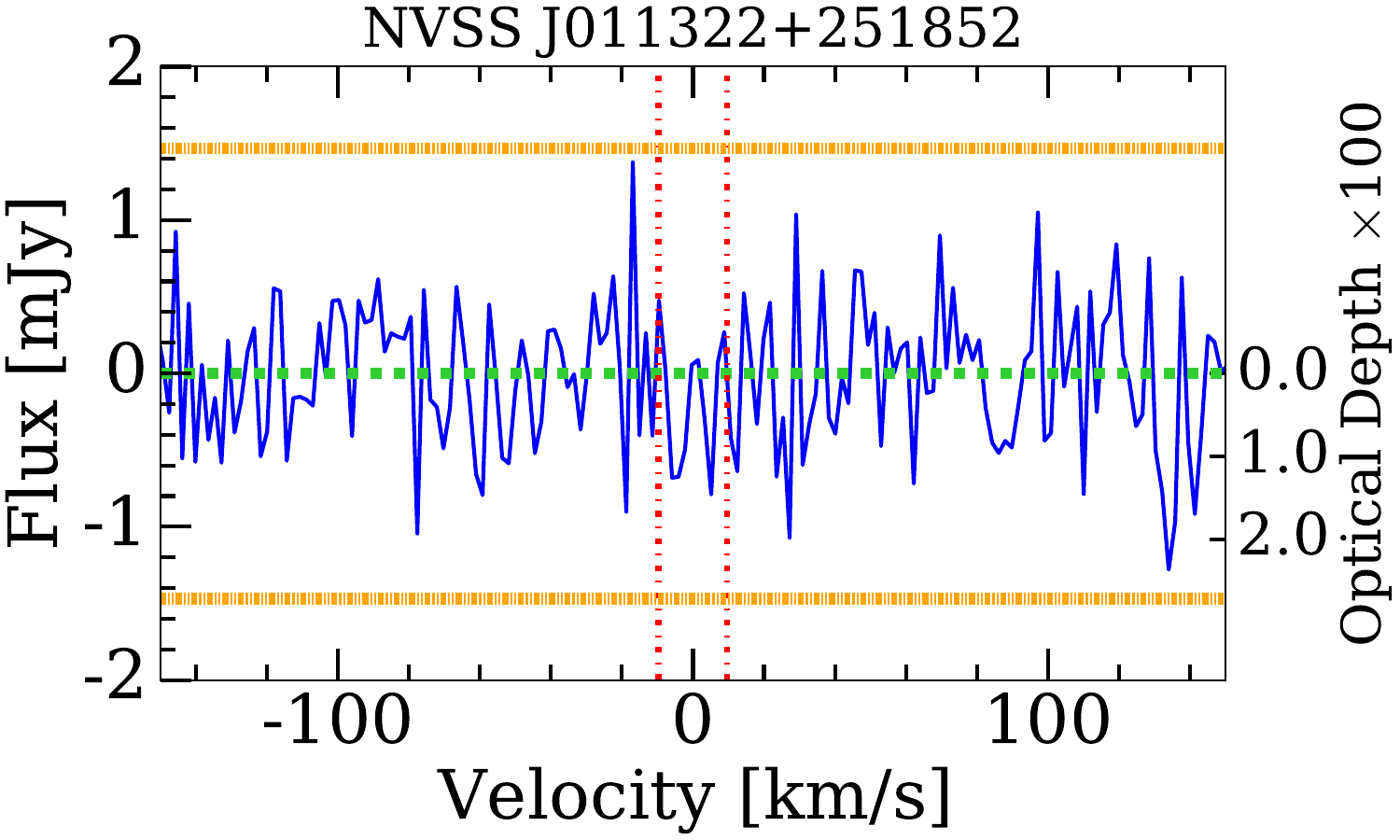}
    \includegraphics[width=0.32\textwidth]{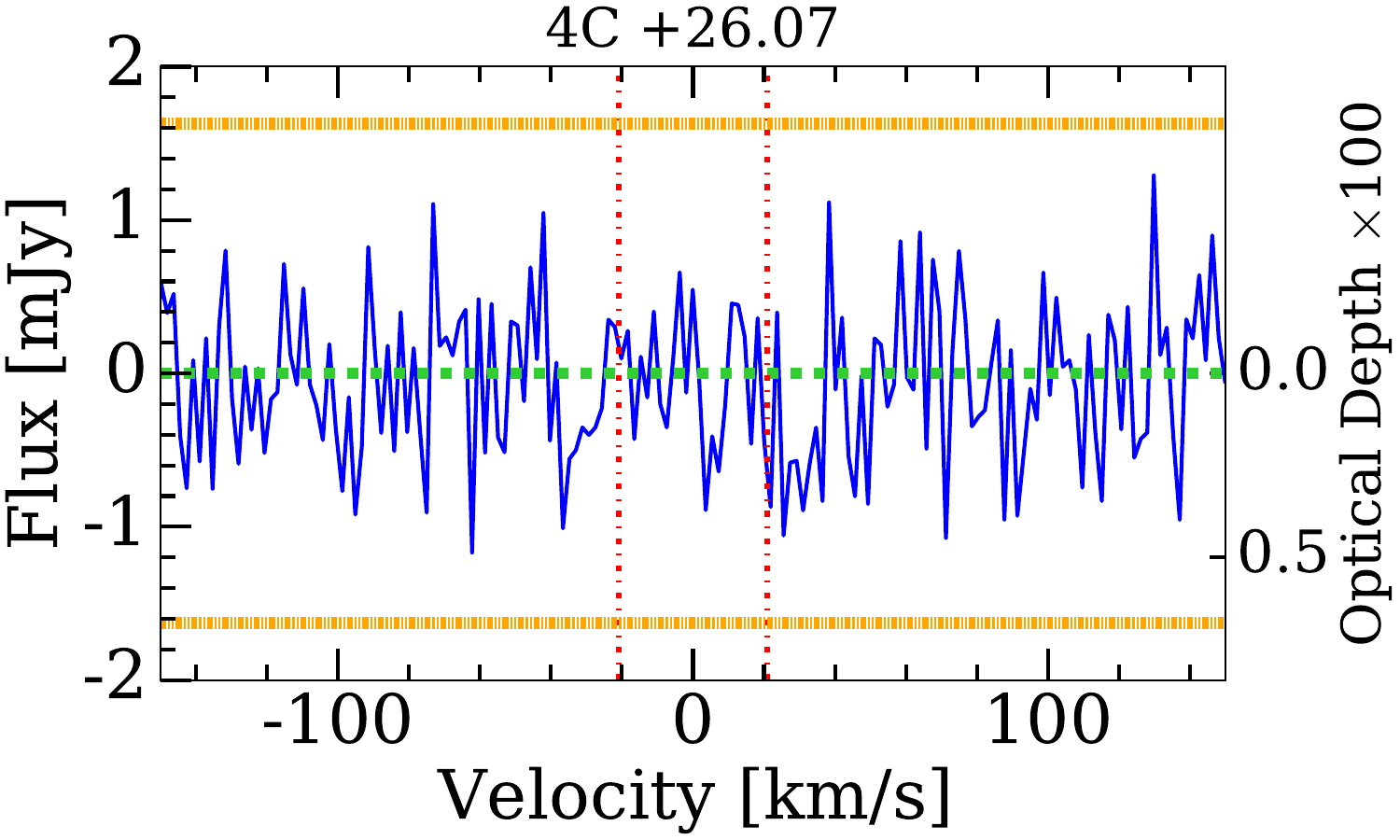}
    \includegraphics[width=0.32\textwidth]{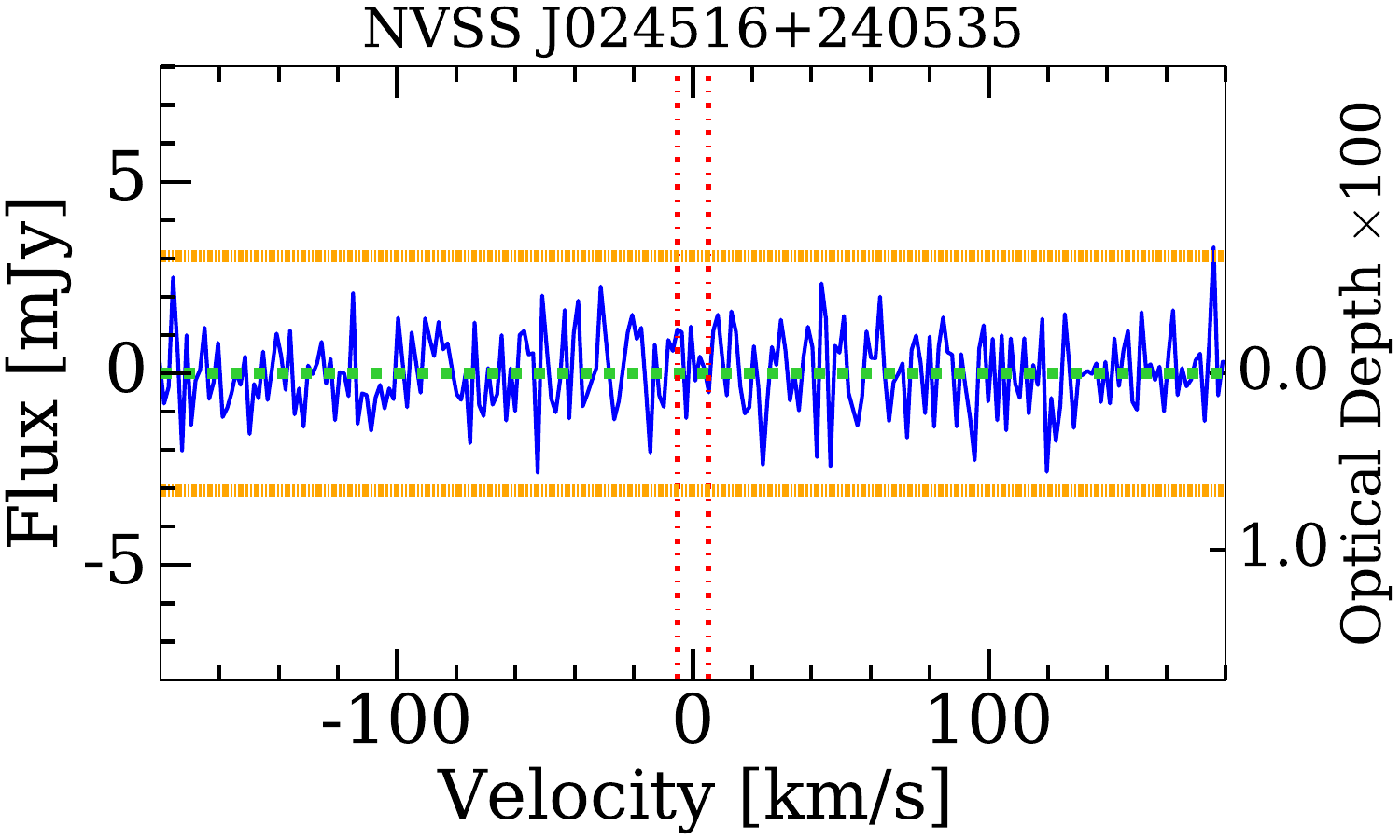}
    \includegraphics[width=0.32\textwidth]{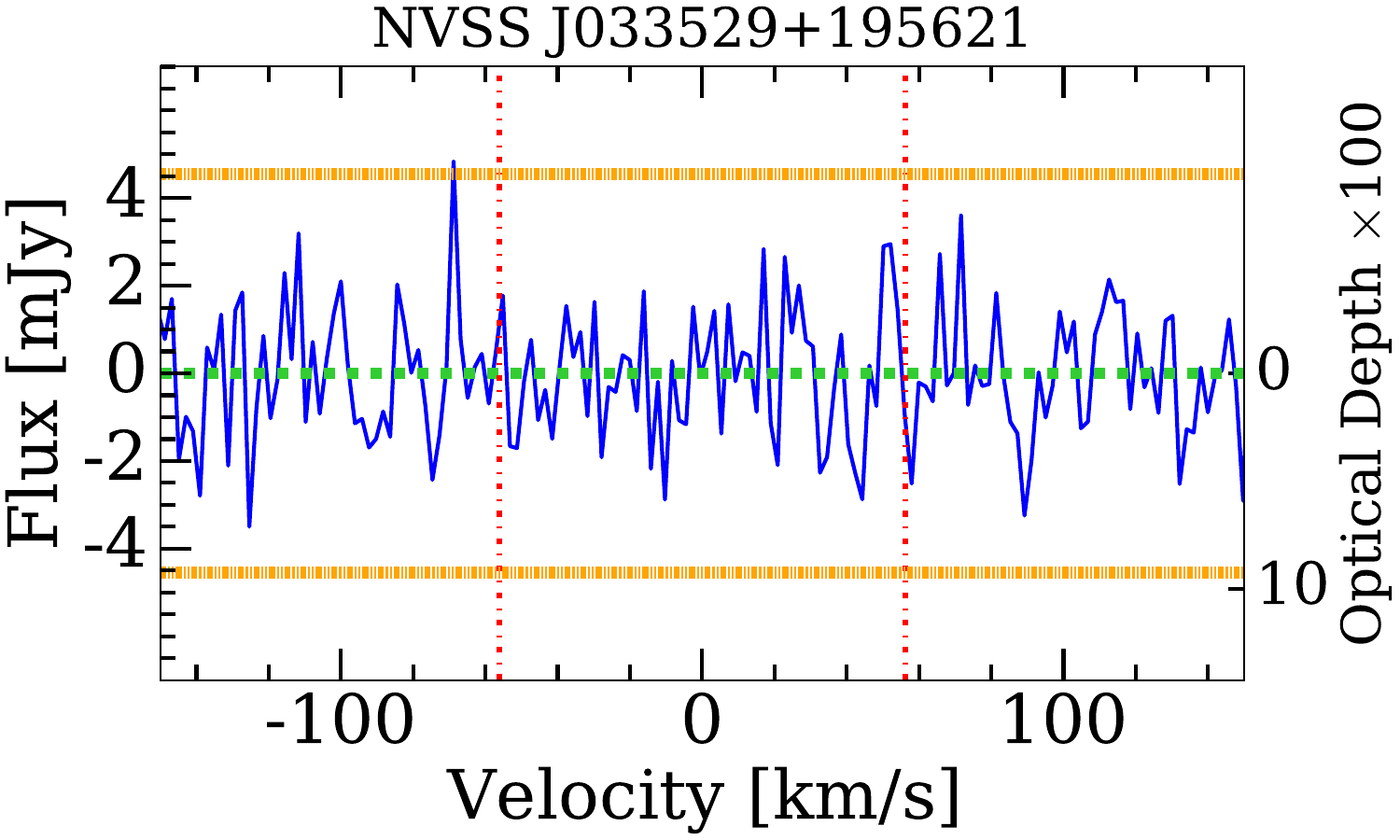}
    \includegraphics[width=0.32\textwidth]{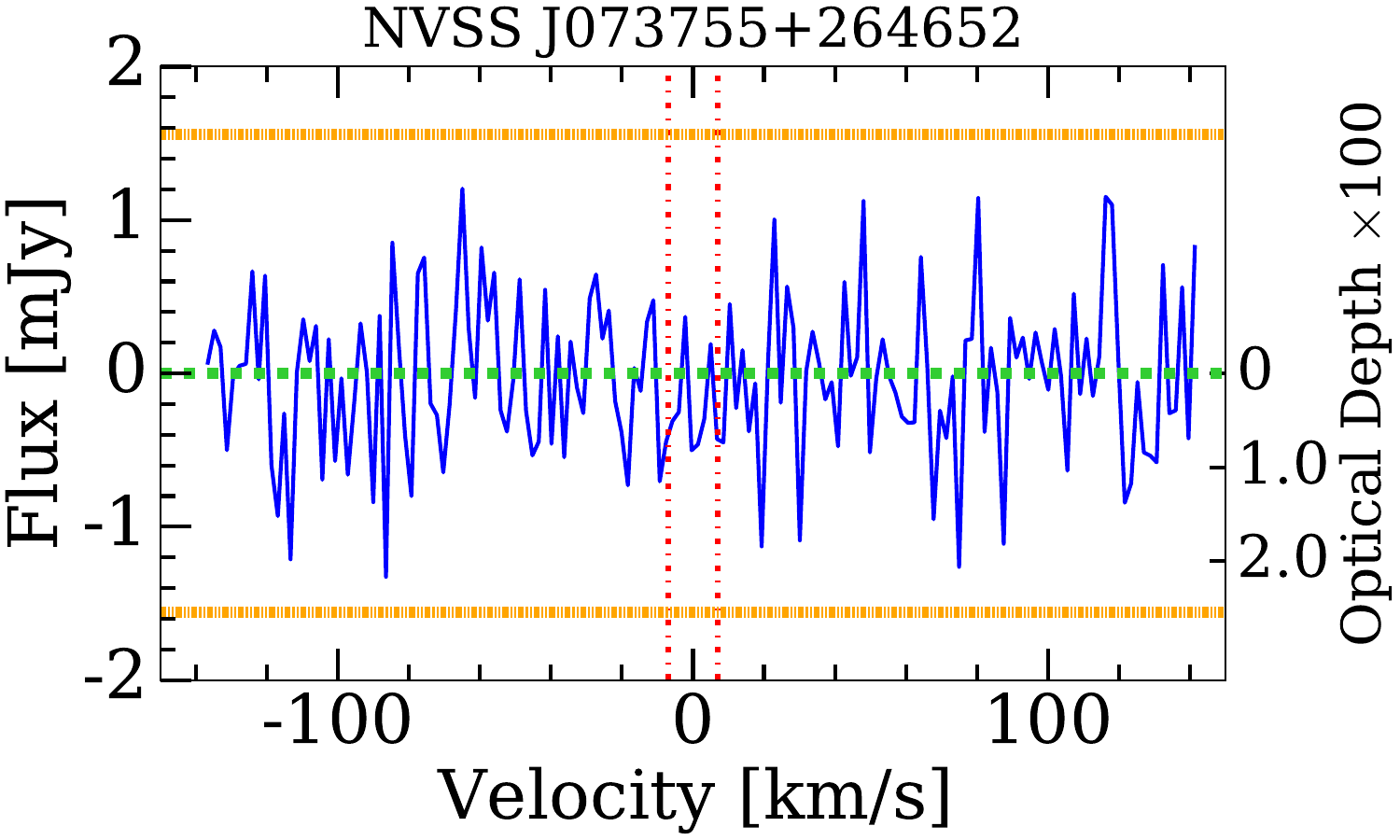}
    \includegraphics[width=0.32\textwidth]{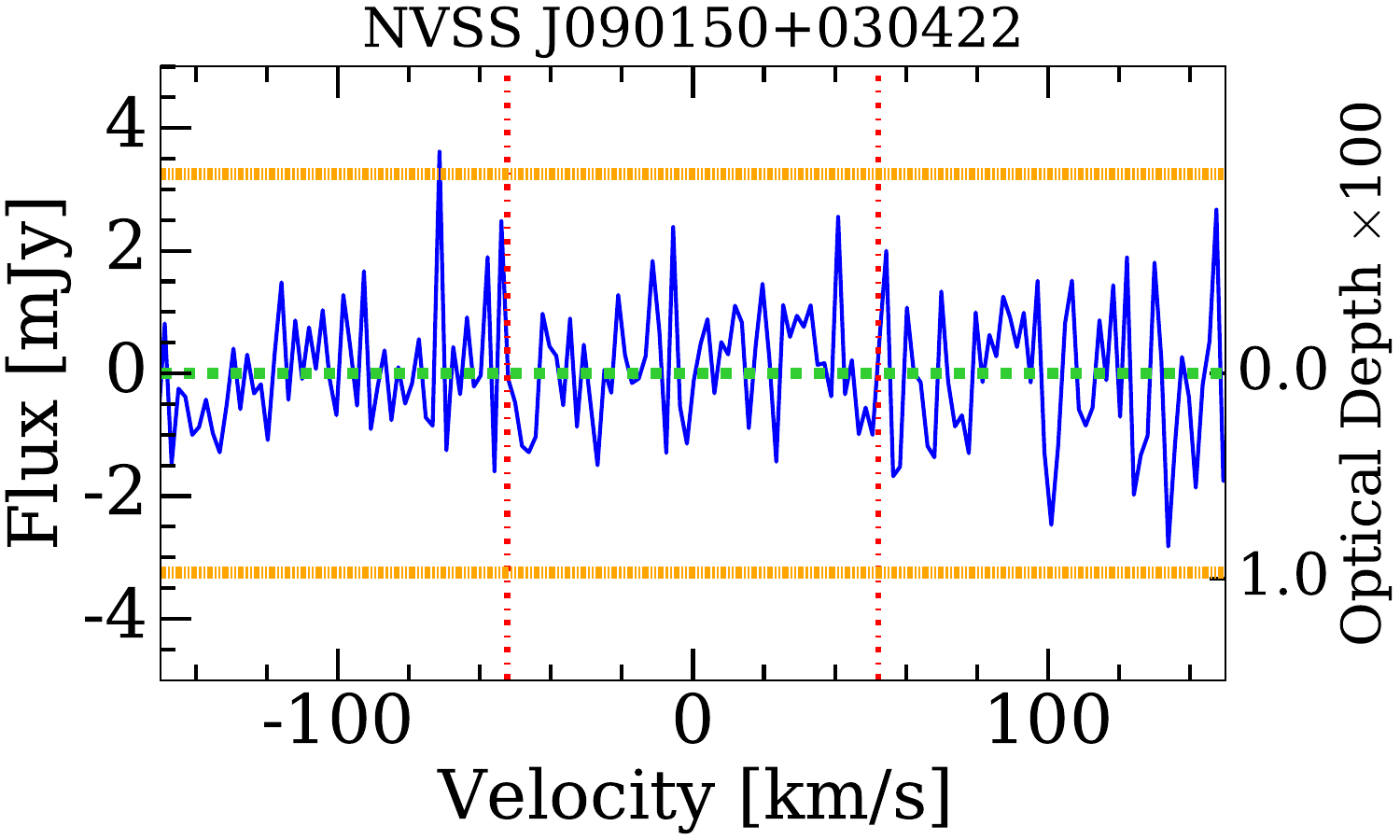}
    \includegraphics[width=0.32\textwidth]{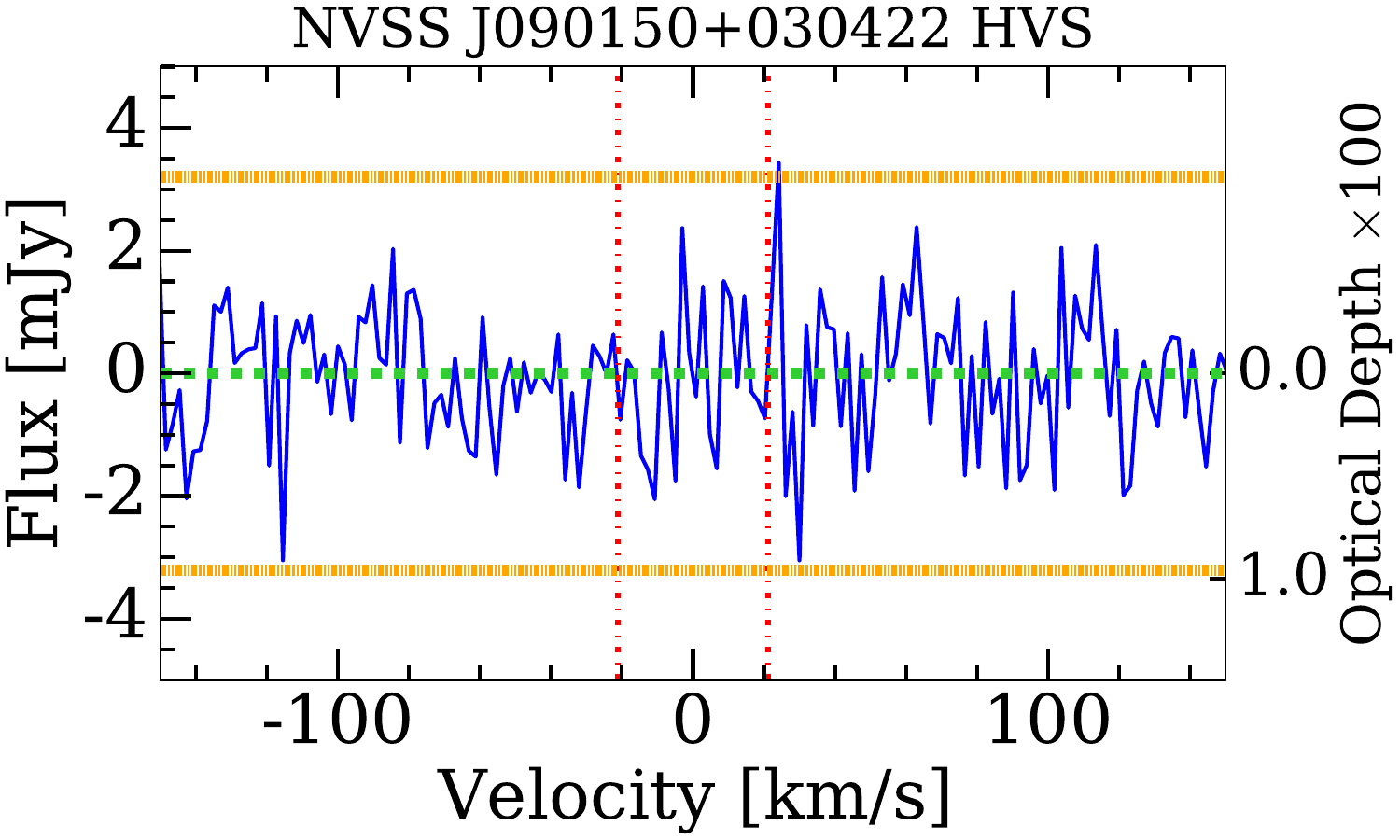}
    \includegraphics[width=0.32\textwidth]{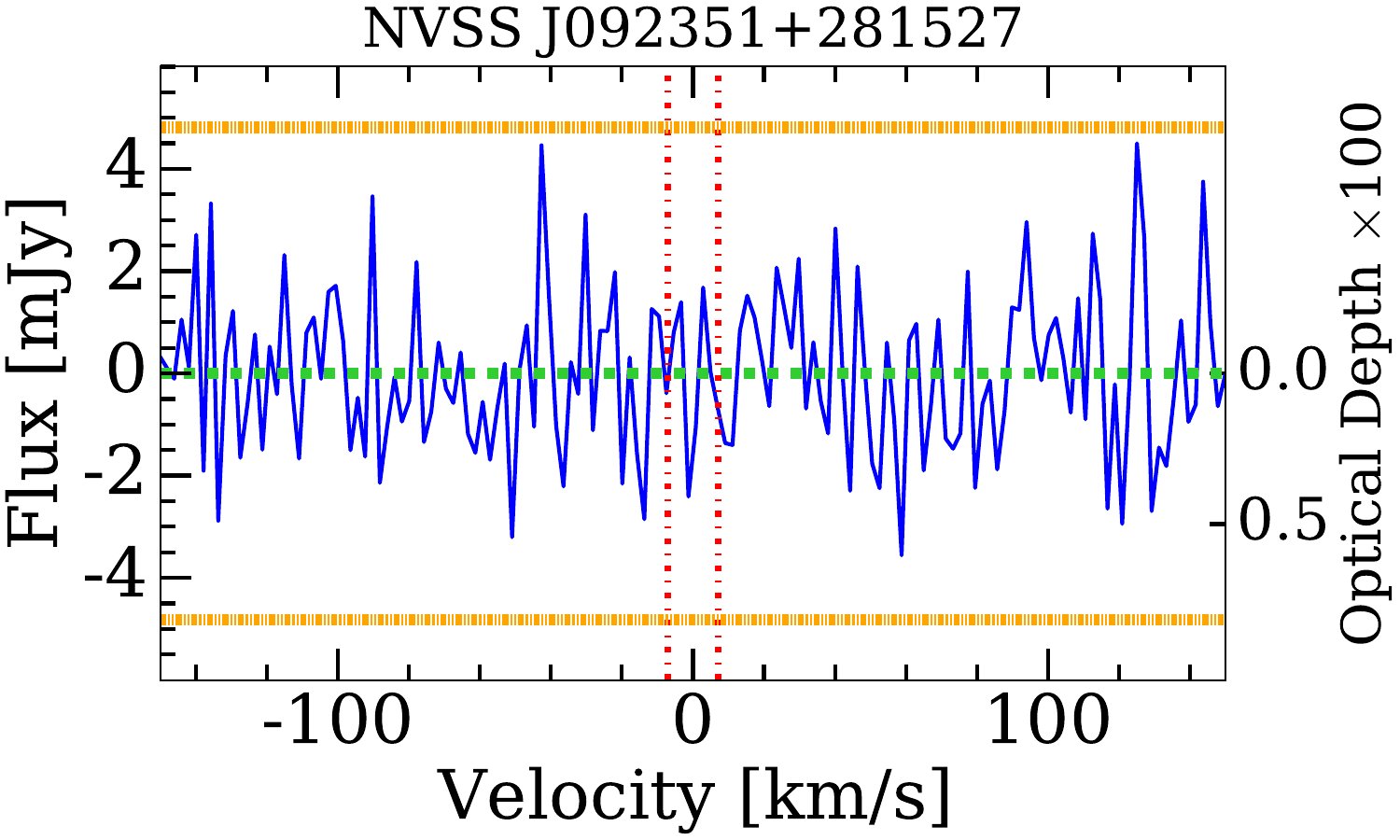}
    \includegraphics[width=0.32\textwidth]{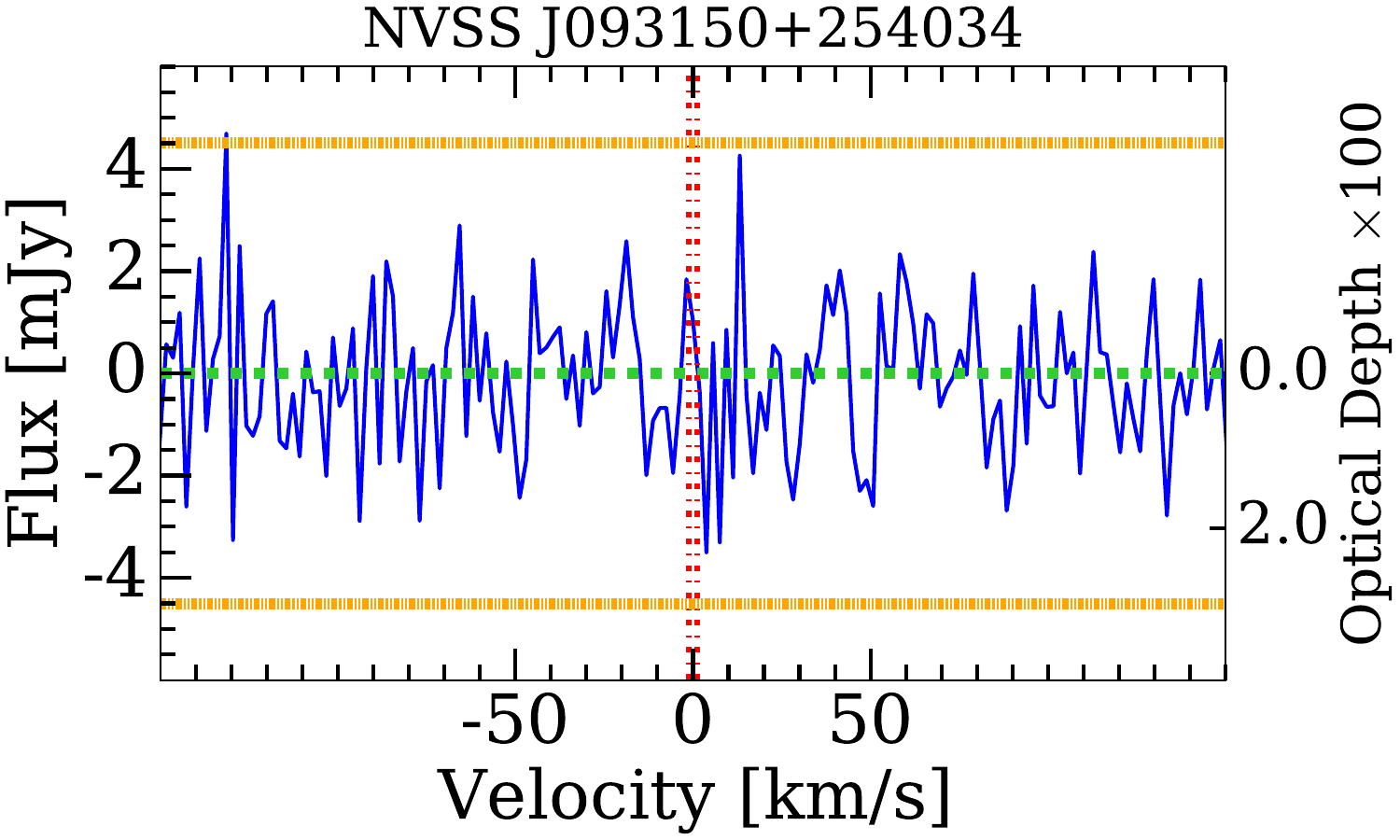}
    \caption{Calibrated flux density spectra centered on the redshifted frequency of the \oh\ 1667 MHz line for radio targets where \hi\ was detected. The red dashed line represents the FWHM of the corresponding \hi\ absorption. The orange densely dashed dotted lines indicate the 3$\sigma$ RMS.}
    \label{OH_abs_1667_1}
\end{figure*}

\begin{figure*}[hbt!]
    \centering
    \includegraphics[width=0.32\textwidth]{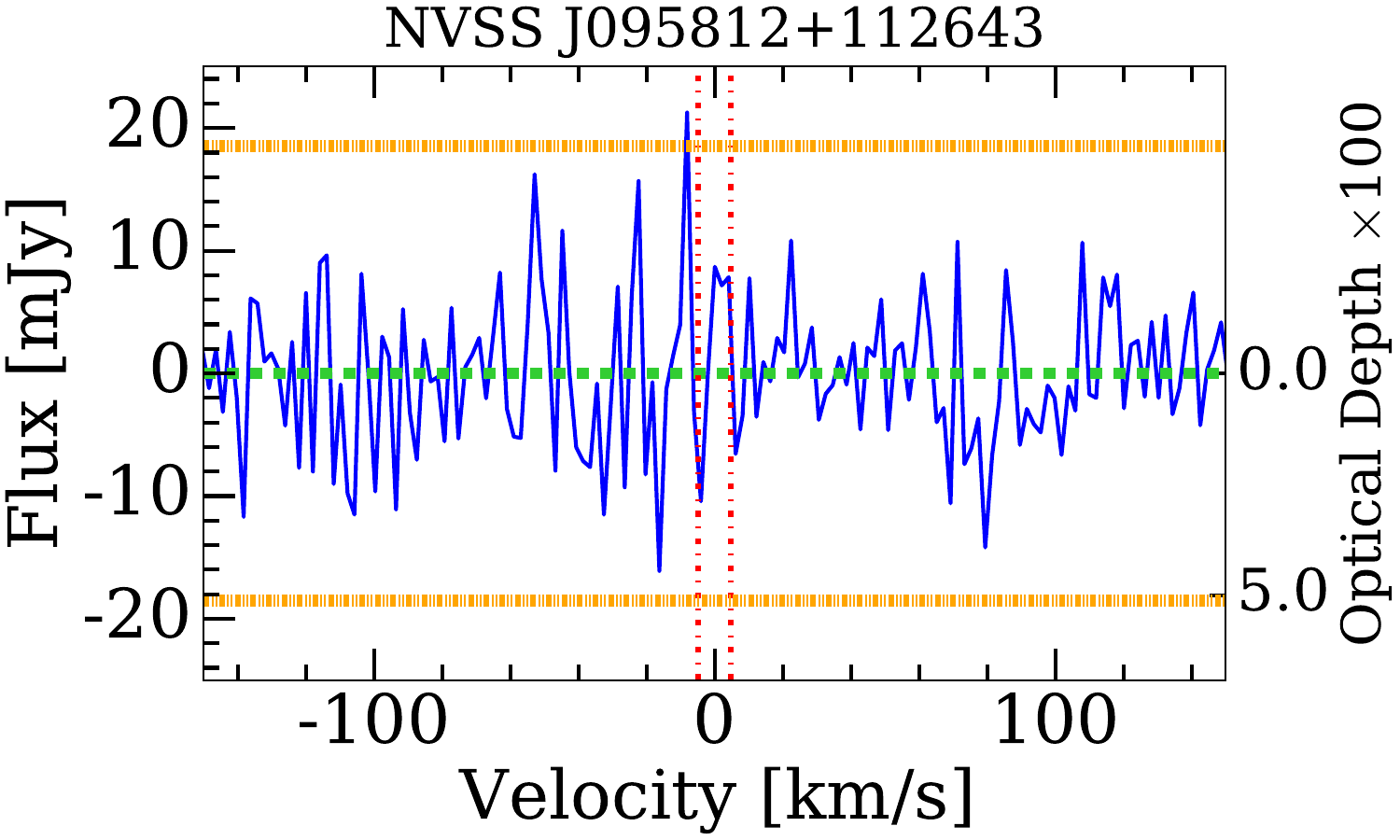}
    \includegraphics[width=0.32\textwidth]{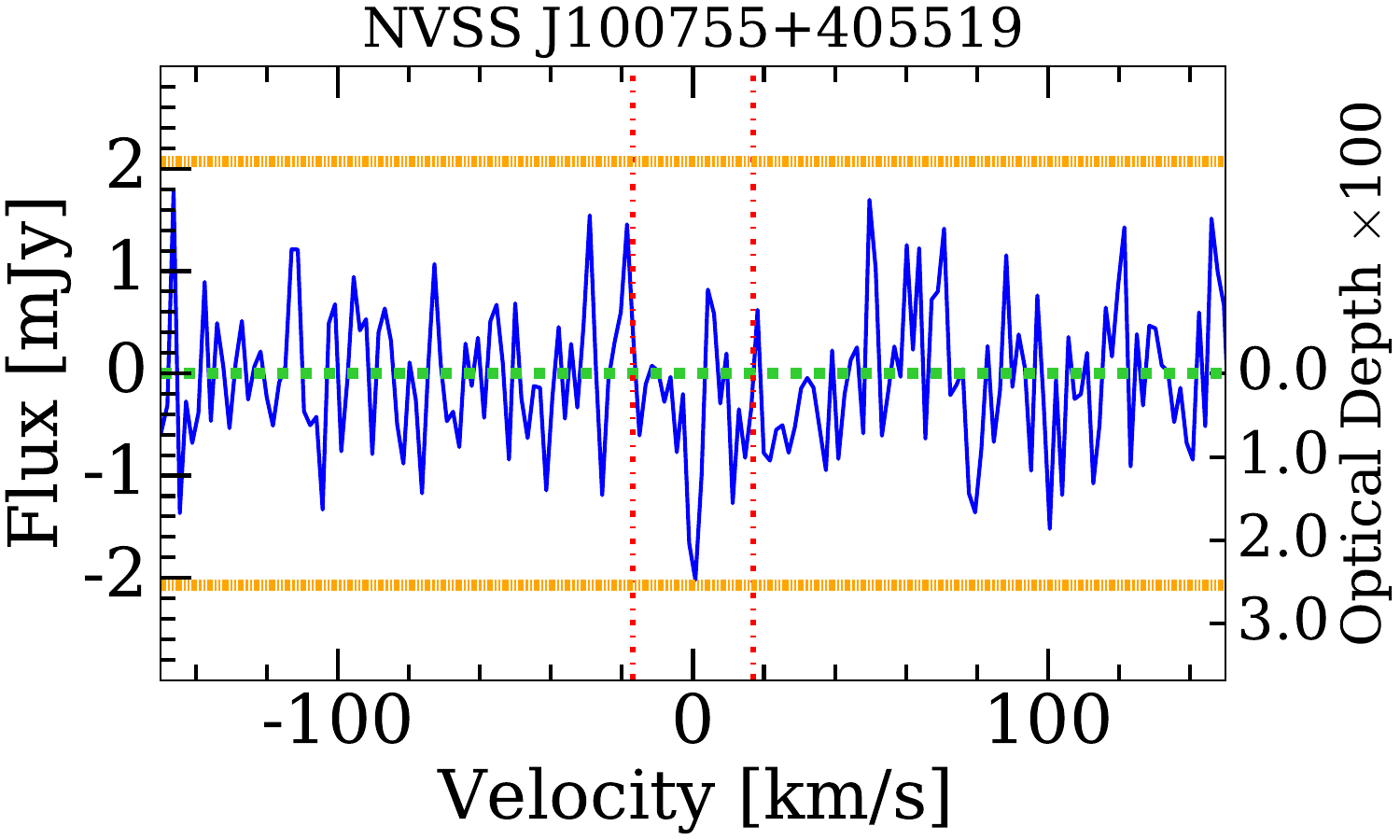}
    \includegraphics[width=0.32\textwidth]{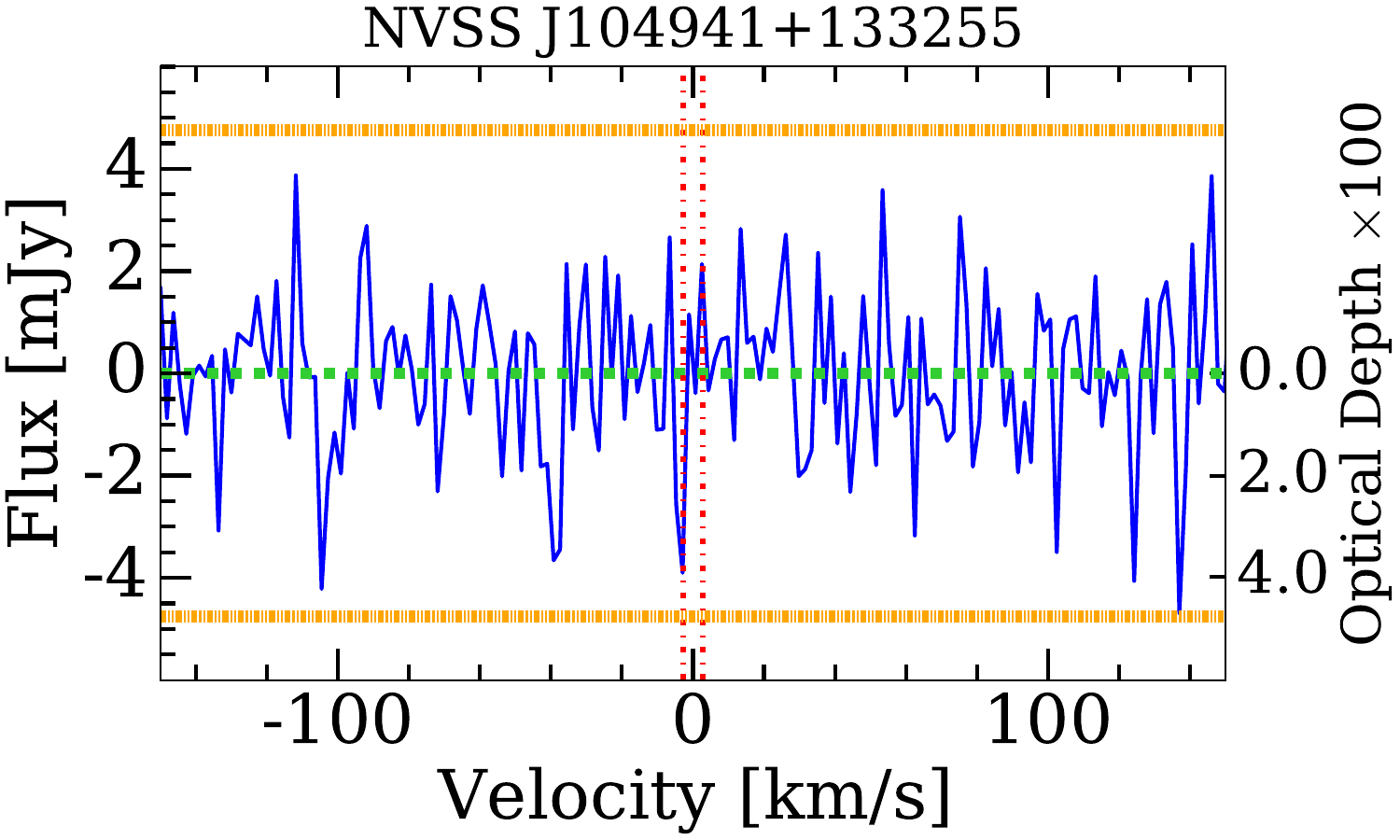}
    \includegraphics[width=0.32\textwidth]{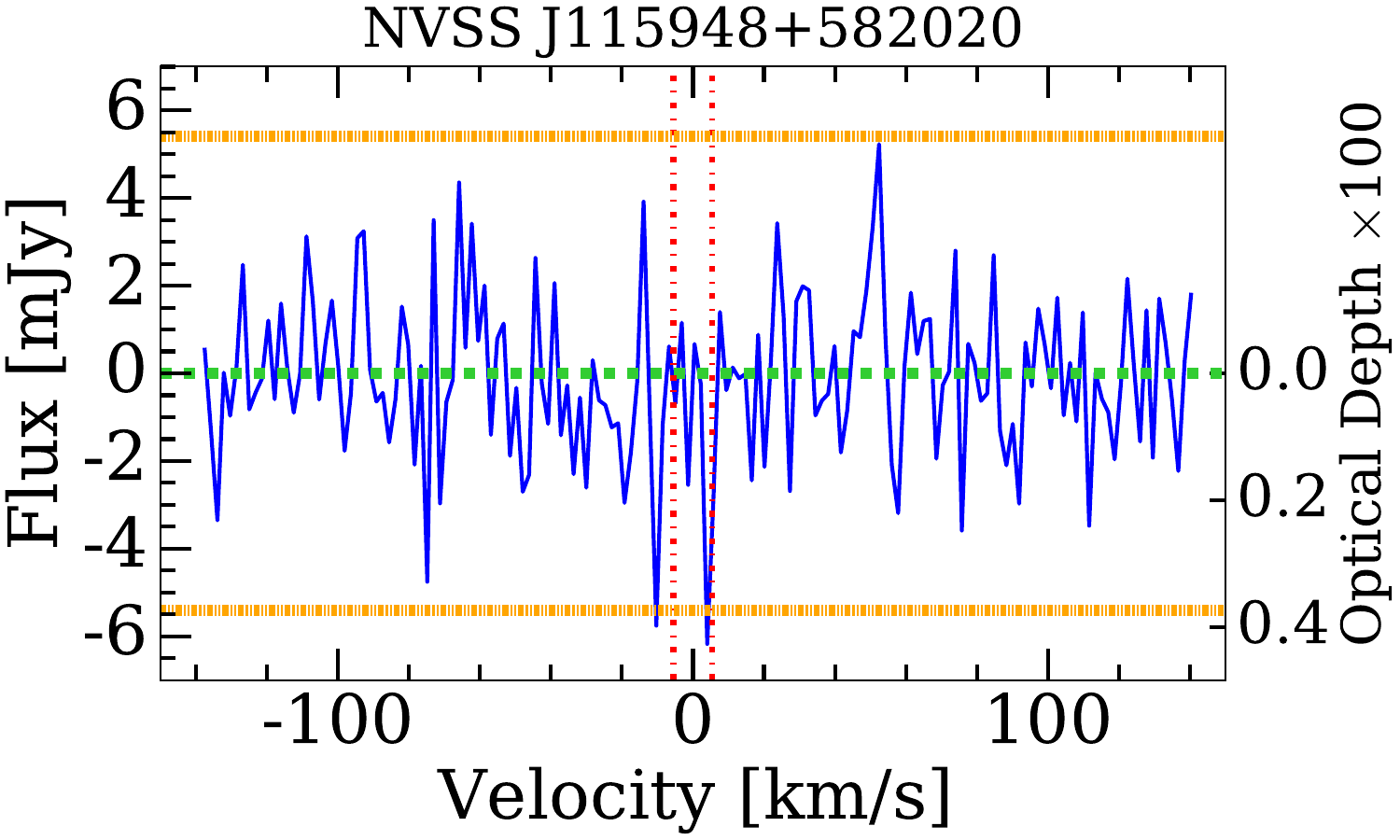}
    \includegraphics[width=0.32\textwidth]{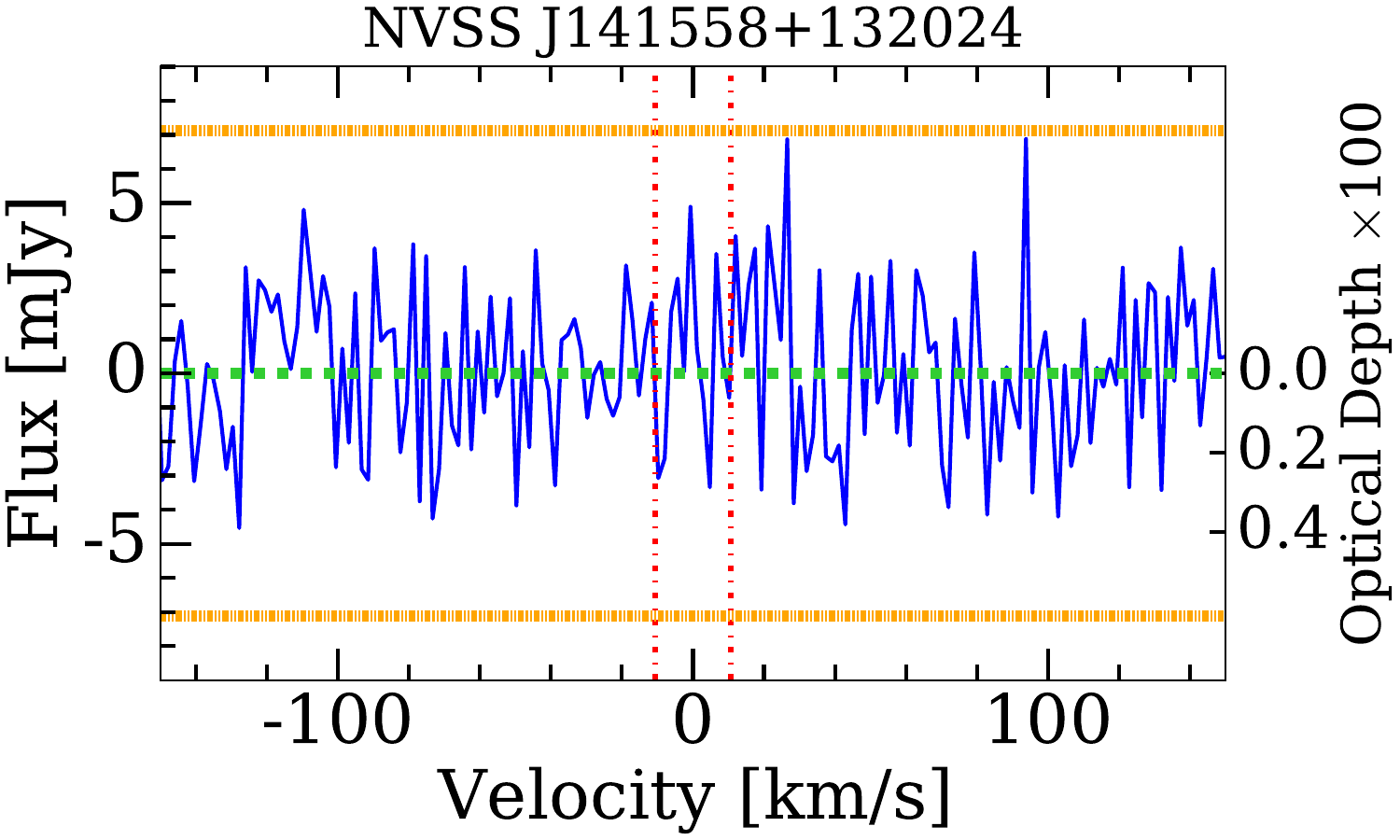}
    \includegraphics[width=0.32\textwidth]{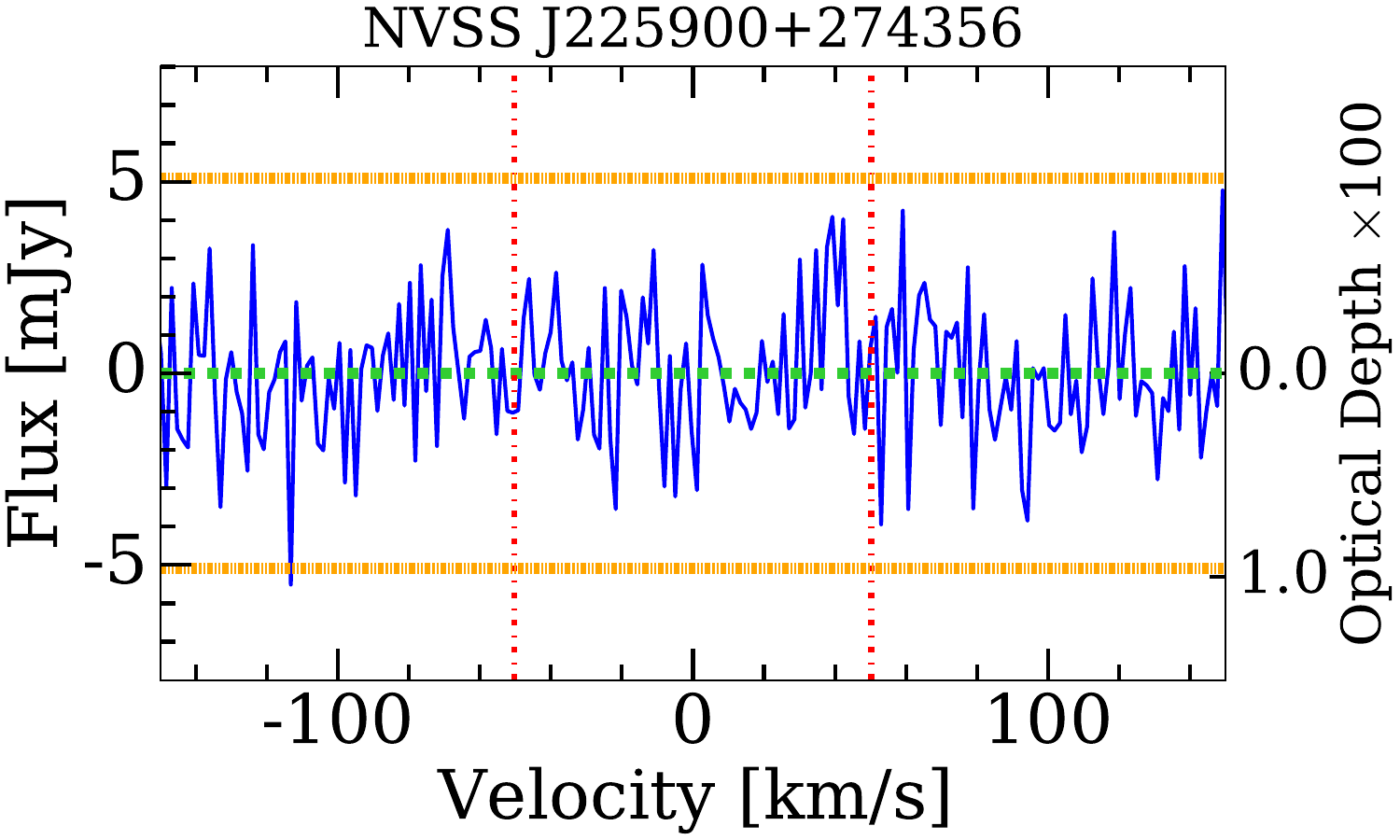}
    \includegraphics[width=0.32\textwidth]{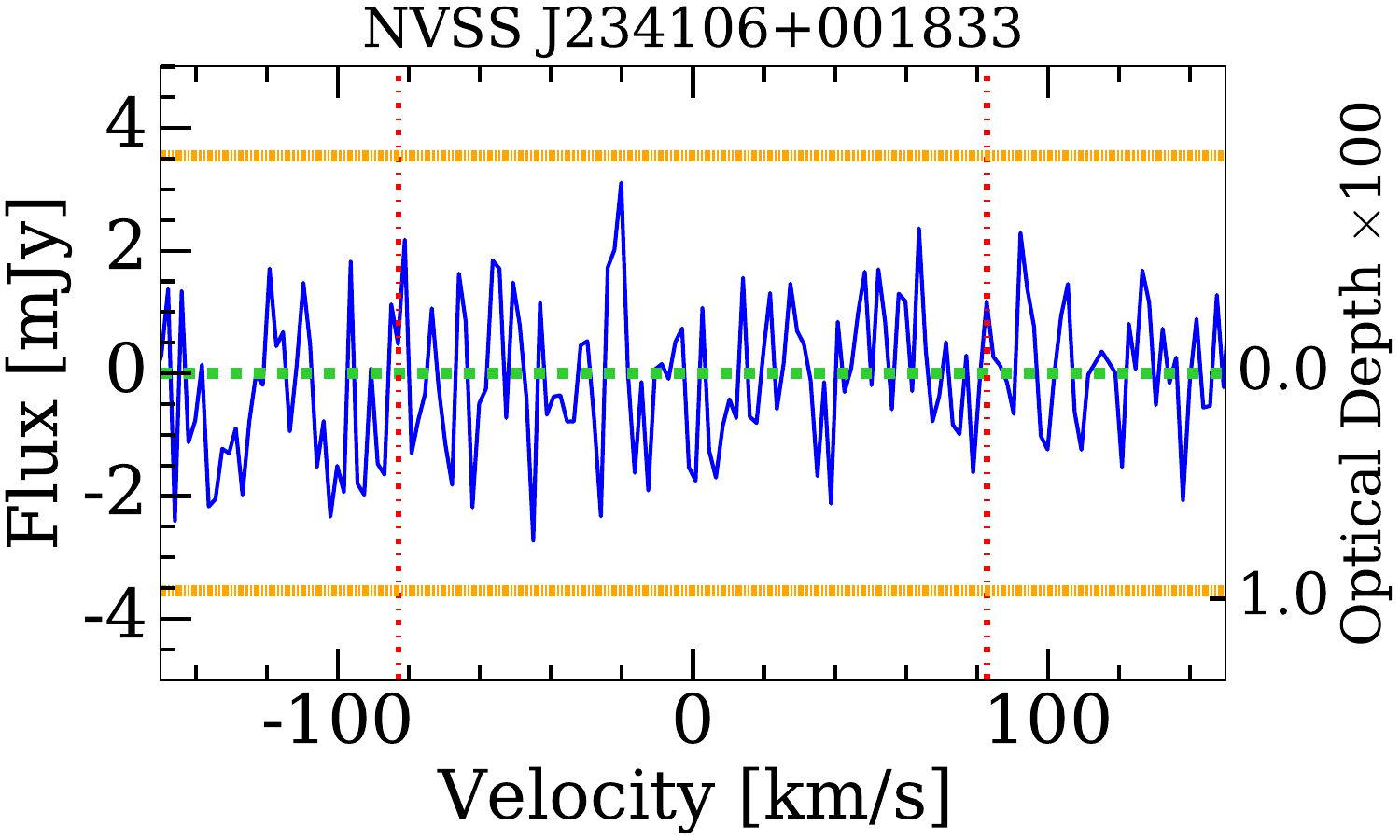}
    \caption{Continuation of Figure~\ref{OH_abs_1667_1}.}
    \label{OH_abs_1667_2}
\end{figure*}

\begin{table*}
    \fontsize{8}{10}\selectfont
	\centering
	\caption{The spectral RMS noise, integration time, t$_{\rm{int}}$, measured velocity integrated optical path or 3$\sigma$ upper limit of the \oh\ 1612, 1665, 1667, and 1720 MHz absorptions for radio targets where \hi\ absorption was detected. The associated, intervening, and unknown types are labeled with stars, diamonds, and crosses, respectively.}
	\label{OH_absorption_table}
	\begin{tabular}{|c|c|c|c|c|c|c|c|c|c|c|}
	\hline
        Radio Source & z$_{\rm{abs}}$ & t$_{\rm{int}}$ & rms$_{1612}$ & $\int\tau_{\rm{1612}} dv$ & rms$_{1665}$ & $\int\tau_{\rm{1665}} dv$ & rms$_{1667}$ & $\int\tau_{\rm{1667}} dv$ & rms$_{1720}$ & $\int\tau_{\rm{1720}} dv$\\
        & & (mins) & (mJy) & (\kms) & (mJy) & (\kms) & (mJy) & (\kms) & (mJy) & (\kms)\\
	  \hline
        NVSS J004219+570836$^{\diamond}$ & 0.264 & 26.3 & 3.05 & $<$0.04 & 1.38 & $<$0.02 & 1.31 & $<$0.02 & 1.52 & $<$0.02\\
        NVSS J010015+201710$^{\diamond}$ & 0.266 & 17.3 & RFI & RFI & 0.61 & $<$0.09 & 0.70 & $<$0.10 & 0.64 & $<$0.09\\
        4C +56.02$^{\otimes}$ & 0.265 & 26.5 & 3.04 & $<$0.04 & 1.22 & $<$0.01 & 1.29 & $<$0.02 & 1.19 & $<$0.01\\
        NVSS J011322+251852$^{\diamond}$ & 0.255 & 26.4 & 0.55 & $<$0.23 & 0.51 & $<$0.21 & 0.49 & $<$0.20 & 0.46 & $<$0.18\\
        4C +26.07$^{\otimes}$ & 0.251 & 26.1 & RFI & RFI & 0.52 & $<$0.05 & 0.53 & $<$0.05 & 0.52 & $<$0.05\\
        NVSS J033529+195621$^{\otimes}$ & 0.292 & 19.1 & 0.59 & $<$0.23 & 1.41 & $<$0.62 & 1.51 & $<$0.67 & 0.64 & $<$0.23\\
        NVSS J073755+264652$^{\star}$ & 0.239 & 24.3 & 0.63 & $<$0.23 & 0.53 & $<$0.19 & 0.51 & $<$0.18 & 0.47 & $<$0.17\\
        NVSS J090150+030422$^{\star}$ & 0.289 & 17.1 & 1.13 & $<$0.08 & 1.05 & $<$0.07 & 1.08 & $<$0.07 & 0.97 & $<$0.06 \\
        NVSS J090150+030422 HVC$^{\star}$ & 0.287 & 17.1 & 1.00 & $<$0.06 & 1.40 & $<$0.10 & 1.07 & $<$0.07 & 1.05 & $<$0.07\\
        NVSS J092351+281527$^{\diamond}$ & 0.331 & 4.5 & 2.64 & $<$0.11 & 1.71 & $<$0.07 & 1.60 & $<$0.06 & 1.99 & $<$0.08\\
        NVSS J093150+254034$^{\diamond}$ & 0.268 & 4.0 & RFI & RFI & 1.46 & $<$0.22 & 1.50 & $<$0.22 & 1.53 & $<$0.22\\
        NVSS J095812+112643$^{\otimes}$ & 0.320 & 24.8 & 1.21 & $<$0.09 & 4.43 & $<$0.33 & 6.18 & $<$0.39 & 0.84 & $<$0.06\\
        NVSS J100755+405519$^{\otimes}$ & 0.224 & 26.8 & 0.59 & $<$0.16 & 0.65 & $<$0.17 & 0.70 & $<$0.18 & 0.65 & $<$0.17\\
        NVSS J104941+133255$^{\diamond}$ & 0.247 & 4.4 & 1.96 & $<$0.43 & 1.85 & $<$0.39 & 1.58 & $<$0.35 & 1.70 & $<$0.35\\
        NVSS J115948+582020$^{\diamond}$ & 0.238 & 12.3 & 1.95 & $<$0.03 & 1.90 & $<$0.03 & 1.80 & $<$0.03 & 1.85 & $<$0.03\\
        NVSS J141558+132024$^{\diamond}$ & 0.247 & 4.5 & 2.87 & 0.07 & 2.39 & $<$0.05 & 2.38 & $<$0.05 & 2.40 & -0.09 \\
        NVSS J225900+274356$^{\star}$ & 0.143 & 4.4 & 1.82 & $<$0.07 & 1.68 & $<$0.06 & 1.70 & $<$0.06 & Off Band & Off Band \\
        NVSS J234106+001833$^{\star}$ & 0.277 & 13.1 & RFI & RFI & 1.56 & $<$0.10 & 1.18 & $<$0.07 & 0.90 & $<$0.05\\
        NVSS J024516+240535$^{\diamond}$ & 0.141 & 12.6 & 0.99 & $<$0.05 & 0.99 & $<$0.04 & 1.04 & $<$0.05 & Off Band & Off Band \\
        \hline
        \end{tabular}
\end{table*}

\section{\oh\ abundance}
\label{sec:OH_abundance}

\subsection{Correlation Between $N_{\hi}$ and $N_{\rm{OH}}$}

Using survival statistics, \citet{2020MNRAS.499.3085Z} investigated the relationship between \oh\ and \hi\ abundances. They reported that both associated and intervening absorbers exhibit a weak positive correlation between $N_{\hi}$ and $N_{\rm{OH}}$ (generalized Spearman’s $\rho$=0.68 and
0.12 for associated and intervening absorbers), as well as between redshift and $N_{\rm{OH}}$/$N_{\hi}$ (generalized Spearman’s $\rho$=0.56 and
0.65 for associated and intervening absorbers). However, the reliability of those statistics was limited by the sample selection bias (see more discussion in Section~\ref{sec:sample_bias}). By incorporating the detection and upper limits derived from our FAST search, we restrict the similar analysis to the FAST sample and refine these measurements using an unbiased sample.

\begin{table*}
    \fontsize{6.8}{8}\selectfont
	\centering
	\caption{Basic observed parameters of the six known extragalactic \oh\ absorbers. From left to right: source name, redshift of the absorption($z_{\rm{abs}}$), covering factor for \oh\ absorption($c_{\rm f,\oh}$), excitation temperature(T$_{ex}$), integrated optical path for 1667 MHz \oh\ absorption ($\int\tau_{1667}dv$), \oh\ column density($N_{\mathrm{\oh}}$), covering factor for \hi\ absorption($c_{\rm f,\hi}$), spin temperature(T$_{s}$), integrated optical path for \hi\ absorption($\int\tau_{\mathrm{\hi}}dv$), \hi\ column density($N_{\mathrm{\hi}}$), and reference literature. The associated and intervening types are labeled with stars and diamonds, respectively. Reference abbreviations are as follows: C99: \citet{1999A&A...343L..79C}, KC02:\citet{2002A&A...381L..73K}, K03:\citet{2003MNRAS.345L...7K},KB03:\citet{2003A&A...412L..29K}, C03:\citet{2003ApJ...583...67C}, C97:\citet{1997ApJ...474L..89C}, C92:\citet{1992ApJ...400L..13C}, Z20:\citet{2020MNRAS.499.3085Z}, H25:\citet{2025ApJS..277...25H}, K12:\citet{2012ApJ...746L..16K}, G18:\citet{2018ApJ...860L..22G}.}
	\label{known_OH_abs_table}
	\begin{tabular}{|c|c|c|c|c|c|c|c|c|c|c|}
	\hline
	Source Name & $z_{\rm{abs}}$ & $c_{\rm f,\oh}$ & T$_{ex}$ & $\int\tau_{1667}dv$ & $N_{\mathrm{\oh}}$ & $c_{\rm f,\hi}$ & T$_{s}$ & $\int\tau_{\mathrm{\hi}}dv$ & $N_{\mathrm{\hi}}$ & Refs\\
	& & & (K) & (\kms) & (10$^{15}$cm$^{-2}$) & & (K) & (\kms) & (10$^{20}$cm$^{-2}$) & \\
    \hline
    PKS 1830-211$^{\diamond}$   & 0.88582 & 0.36 & 10 & 1.83 & 11.39 & 0.33 & 100 & 5.8 & 31.72 & C99,KC02\\
    B0218+357$^{\diamond}$      & 0.68468 & 0.4 & 10 & 0.40 & 2.24 & 1 & 100 & 2.94 & 5.35 & K03,KB03,C03\\
    B3 1504+377$^{\star}$    & 0.67343 & 0.46 & 10 & 0.448 & 2.18 & 0.74 & 100 & 21 & 51.73 & C97,KC02\\
    PKS 1413+135$^{\diamond}$   & 0.24671 & 0.044 & 10 & 0.023 & 1.17 & 0.1 & 150 & 7.13 & 194.97 &  C92,KC02,Z20,H25\\
    PMN J0134-0931$^{\diamond}$ & 0.7645  & 1 & 10 & 2.19 & 4.9 & 1 & 200 & 7.06 & 25.74 & KB03,K12\\
    G0248+430$^{\diamond}$      & 0.0519  & 1 & 3.5 & 0.08 & 0.063 & 1 & 70 & 0.43 & 0.55 & G18\\
    \hline
    \end{tabular}
\end{table*}

\begin{figure}[hbt]
    \centering
    \includegraphics[width=8.2cm]{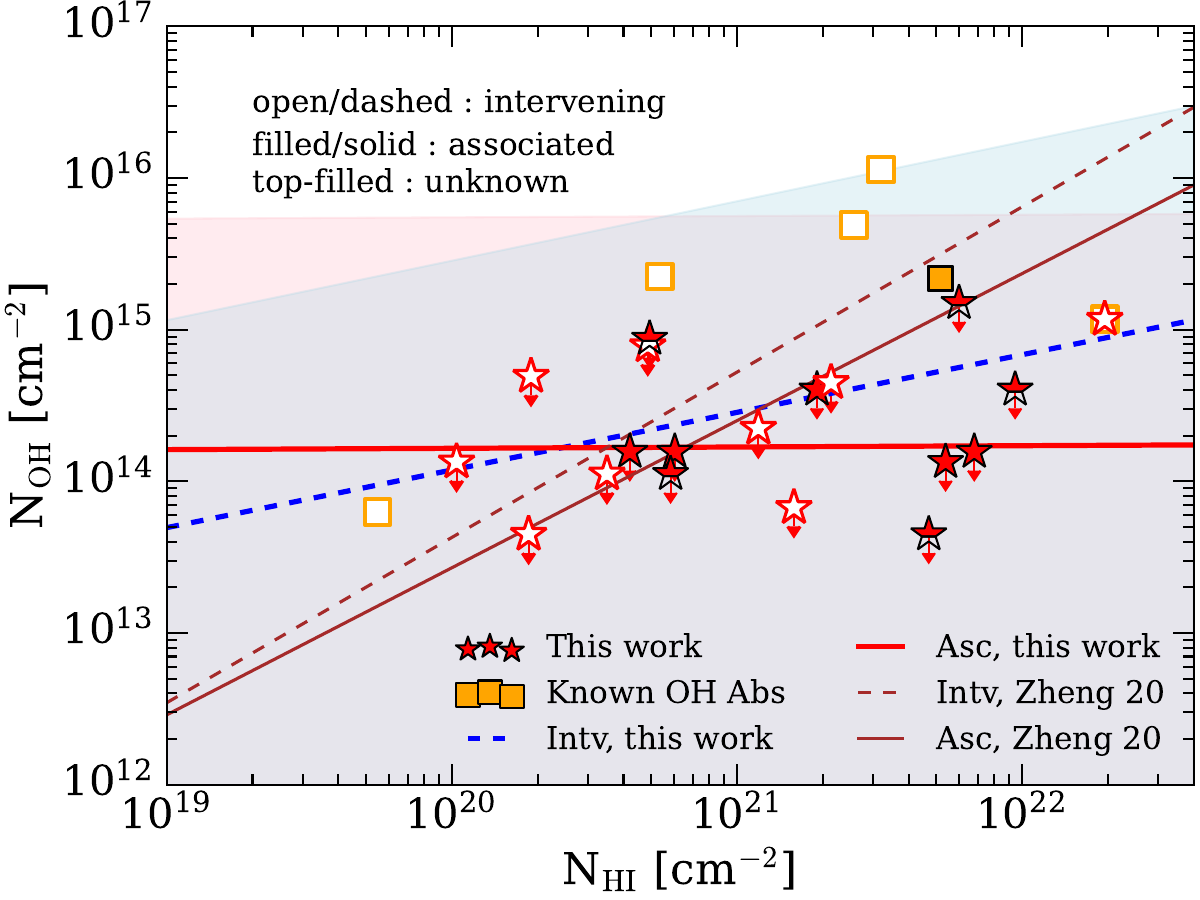}
    \includegraphics[width=8.2cm]{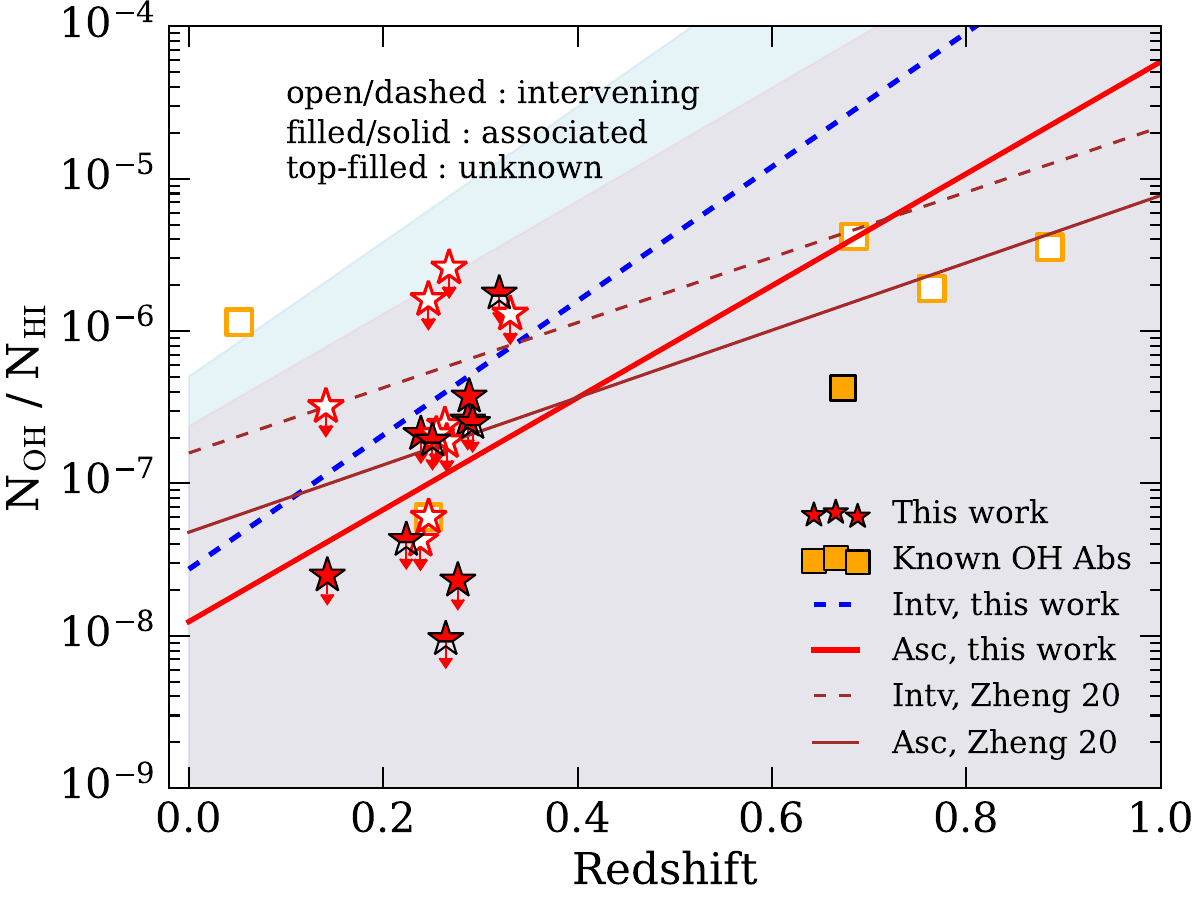}
    \caption{The correlation analysis between $N_{\hi}$ and $N_{\rm{OH}}$. Upper panel: $N_{\hi}$ versus $N_{\rm{OH}}$. Lower panel: $N_{\rm{OH}}$/$N_{\hi}$ ratio as a function of redshift. Red stars represent the detection and 3$\sigma$ upper limits from the sample analyzed in this work. Orange squares mark the six currently known \oh\ absorption detections. The red-solid and blue-dashed lines show the fits for the associated and intervening absorbers from our FAST sample, respectively, obtained using Bayesian censored regression with linmix. The shaded light pink and light blue regions indicate the 1$\sigma$ uncertainties in the fitted relations along the y-axis. The brown lines show the fits reported by \citet{2020MNRAS.499.3085Z}. Open symbols and dashed lines represent intervening absorbers, whereas filled symbols and solid lines represent associated absorbers.}
    \label{N_OH_HI}
\end{figure}

We examine the correlation between $N_{\hi}$ and $N_{\rm{OH}}$, with the results presented in Figure~\ref{N_OH_HI}. For comparison, we overplot the known \oh\ absorbers together with our sample. The adopted observed parameters of the six known extragalactic \oh\ absorptions are presented in Table~\ref{known_OH_abs_table}. Based on the VLBI observations of \citet{2021ApJ...907...61R}, we classify the absorption in PKS 1413\allowbreak+135 as an intervening absorber, in contrast to previous studies that treated it as an associated system. For the other sources, we calculate the corresponding column densities assuming a covering factor of unity and adopting spin temperature of 100 K for \hi\ and an excitation temperature of 10 K for \textsc{OH}. 

Linear regressions were performed separately for associated and intervening absorbers using survival statistics \citep{1985ApJ...293..192F} combined with Bayesian censored regression \citep{2007ApJ...665.1489K}, which allows us to incorporate both detections and upper limits in a statistically robust manner. The regression analysis yields a relation for intervening absorbers of log$N^{\rm{intv}}_{\oh}$=(6.47$\pm$5.95)+(0.38$\pm$0.28)$\times$log$N^{\rm{intv}}_{\hi}$, with a posterior-imputed Spearman correlation coefficient $\rho^{\rm{intv}}$=0.53$\pm$0.20, and a relation for associated absorbers of log$N^{\rm{asc}}_{\oh}$=(14.05$\pm$12.61)+(0.01$\pm$0.62)$\times$log$N^{\rm{asc}}_{\hi}$, with $\rho^{\rm{asc}}$=0.10$\pm$0.50.

For intervening absorbers, the censored-regression fit yields a positive median slope between $N_{\hi}$ and $N_{\rm{OH}}$; however, its uncertainty is large, and the present data do not provide statistically significant evidence for a correlation. For associated absorbers, no corresponding correlation is found. The shallow slopes of the regression fits, together with the low correlation coefficients, indicate that the abundance of \oh\ does not scale strongly with the strength of the neutral hydrogen absorption. This suggests that while the presence of cold \hi\ gas is a necessary condition for \oh\ absorption, it is not sufficient to guarantee its detection. The negligible correlation observed in associated absorbers ($\rho^{\rm{asc}} \approx 0.1$) highlights the complexity of ISM conditions within host galaxies. In such environments, the balance between atomic and molecular phases is strongly influenced by local star formation, feedback processes, and the availability of shielding material. 

We also investigate the redshift evolution of the ratio $N_{\rm{OH}}$/$N_{\hi}$, as shown in the lower panel of Figure~\ref{N_OH_HI}. The regression analysis yields a relation for intervening absorbers of log[$N^{\rm{intv}}_{\oh}$/$N^{\rm{intv}}_{\hi}$]=(-7.56$\pm$1.63)+(4.40$\pm$6.42)$\times z$, with a posterior-imputed Spearman correlation coefficient $\rho^{\rm{intv}}$=0.37$\pm$0.30, and a relation for associated absorbers of log[$N^{\rm{asc}}_{\oh}$/$N^{\rm{asc}}_{\hi}$]=(-7.91$\pm$4.19)+(3.68$\pm$16.25)$\times z$, with $\rho^{\rm{asc}}$=0.50$\pm$0.52.

The analysis finds no statistically significant positive correlation between between $N_{\rm{OH}}$/$N_{\hi}$ and redshift for either intervening or associated absorbers. A possible increase of $N_{\rm{OH}}$/$N_{\hi}$ with redshift would be qualitatively consistent with scenarios in galaxies at higher redshift possess higher molecular gas fractions \citep{2010Natur.463..781T,2020ARA&A..58..157T}, elevated star-formation activity \citep{2014ARA&A..52..415M}, and enhanced conditions for molecule formation, such as increased pressure \citep{2009IAUS..254..289E} and turbulence within the interstellar medium \citep{2009ApJ...706.1364F,2010Natur.463..781T}. Such evolution could increase the molecular gas content relative to the atomic gas content at earlier cosmic times. However, the current sample is too limited, and the inferred relations have large uncertainties; therefore, no conclusion regarding redshift evolution can be drawn. Larger unbiased OH samples will be required to constrain any underlying relation.

%An increase trend in $N_{\rm{OH}}$/$N_{\hi}$} with redshift is consistent with scenarios in which galaxies at higher redshift possess higher molecular gas fractions \citep{2010Natur.463..781T,2020ARA&A..58..157T}, elevated star-formation activity \citep{2014ARA&A..52..415M}, and enhanced conditions for molecule formation, such as increased pressure \citep{2009IAUS..254..289E} and turbulence within the interstellar medium \citep{2009ApJ...706.1364F,2010Natur.463..781T}, resulting in a picture that the molecular content relative to atomic gas becomes more prominent at earlier cosmic times.  However, because of the limited sample size, our inferred relations are subject to large uncertainties, underscoring the need for larger samples to verify the correlation and improve the precision of the fitted relations.}

\subsection{\oh\ abundance relative to \hi}

Using our FAST catalog from the blind survey, we are able, for the first time, to place constraints on the unbiased \oh\ abundance relative to \hi, i.e., the upper limits on [\rm{OH}]/[\hi]. The average $N_{\hi}$ and the corresponding 3$\sigma$ upper limits on $N_{\oh}$ are derived through the spectral stacking method \citep{2019MNRAS.489.1619H}. The stacked optical depth spectrum is calculated as $\langle \tau(\nu)\rangle$=$\sum_{i=1}^{n}w_{i}\tau_{i}/\sum_{i=1}^{n}w_{i}$, where $n$ is the number of the stacked systems, $w_{i}$ is the weight factor, defined as 1/[$C(W_{i}, \mathrm{SN}_{i}, \nu_{i})\times \mathrm{rms}^{2}_{i}$]. Here, $\mathrm{rms}_{i}$ is the spectra noise level, $C$ denotes the completeness for $i$-th \hi\ absorbers with FWHM $W_{i}$, detected S/N $\mathrm{SN}_{i}$, and frequency $\nu_{i}$. The completeness correction accounts for absorption dilution caused by emission, RFI, and standing waves. It is estimated by injecting mock absorption lines into real data and measuring the detection fraction based on our search method and selection threshold. 

\begin{figure*}[hbt]
    \centering
    \includegraphics[width=0.24\textwidth]{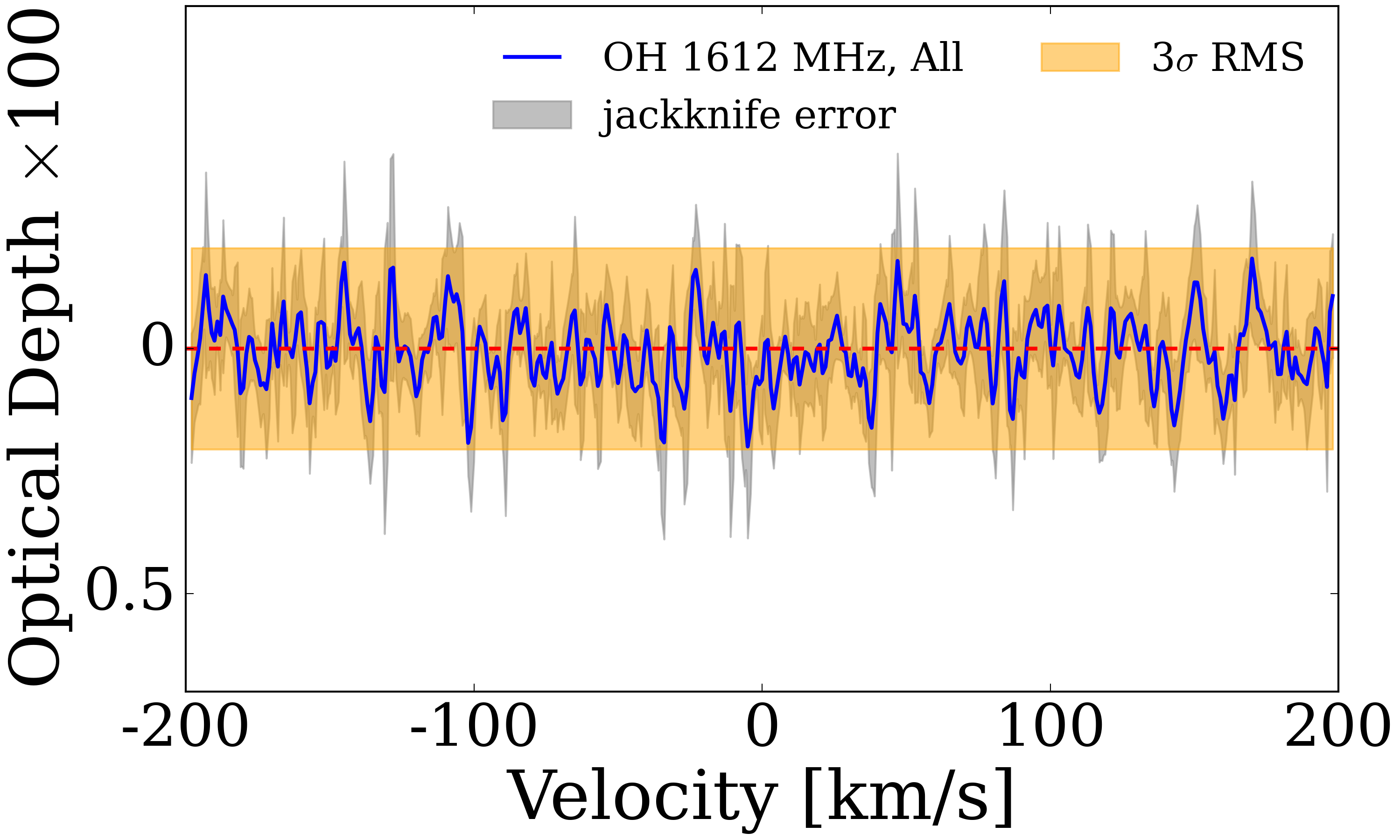}
    \includegraphics[width=0.24\textwidth]{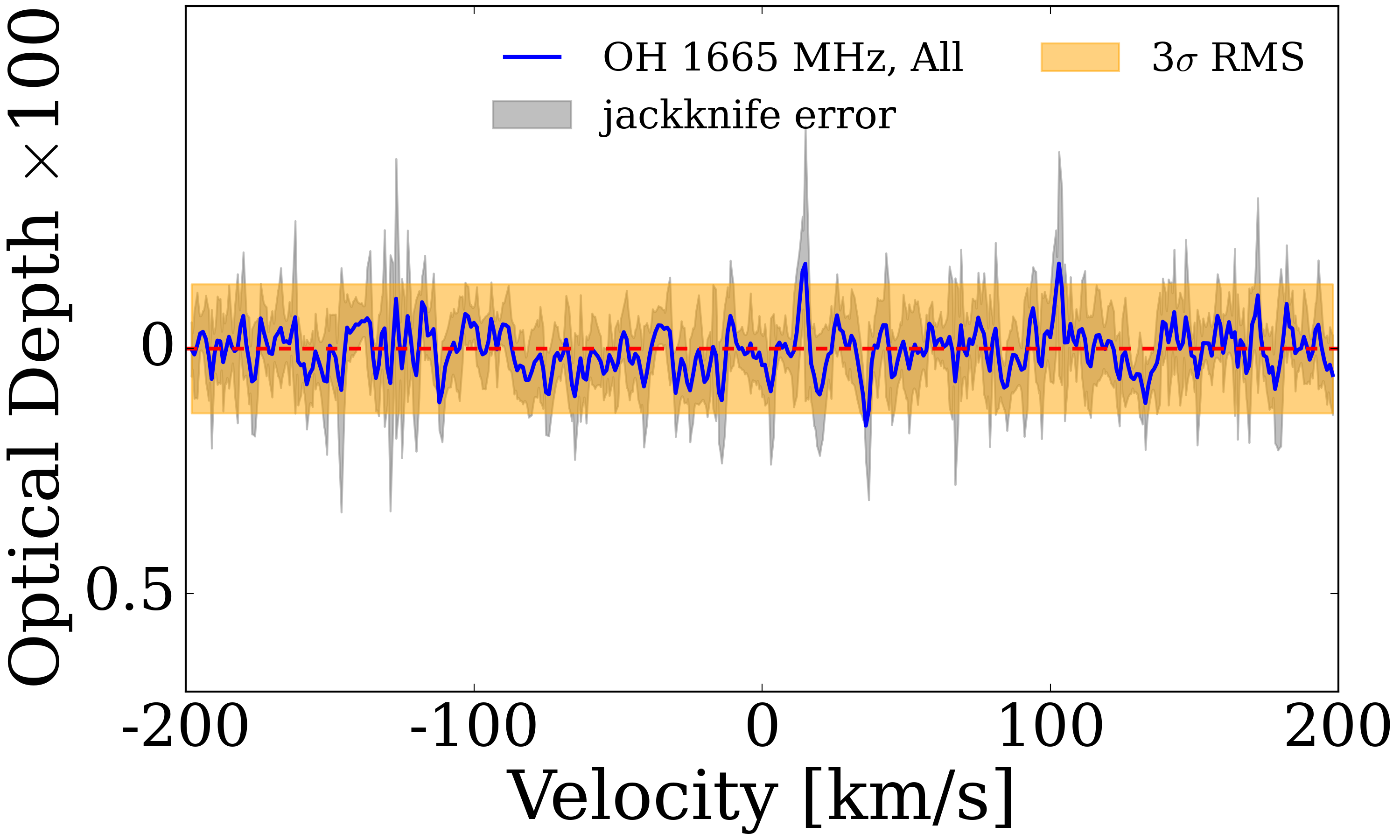}
    \includegraphics[width=0.24\textwidth]{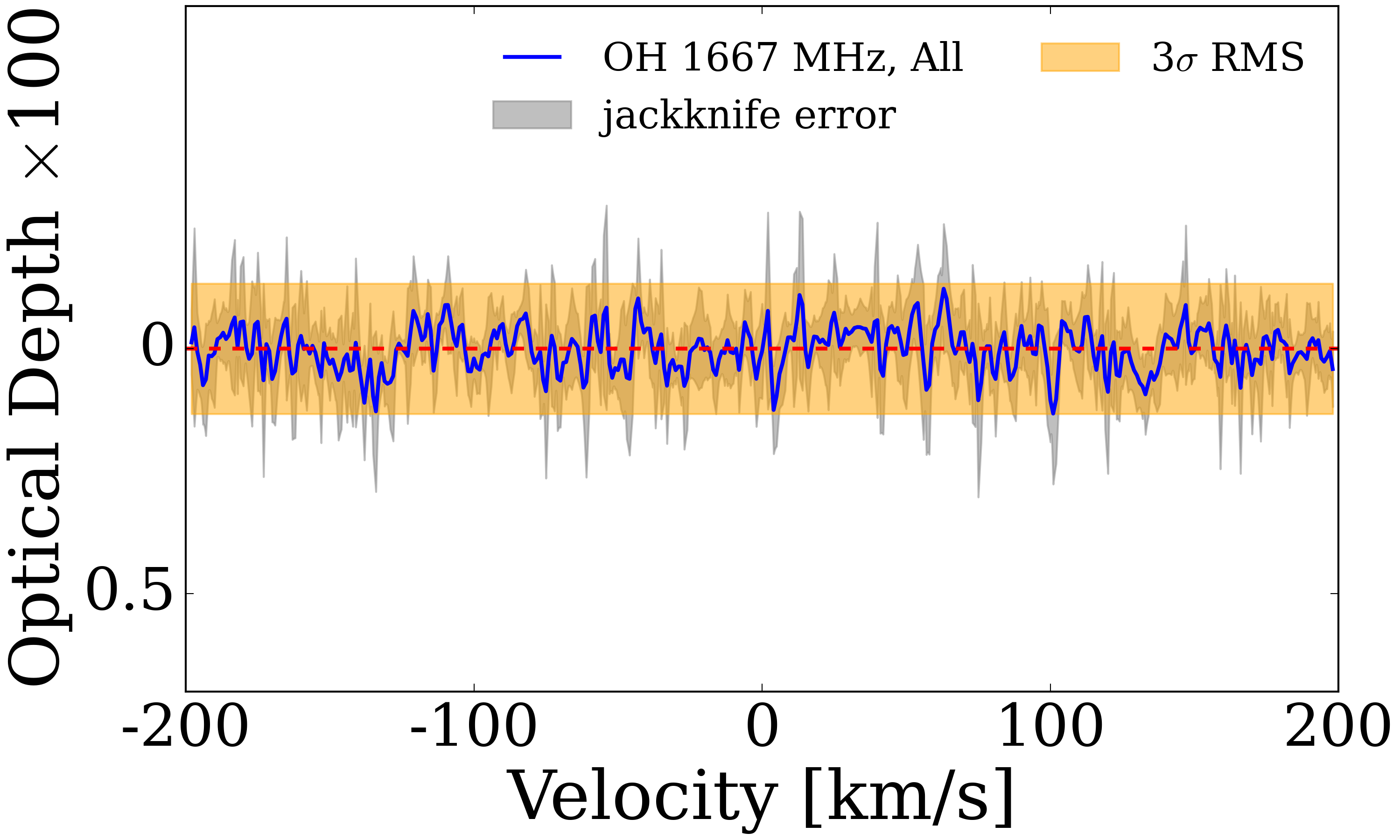}
    \includegraphics[width=0.24\textwidth]{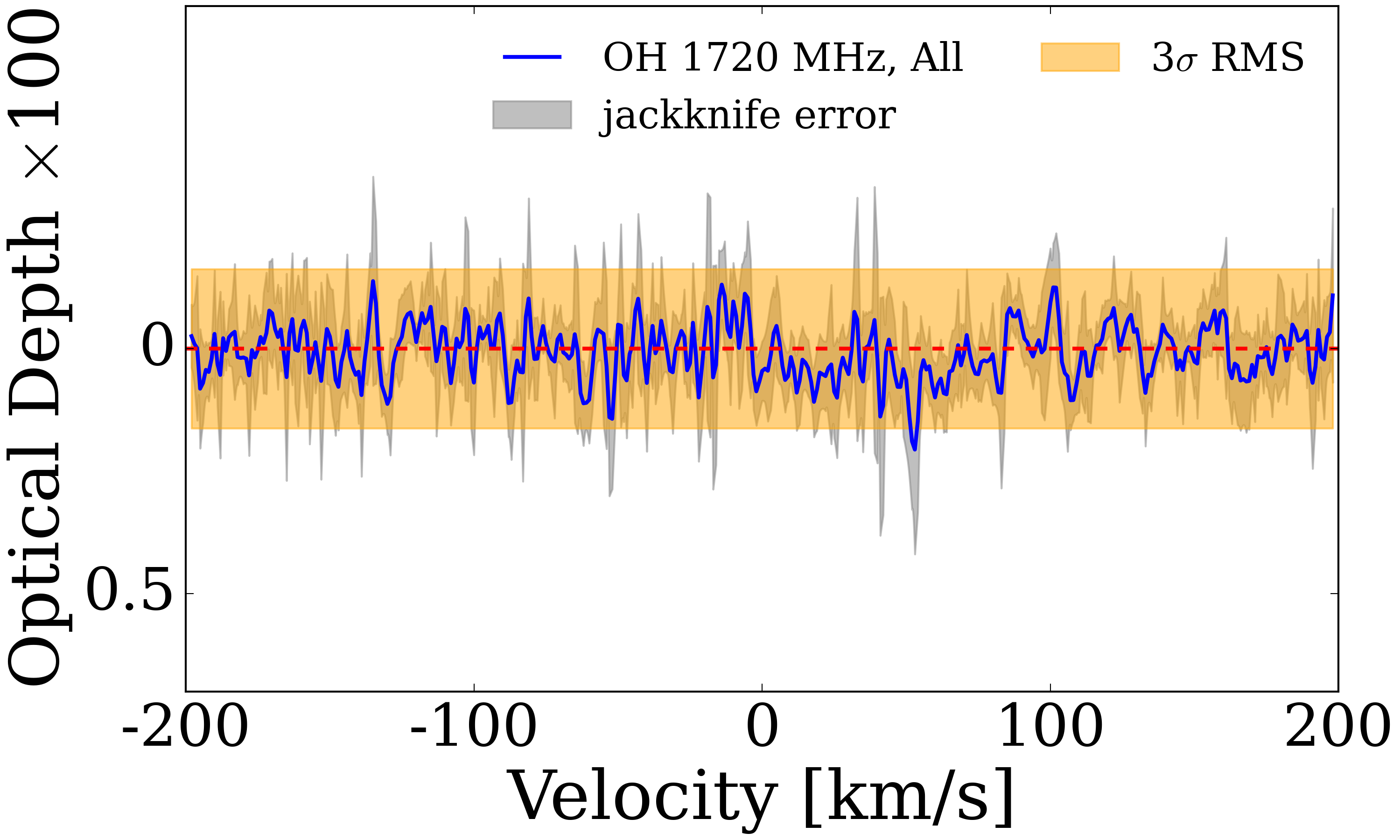}
    \includegraphics[width=0.24\textwidth]{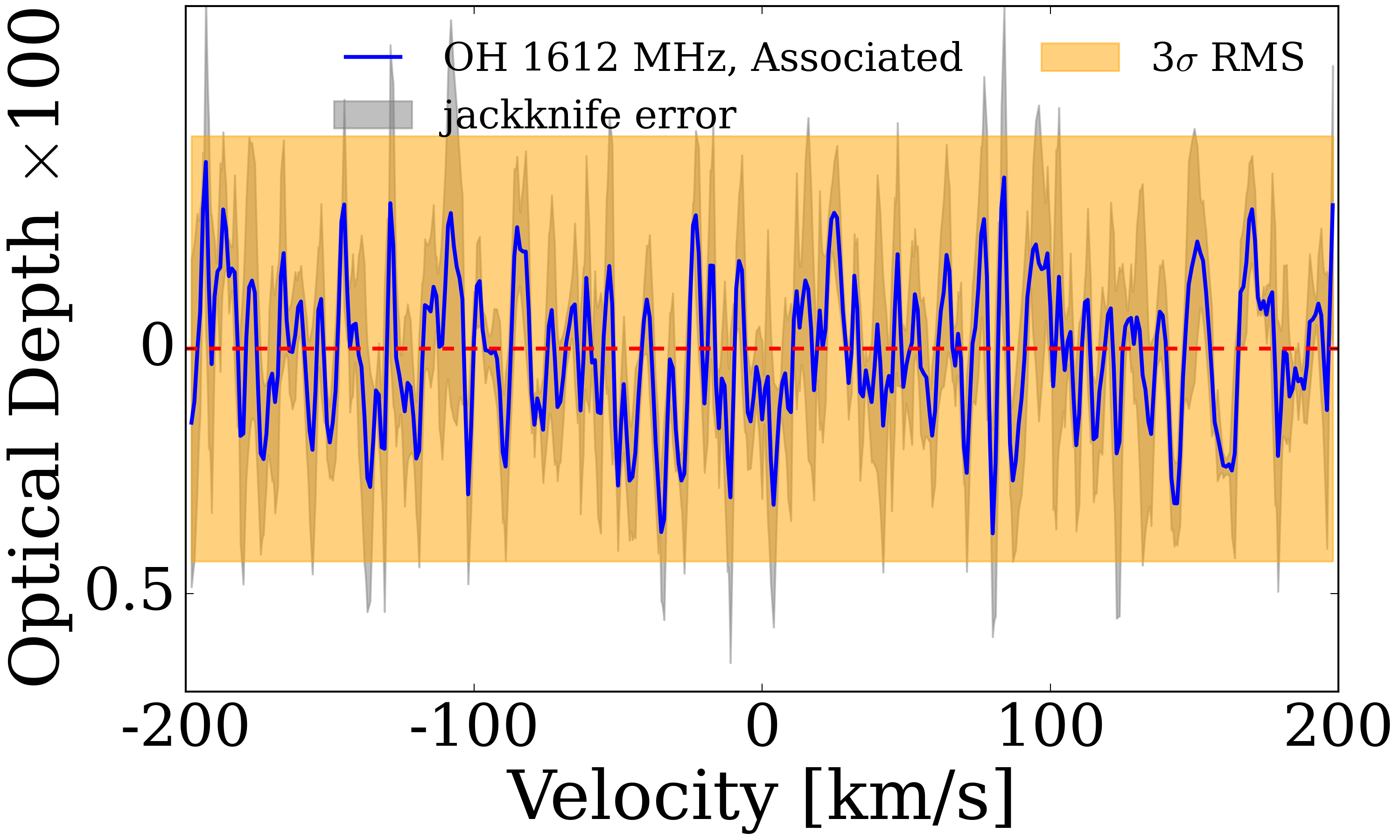}
    \includegraphics[width=0.24\textwidth]{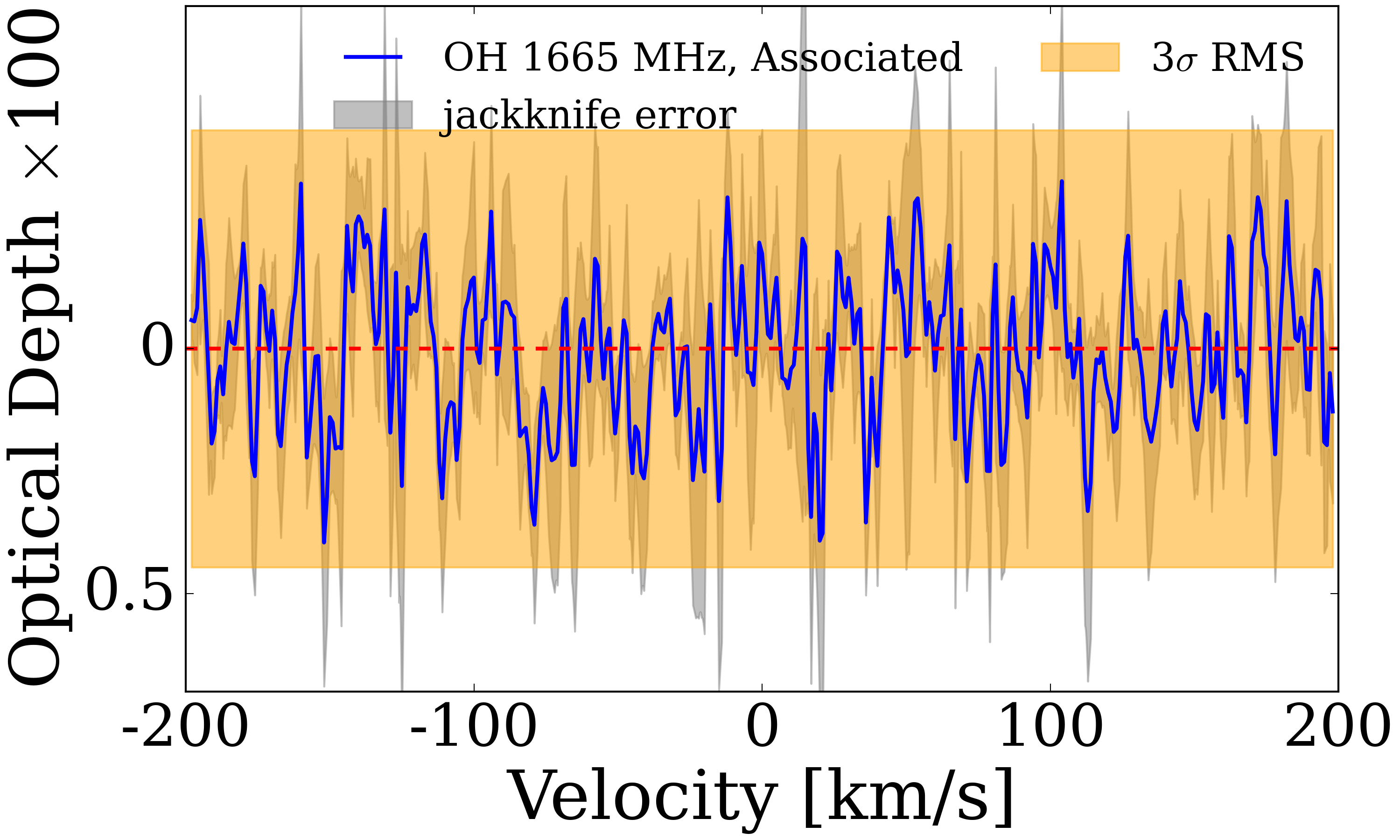}
    \includegraphics[width=0.24\textwidth]{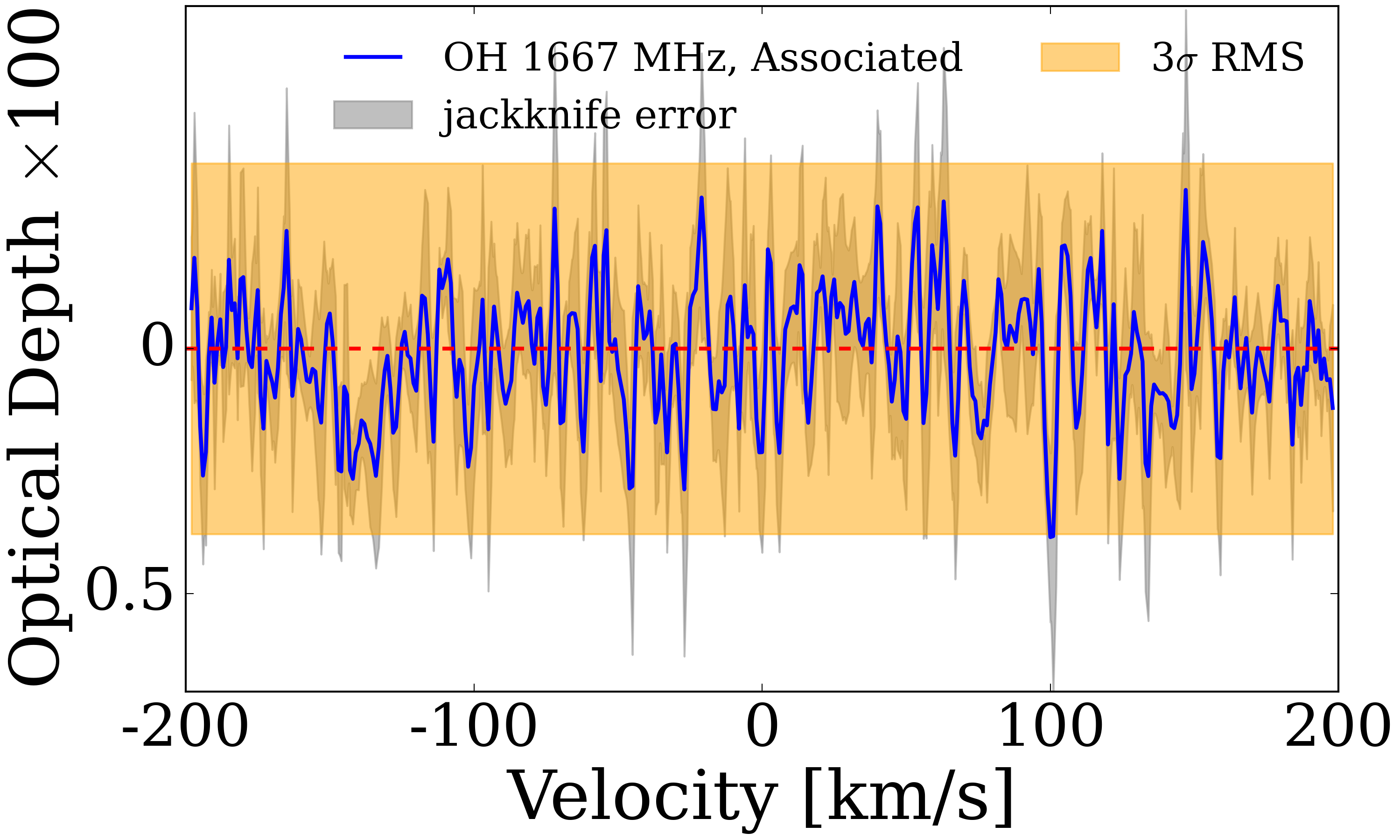}
    \includegraphics[width=0.24\textwidth]{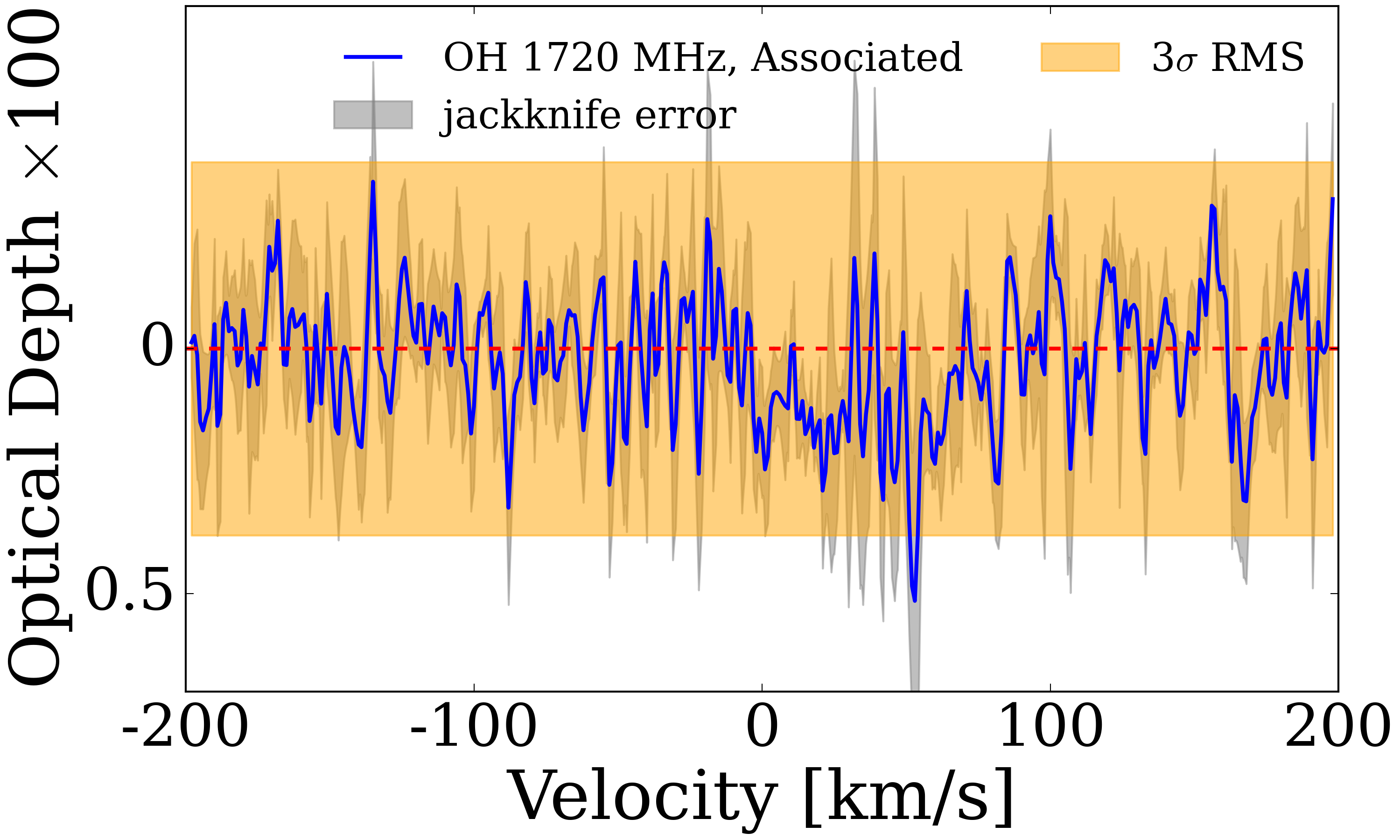}
    \includegraphics[width=0.24\textwidth]{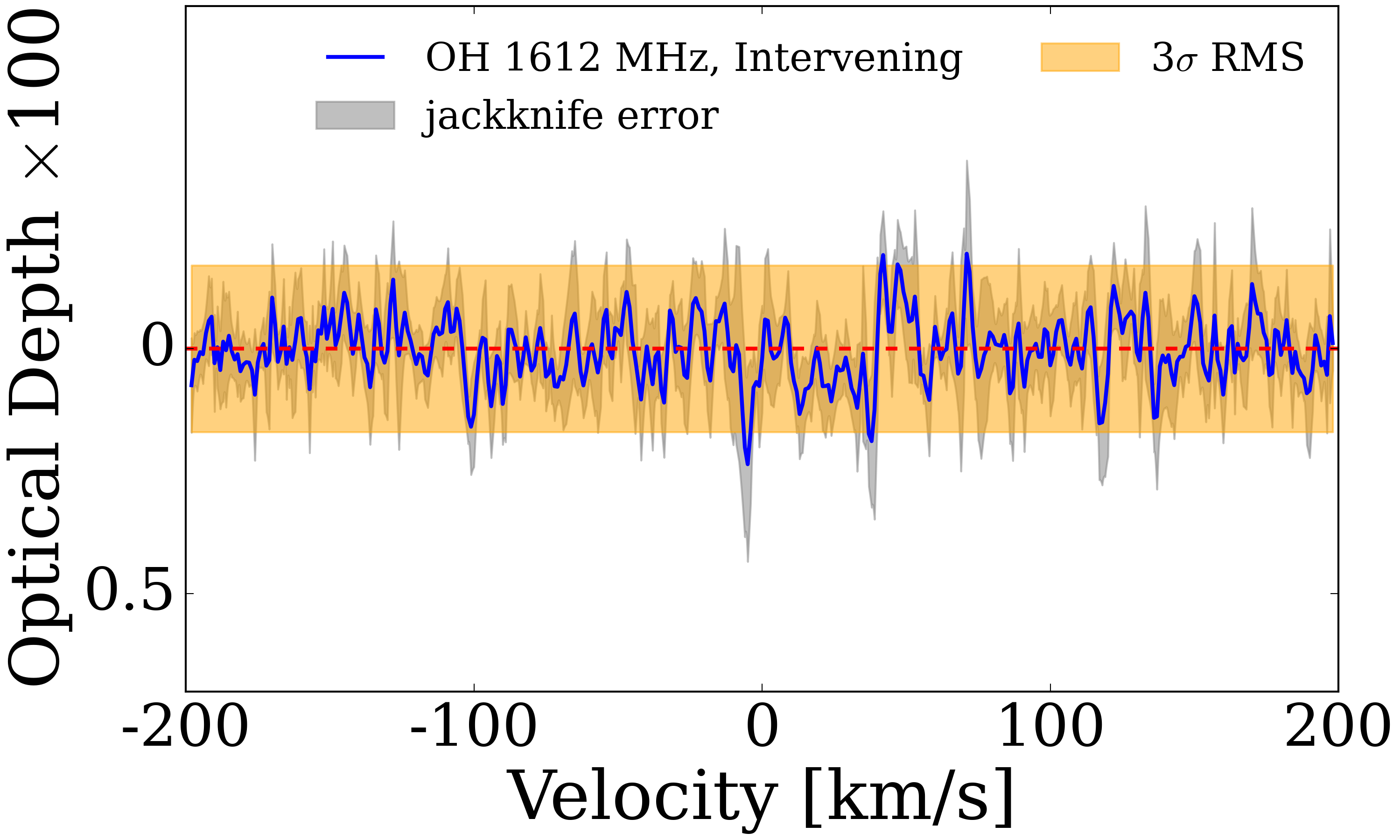}
    \includegraphics[width=0.24\textwidth]{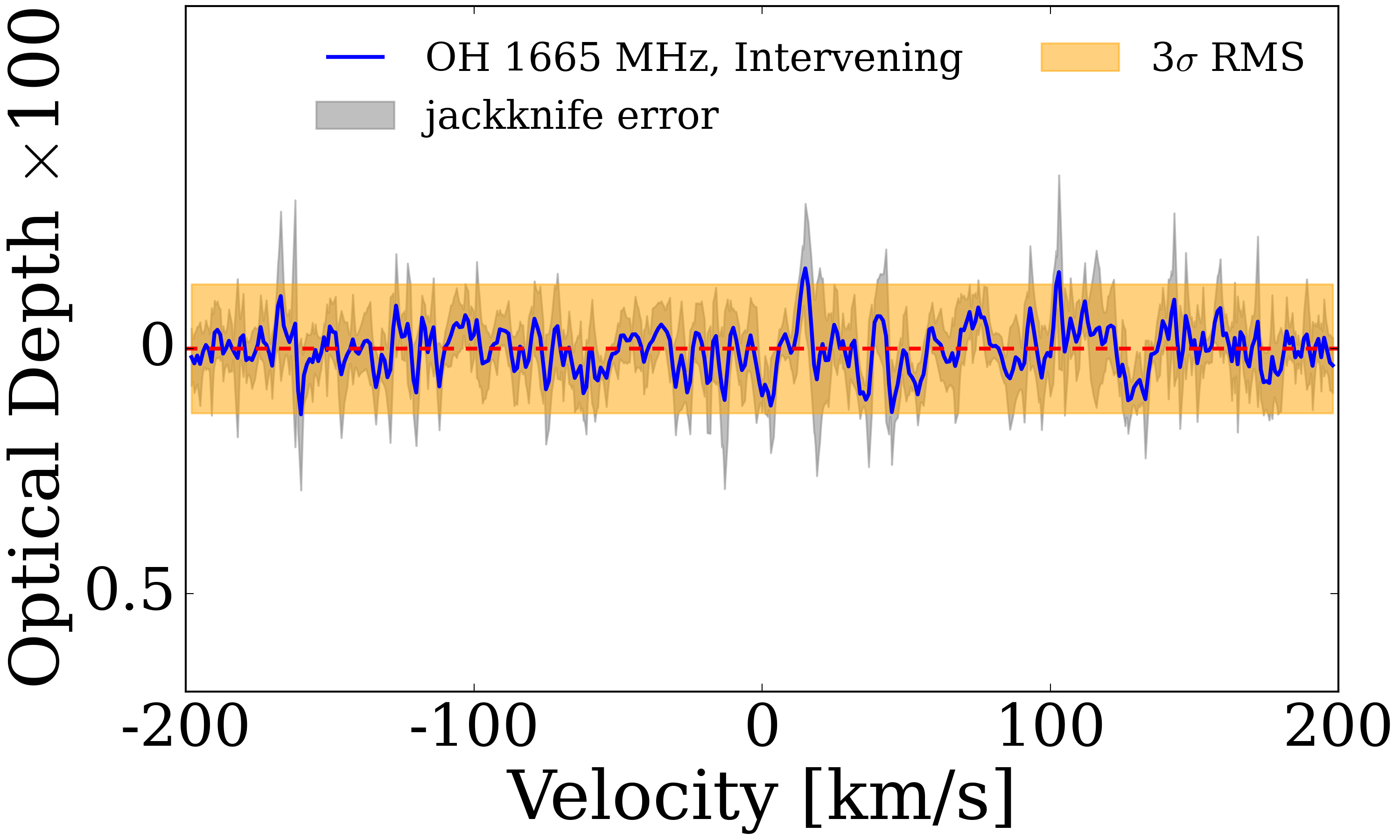}
    \includegraphics[width=0.24\textwidth]{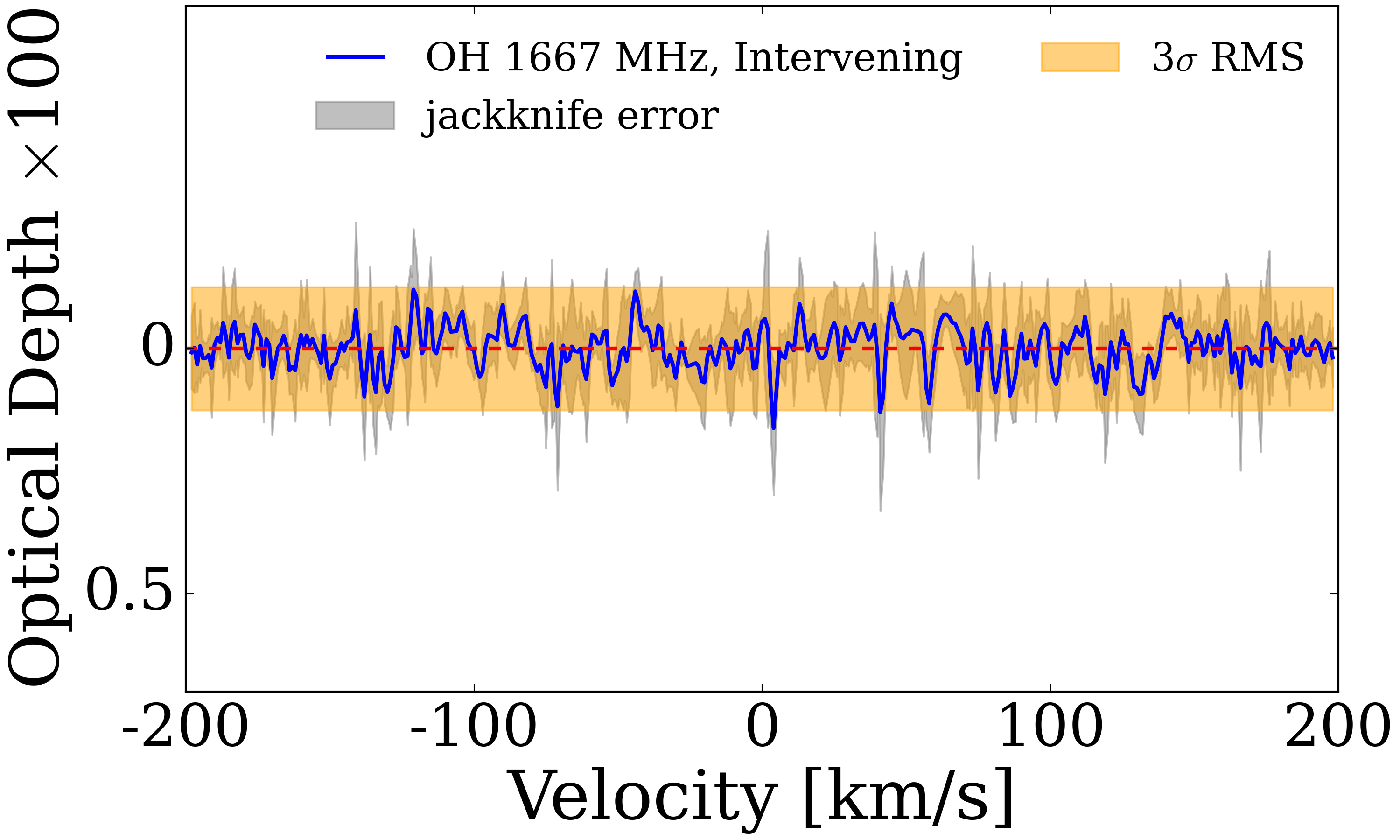}
    \includegraphics[width=0.24\textwidth]{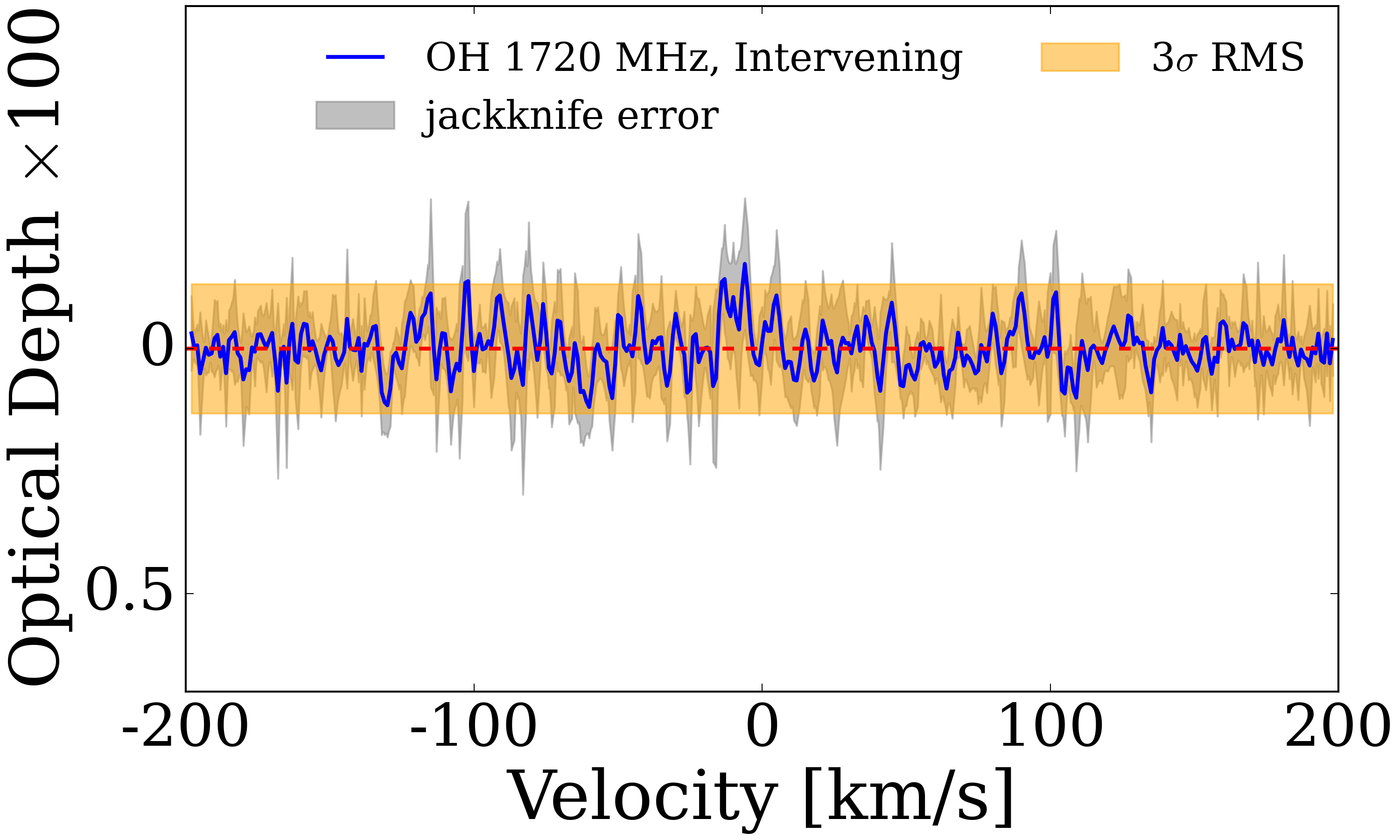}
    \caption{Stacked optical depth spectra centered on the redshifted \oh\ lines at 1612 MHz, 1665 MHz, 1667 MHz, and 1720 MHz are presented for all systems (top row), associated systems (middle row), and intervening systems (bottom row). The grey shaded regions represent the jackknife errors, while the orange region indicates the 3$\sigma$ RMS.}
    \label{OH_abs_stacked}
\end{figure*}

\begin{figure}[hbt]
    \centering
    \includegraphics[width=8.2cm]{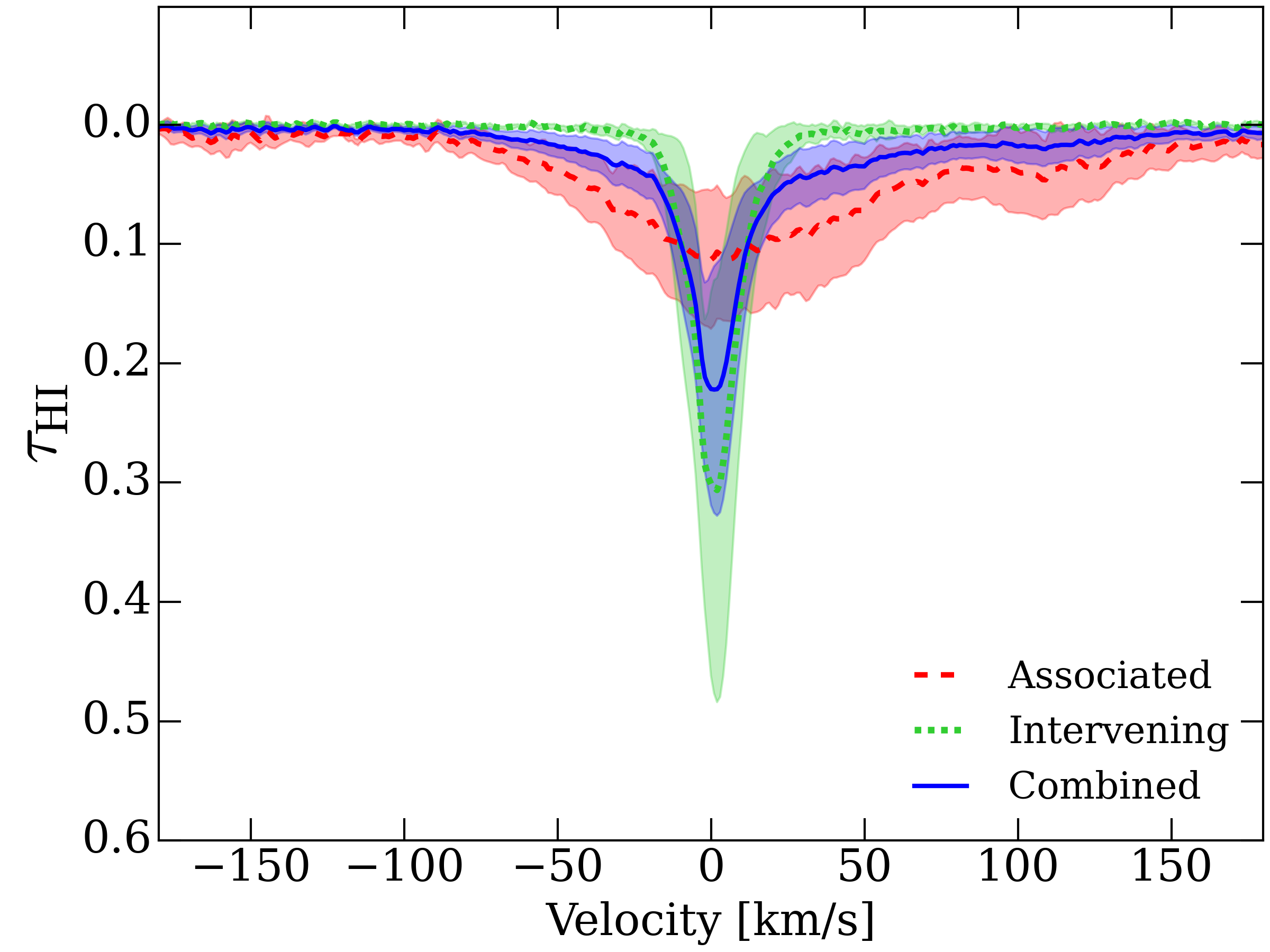}
    \caption{Stack of \hi\ absorption spectra of the associated (red dashed line), intervening (green dotted line), and combined (blue solid line) \hi\ absorption samples that are used for \oh\ absorption search. The shaded regions represent the jackknife errors.}
    \label{HI_abs_stacked}
\end{figure}

The stacking analysis was performed separately for the associated sample, the intervening sample, and the combined set of both. The stacking results for the \oh\ and \hi\ optical path spectra are presented in Figure~\ref{OH_abs_stacked} and Figure~\ref{HI_abs_stacked}, respectively. From stacking, we calculate the RMS of the stacked optical path spectra (at a velocity resolution of 1\kms) for the \oh\ 1667 line are (1.26$\pm$0.14)$\times$10$^{-3}$, (0.42$\pm$0.16)$\times$10$^{-3}$, and (0.44$\pm$0.20)$\times$10$^{-3}$ for associated, intervening and combined absorption samples. The error bars are estimated using jackknife sampling. These optical path RMS corresponding to a 3$\sigma$ upper limit of $N_{\oh}$ of 4.93, 1.64, and 1.72$T_{\rm{ex}}\times$10$^{12}$cm$^{-2}$K$^{-1}$ for associated, intervening, and all \hi\ absorption samples, assuming an \oh\ absorption velocity width of 30 \kms (Eq.~(\ref{optical_path_int})). The mean \hi\ column density $\langle N_{\hi}\rangle$ is measured to be 2.97$\pm$1.84,  1.15$\pm$0.77, and 1.92$\pm$0.78 $T_{\rm{s}}\times$10$^{19}$cm$^{-2}$K$^{-1}$ for associated, intervening and combined \hi\ absorption samples. Assuming $T_{\rm{ex}}$ for \oh\ is 10K and $T_{\rm{s}}$ for \hi\ is 100K, we obtain an \oh\ abundance upper limit [\rm{OH}]/[\hi] $<$(1.66$\pm$1.05)$\times$10$^{-8}$, (1.42$\pm$1.10)$\times$10$^{-8}$, and (0.90$\pm$0.54)$\times$10$^{-8}$ for associated, intervening and combined \hi\ absorption samples.

For comparison, diffuse Galactic sightlines typically exhibit [\rm{OH}]/[\hi] $\sim$ a few$\times$10$^{-8}$ \citep{1996A&A...314..917L,2018ApJ...862...49N}, consistent with the level of our upper limits. By contrast, extragalactic OH absorbers, though rare and preferentially associated with molecule-rich environments, show substantially higher ratios. For example, in the $z$=0.6847 absorber toward B0218\allowbreak+357, [\rm{OH}]/[\hi]$\sim$6$\times$10$^{-6}$ \citep{1993ApJ...412L..59C,2003MNRAS.345L...7K}, assuming $T_{\rm{ex}}$=10K and $T_{\rm{s}}$=100K. Similarly PKS 1830\allowbreak–211 yields a ratio of 4$\times$10$^{-6}$\citep{1999A&A...343L..79C}, and the $z\sim0.05$ absorber toward Q0248\allowbreak+430 shows [\rm{OH}]/[\hi] $\sim$2$\times$10$^{-6}$ \citep{2018ApJ...860L..22G}, under the same assumptions.

Thus, our stacked upper limits are comparable to or below the ratios measured along diffuse Galactic sightlines, but remain substantially lower than the values reported for molecule-rich extragalactic OH absorbers. This strongly suggests that the typical \hi\ absorbers probed here contain little \rm{OH}, either because molecular gas is confined to small, low-filling-factor clumps, or because reduced metallicity/dust shielding suppresses the \oh\ abundance. Alternatively, the assumed excitation temperature or OH covering-factor parameters may systematically bias the inferred abundances.

\section{Discussions}
\label{sec:Discussion}

\subsection{Detection Rate of \oh\ Absorption}

We searched for \oh\ absorption in a catalog of 19 \hi\ 21-cm absorbers from a blind survey and re-detected OH absorption towards PKS 1413\allowbreak+135. In comparison, \citet{2008MNRAS.391..765C} conducted a $z$$\gtrsim$3 survey of seven radio quasars from the Parkes quarter-Jansky flat-spectrum sample and reported no OH detections. \citet{2011MNRAS.413.1165C} searched for redshifted \hi\ 21-cm and \oh\ 18-cm absorption in type-2 AGNs and reddened flat-spectrum sources; among the 14 targets with usable \oh\ data, none showed absorption. \citet{2019ApJS..245....3G} surveyed 145 compact radio sources over 0.02$<$$z$$<$3.8 for intrinsic \hi\ 21-cm and \oh\ 18-cm absorption and found no OH detections. Based on the WSRT \hi\ absorption survey \citep{2017A&A...604A..43M}, \citet{2020MNRAS.499.3085Z} selected eight \hi\ 21-cm absorbers whose \oh\ 18-cm main lines fall within the FAST frequency coverage and conducted observations searching for OH absorption, but also reported no detections. Finally, \citet{2024ApJ...973...48C} searched for \hi\ and \oh\ absorbers towards 40 low to intermediate-luminosity radio sources ($\sim$10$^{23}$–10$^{26}$ W Hz$^{-1}$ at 1.4 GHz) with red mid-IR colors (W2–W3$>$2.5 mag) out to $z$=0.35, but detected no OH absorption. 

\citet{2021MNRAS.508.1165C} suggest that selecting optically bright sources as targets introduces a bias against dustier, more molecule-rich sightlines. This optical redshift bias may explain the scarcity of millimetre-band absorption detections. Previous studies \citep{2006MNRAS.371..431C, 2021MNRAS.508.1165C} have found that the strength of redshifted \oh\ absorption correlates with the optical–near-infrared colour of the sightline. Notably, at least five (out of six) known \oh\ absorbers are located along very red sightlines, with optical–near-infrared colours of V-K$\gtrsim$5, whereas the non-detections tended to be bluer. This reddening is interpreted as due to dust, which shields the molecules from the ambient ultraviolet radiation. Consequently, these studies suggest that sightlines with V-K$\gtrsim$6 that are detected in 21-cm absorption should be prioritized to increase the number of known molecular absorbers. Our successful re-detection of OH absorption toward PKS 1413\allowbreak+135 is consistent with this strategy. Our survey represents the first unbiased survey to successfully detect OH absorption, highlighting the importance of prior \hi\ absorption in guiding \oh\ absorption searches, which were previously identified only through long-term targeted observations.

\subsection{Survival Statistics}
\citet{2020MNRAS.499.3085Z} analyzed the relationship between \oh\ and \hi\ abundances using survival statistics with Schmitt binning\citep{1985ApJ...293..178S}, where nearly 70 percent of the data were censored. In our study, we incorporate additional upper limits, increasing the censored fraction to $\sim$85 percent, a regime where Bayesian linear regression provides a more reliable treatment of censored data. While \citet{2020MNRAS.499.3085Z} reported that both associated and intervening absorbers show an increase of $N_{\rm{OH}}$/$N_{\hi}$ with redshift, our results reveal a negligible correlation for associated systems and a shallower increase for intervening systems. Similarly, although \citet{2020MNRAS.499.3085Z} found weak positive correlations between \oh\ and \hi\ column density, our analysis yields no significant correlation for associated absorbers and a weaker trend for intervening absorbers. These differences arise from the inclusion of larger samples and the application of improved statistical methods.

The inclusion of upper limits is essential, as most searches for \oh\ absorption result in non-detections. Our analysis demonstrates that incorporating censored data prevents overestimation of the correlation strength, which would otherwise be biased by a handful of detections. The weak trends observed here suggest that large, statistically complete samples will be required to establish a more definitive connection between \oh\ and \hi\ absorption.

\subsection{Sample Selection Bias}
\label{sec:sample_bias}

\subsubsection{Correlation Between $N_{\hi}$ and $N_{\rm{OH}}$}

\citet{2020MNRAS.499.3085Z} investigated the relationship between \oh\ and \hi\ abundances, reporting a weak positive correlation between $N_{\hi}$ and $N_{\rm{OH}}$ for both associated and intervening absorbers, as well as an increase in $N_{\rm{OH}}$/$N_{\hi}$ with redshift. However, their sample is affected by selection biases. It combined three heterogeneous subsets: a preselected sample from the WSRT \hi\ absorption survey, a sample from \citet{2018ApJ...860L..22G} selected toward specific quasar-galaxy pairs, and the six previously known extragalactic \oh\ absorbers. In particular, all four known high-redshift \oh\ absorbers ($z\sim0.6–1.0$) were identified through targeted observations rather than blind surveys. These systems were selected for \oh\ search because they exhibit strong \hi\ 21 cm absorption, are gravitational lenses, or are red quasars (or combinations of these properties). Consequently, the sample is biased toward sources with high dust content and large \hi\ column densities. In contrast, the low-redshift systems from our FAST survey are drawn from an unbiased sample.

For comparison, we overplot the fitted relations from \citet{2020MNRAS.499.3085Z} as brown lines in Figure~\ref{N_OH_HI}. The distributions of the FAST sample and the literature compilation differ, particularly in their redshift coverage, sample-selection criteria, and relative fraction of censored measurements. Within the FAST sample, the censored-regression fits do not provide statistically significant evidence for either a correlation between $N_{\hi}$ and $N_{\rm{OH}}$ or an increase of $N_{\rm{OH}}$/$N_{\hi}$ with redshift. The differences between our results and those of \citet{2020MNRAS.499.3085Z} indicate that the current constraints remain limited by the small sample size and possible selection effects. A larger sample of \oh\ absorbers is therefore needed to better quantify these correlations.

%\begin{table*}
%    \fontsize{8}{10}\selectfont
%    \centering
%    \caption{Slope ($\alpha$), intercept ($\beta$), and Spearman’s correlation coefficient ($\rho$) from the regression analysis, assuming a relation of $Y=\alpha X+\beta$. Results are shown for three samples: (1) the sample in \citet{2020MNRAS.499.3085Z} (2) the sample from our FAST survey.}
%    \label{survival_fit_table}
%    \begin{tabular}{@{}c c ccc ccc @{}} % Using 'cc' pairs for visual grouping
%    \toprule
%    & & \multicolumn{3}{c}{Zheng 20} & \multicolumn{3}{c}{This Work} \\
%    \cmidrule(lr){3-5} \cmidrule(lr){6-8} 
%    $X$ & $Y$ & $\alpha$ & $\beta$ & $\rho$ & $\alpha$ & $\beta$ & $\rho$ \\
%    \midrule
%    log$N^{\rm{intv}}_{\hi}$ & log$N^{\rm{intv}}_{\oh}$ & 1.09$\pm$0.30 & -8.17$\pm$6.16 & 0.12 & 0.38$\pm$0.28 & 6.47$\pm$5.95 & 0.53\\
%    log$N^{\rm{asc}}_{\hi}$ & log$N^{\rm{asc}}_{\oh}$ & 0.97$\pm$0.27 & -5.97$\pm$5.61 & 0.68 & 0.01$\pm$0.62 & 14.05$\pm$12.61 & 0.10\\
%    $z$ & log[$N^{\rm{intv}}_{\oh}$/$N^{\rm{intv}}_{\hi}$] & 2.14$\pm$0.43 & -6.80$\pm$0.24 & 0.65 & 4.40$\pm$6.42 & -7.56$\pm$1.63 & 0.37\\
%    $z$ & log[$N^{\rm{asc}}_{\oh}$/$N^{\rm{asc}}_{\hi}$] & 2.21$\pm$6.36 & -7.32$\pm$1.39 & 0.56 & 3.68$\pm$16.25 & -7.91$\pm$4.19 & 0.50\\
%    \bottomrule
%    \end{tabular}
%\end{table*}

\subsubsection{\oh\ abundance relative to \hi}

\citet{2024ApJ...973...48C} carried out a combined \hi\ and \oh\ absorption survey of 40 radio sources with low–intermediate radio luminosities($\sim$10$^{23}$-10$^{26}$ W Hz$^{-1}$ at 1.4 GHz) and red mid-infrared colors (W2[\SI{4.6}{\micro\metre}]-W3[\SI{12}{\micro\metre}]$>$2.5mag). By stacking the spectra of seven associated \hi\ absorbers, they derived a 3$\sigma$ upper limit of $N_{\oh}$$<$3.47$\times$10$^{14}(T_{\rm{ex}}$/10K)$c^{-1}_{\rm{f,\oh}}$cm$^{-2}$, corresponding to [\oh]/[\hi]$<$1.78$\times$10$^{-7}$. Similarly, \citet{2020MNRAS.499.3085Z}, using three associated systems selected from the WSRT \hi\ absorption survey \citep{2014A&A...569A..35G,2017A&A...604A..43M}, obtained a 3$\sigma$ upper limit of $N_{\rm{OH}}$$<$1.57$\times$10$^{14}$($T_{\rm{ex}}$/10K)$c^{-1}_{\rm{f,OH}}$cm$^{-2}$, or [\oh]/[\hi]$<$5.45$\times$10$^{-8}$. In our work, we conducted an \hi\ absorption survey targeting radio sources brighter than 12 mJy at 1.4 GHz and subsequently searched for OH absorption in the detected \hi\ systems through follow-up observations. Stacking five associated absorbers, we achieved a substantially more stringent 3$\sigma$ upper limit of $N_{\oh}$$<$4.93$\times$10$^{13}(T_{\rm{ex}}$/10K)$c^{-1}_{\rm{f,\oh}}$cm$^{-2}$, corresponds to [\oh]/[\hi]$<$1.66$\times$10$^{-8}$ (assuming $T_{\rm{ex}}$=10K and $T_{\rm{s}}$=100K).

The differences across studies mainly reflect sample selection. Targeted or color-selected surveys (e.g., high radio luminosity, red mid-IR sources, or WSRT preselection) are biased toward dustier, molecule-rich galaxies, while our flux-limited, blind \hi\ survey provides a more representative view. The much lower [\rm{OH}]/[\hi] limits we obtain show that most \hi\ absorbers are molecule-poor, underscoring the value of blind surveys for constraining the true molecular fraction and highlighting the rarity of OH-rich systems.

\subsection{Implications for Future Surveys}

All six known \oh\ absorbers are found in \hi\ absorption systems, confirming that \hi\ selection is essential for detecting \textsc{OH}. Yet, our low [\rm{OH}]/[\hi] limits show that most \hi\ absorbers are molecule-poor, unlike the rare \rm{OH}-rich cases likely associated with compact, dusty, star-forming environments (e.g., gravitational lenses, starbursts). This indicates that \oh\ absorption is a selective tracer of molecular gas, valuable for probing specific environments rather than broadly tracing neutral hydrogen. Our detection of stimulated \hi\ absorption, as well as both \oh\ emission and absorption in PKS 1413\allowbreak+135, demonstrates the feasibility of conducting simultaneous searches for \hi\ and \oh\ absorption. Instantaneous wide bandwidth coverage will be essential for enabling such joint \hi\ and \oh\ absorption surveys. Deeper, wide-area surveys with broad instantaneous bandwidths enabled by next-generation facilities (e.g., FAST Core Array \citep{2024AstTI...1...84J}, SKA \citep{2009IEEEP..97.1482D} and its precursors\citep{2008ExA....22..151J,2016mks..confE...1J}) will enable a comprehensive census of these absorbers across a broad redshift range ($0<z<6$). Such surveys will reveal the physical drivers of the \rm{OH}–\hi\ connection, probe galaxy evolution cross cosmic time, and place the highest-precision constraints on the stability of fundamental physical constants.

\section{Summary}
\label{sec:Summary}

We extended the FAST blind \hi\ absorption search using 2024 and part of 2025 CRAFTS and FASHI data (394.4 hr, 1622.1 deg$^{2}$) together with FATHOMER observations, detecting four previously known and three new \hi\ absorbers (NVSS J010015\allowbreak+201710, NVSS J024516\allowbreak+240535, and NVSS J053538\allowbreak+643141). Combined with our earlier search, this yields a catalog of 41 \hi\ absorption systems. OH absorption searches were carried out for 19 of these systems with redshifted OH lines within the FAST band, re-detecting the known absorber in PKS 1413\allowbreak+135 but finding no new OH absorption systems. Our survey is the first unbiased \hi\ search to detect \oh\ absorption, demonstrating that telescopes with broad instantaneous bandwidth can detect \hi\ and \oh\ absorption simultaneously.

Using survival analysis on the sample from our FAST blind search, we find no statistically significant correlation between $N_{\hi}$ and $N_{\rm{OH}}$, and between the $N_{\rm{OH}}$/$N_{\hi}$ ratio with redshift, for either intervening or associated absorbers. The uncertainties of our analysis remain large; larger surveys are therefore needed to confirm these possible trends and improve the statistical precision. Comparison with previous results suggests that earlier samples may be affected by selection biases.

Stacking analyses place stringent constraints on undetected \oh, yielding 3$\sigma$ column density upper limits of 4.93, 1.64, and 1.72 $T_{\rm{ex}}$/$c_{\rm{f}}\times$10$^{12}$cm$^{-2}$K$^{-1}$ for associated, intervening, and combined samples, corresponding to [\rm{OH}]/[\hi] ratios of $<$1.66$\times$10$^{-8}$, $<$1.42$\times$10$^{-8}$, and $<$0.90$\times$10$^{-8}$ (for $T_{\rm{ex}}$=10K and $T_{\rm{s}}$=100K), which are the lowest ever achieved. 

\section*{Data Availability}
The radio data analyzed in this work can be accessed by sending a request to the FAST Data Centre or the corresponding authors of this paper.

\appendix
\section{Gaussian components for \hi\ absorptions}
Some \hi\ absorption profiles detected in our 2024 and 2025 FAST \hi\ absorption surveys can be well described by a multi-Gaussian model. The fitted parameters of the individual Gaussian components are listed in Table~\ref{HI_absorption_gaucomp_table}.

\begin{table*}
    \fontsize{6.8}{8}\selectfont
	\centering
	\caption{Same as Table~\ref{HI_absorption_table}, but for individual Gaussian components.}
	\label{HI_absorption_gaucomp_table}
	\begin{tabular}{|c|c|c|c|c|c|c|c|c|}
	\hline
	Radio Source & Comp & cz$_{\rm peak}$ & FWHM & $S_{\hi, \rm peak}$ & $\int S_{\hi}dv$ & $\tau_{\rm peak}\times10^{2}$ & $\int\tau dv$ & $N_{\hi}/T_{\rm{s}}$\\
	& & (\kms) & (\kms) & (mJy) & (mJy\kms) &  & (\kms) & (10$^{18}$cm$^{-2}$K$^{-1}$)\\
        \hline
        
        & 1 & 79563.17$\pm$0.64 & 4.91$\pm$1.67 & -3.31$\pm$0.92 & -17.30$\pm$7.59 & 1.95$\pm$0.54 & 0.10$\pm$0.04 & 0.18$\pm$0.08\\
        
        & 2 & 79609.52$\pm$2.57 & 53.27$\pm$5.18 & -9.43$\pm$2.53 & -535.01$\pm$152.47 & 5.66$\pm$1.56 & 3.18$\pm$0.92 & 5.79$\pm$1.68\\
        
        & 3 & 79595.97$\pm$0.75 & 7.85$\pm$2.50 & -4.25$\pm$1.39 & -35.47$\pm$16.25 & 2.51$\pm$0.83 & 0.21$\pm$0.10 & 0.38$\pm$0.17\\
        
        & 4 & 79612.13$\pm$1.66 & 12.75$\pm$6.32 & -4.16$\pm$2.64 & -56.52$\pm$45.46 & 2.46$\pm$1.57 & 0.33$\pm$0.27 & 0.60$\pm$0.48\\
        
        NVSS J010015+201710$^{\diamond}$ & 5 & 79622.79$\pm$0.39 & 6.27$\pm$0.89 & -17.37$\pm$2.48 & -116.02$\pm$23.36 & 10.68$\pm$1.60 & 0.70$\pm$0.14 & 1.28$\pm$0.26 \\

        & 6 & 79630.87$\pm$0.51 & 9.08$\pm$1.47 & -18.55$\pm$1.88 & -179.40$\pm$34.34 & 11.45$\pm$1.22 & 1.09$\pm$0.21 & 1.98$\pm$0.38\\

        & 7 & 79640.93$\pm$0.23 & 4.90$\pm$0.68 & -13.30$\pm$1.26 & -69.35$\pm$11.67 & 8.07$\pm$0.79 & 0.42$\pm$0.07 & 0.76$\pm$0.12\\

        & 8 & 79647.13$\pm$0.17 & 3.83$\pm$0.44 & -14.06$\pm$1.17 & -57.36$\pm$8.18 & 8.55$\pm$0.73 & 0.34$\pm$0.05 & 0.63$\pm$0.08\\
        
        \hline

        NVSS J053538+643141$^{\otimes}$ & 1 & 24078.27$\pm$2.38 & 80.42$\pm$9.39 & -12.97$\pm$5.23 & -1110.26$\pm$466.30 & 25.39$\pm$11.68 & 20.94$\pm$9.61 & 38.10$\pm$17.48\\
        
        & 2 & 24083.29$\pm$2.09 & 44.01$\pm$10.59 & -8.95$\pm$5.10 & -419.40$\pm$259.28 & 16.81$\pm$10.43 & 7.69$\pm$5.01 & 13.99$\pm$9.12\\

        \hline

        & 1 & 29391.89$\pm$0.63 & 28.83$\pm$1.94 & -18.15$\pm$1.16 & -556.88$\pm$52.04 & 13.67$\pm$0.92 & 4.11$\pm$0.39 & 7.48$\pm$0.71\\
        
        NVSS J080601+190611$^{\star}$ & 2 & 29431.71$\pm$8.59 & 89.37$\pm$12.20 & -8.91$\pm$0.54 & -847.88$\pm$126.96 & 6.48$\pm$0.40 & 6.11$\pm$0.91 & 11.12$\pm$1.66\\
        
        & 3 & 29513.38$\pm$7.81 & 68.75$\pm$12.27 & -5.71$\pm$0.95 & -417.71$\pm$101.88 & 4.10$\pm$0.69 & 2.98$\pm$0.73 & 5.43$\pm$1.33\\

        \hline
        
        & 1 & 82848.20$\pm$1.15 & 40.28$\pm$3.10 & -12.38$\pm$0.77 & -530.94$\pm$52.75 & 3.03$\pm$0.19 & 1.29$\pm$0.13 & 2.35$\pm$0.23\\
        
        & 2 & 83059.43$\pm$23.01 & 346.38$\pm$31.26 & -8.49$\pm$2.52 & -3130.87$\pm$970.77 & 2.07$\pm$0.62 & 7.60$\pm$2.37 & 13.84$\pm$4.31\\
        
        & 3 & 83004.20$\pm$2.43 & 70.84$\pm$5.04 & -54.15$\pm$11.02 & -4082.64$\pm$880.89 & 13.99$\pm$3.05 & 10.33$\pm$2.33 & 18.81$\pm$4.23\\
        
        NVSS J234106+001833$^{\star}$ & 4 & 83087.51$\pm$23.29 & 130.20$\pm$40.90 & -27.30$\pm$3.29 & -3783.58$\pm$1273.32 & 6.81$\pm$0.84 & 9.34$\pm$3.15 & 17.00$\pm$5.73\\
        
        & 5 & 83048.10$\pm$1.25 & 35.57$\pm$3.79 & -22.82$\pm$3.93 & -864.12$\pm$175.05 & 5.66$\pm$1.00 & 2.13$\pm$0.44 & 3.87$\pm$0.79\\
        
        & 6 & 83119.12$\pm$1.05 & 34.60$\pm$4.41 & -17.83$\pm$3.36 & -656.67$\pm$149.44 & 4.39$\pm$0.84 & 1.61$\pm$0.37 & 2.93$\pm$0.67\\
        
        & 7 & 83266.39$\pm$2.54 & 94.38$\pm$7.17 & -15.57$\pm$1.24 & -1564.58$\pm$172.46 & 3.83$\pm$0.31 & 3.82$\pm$0.42 & 6.96$\pm$0.77\\
        \hline

        NVSS J024516+240535$^{\diamond}$ & 1 & 42366.41$\pm$0.25 & 5.13$\pm$0.51 & -29.61$\pm$2.80 & -161.62$\pm$22.32 & 17.52$\pm$0.32 & 0.33$\pm$0.05 & 0.59$\pm$0.08\\
        
        & 2 & 42373.79$\pm$0.15 & 8.73$\pm$0.33 & -81.55$\pm$1.85 & -757.63$\pm$33.40 & 6.01$\pm$0.58 & 1.59$\pm$0.07 & 2.89$\pm$0.12\\
        \hline

        & 1 & 19825.32$\pm$2.79 & 18.16$\pm$4.32 & -5.38$\pm$0.81 & -103.97$\pm$29.38 & 8.58$\pm$1.36 & 1.64$\pm$0.47 & 2.98$\pm$0.85\\
        
        NVSS J150034+364844$^{\star}$ & 2 & 19842.27$\pm$1.30 & 16.56$\pm$1.92 & -10.37$\pm$1.06 & -182.80$\pm$28.47 & 17.27$\pm$1.92 & 2.97$\pm$0.47 & 5.40$\pm$0.86\\
        
        & 3 & 19847.53$\pm$1.47 & 137.29$\pm$3.74 & -7.16$\pm$0.26 & -1047.06$\pm$50.44 & 11.61$\pm$0.43 & 16.67$\pm$0.79 & 30.35$\pm$1.44\\
        \hline
        \end{tabular}
\end{table*}

\section*{Acknowledgements}

This work made use of the data from FAST (Five-hundred-meter Aperture Spherical radio Telescope)(https://cstr.cn/31116.02.FAST). FAST is a Chinese national mega-science facility, operated by National Astronomical Observatories, Chinese Academy of Sciences. This research has made use of the NASA/IPAC Extragalactic Database (NED), which is operated by the Jet Propulsion Laboratory, California Institute of Technology, under contract with the National Aeronautics and Space Administration.  

This work is supported by the National SKA Program of China (Nos. 2022SKA0110100 and 2022SKA0110101), the NSFC International (Regional) Cooperation and Exchange Project (No. 12361141814), the Specialized Research Fund for State Key Laboratory of Radio Astronomy and Technology, and the National Astronomical Observatories, Chinese Academy of Science (No. E5ZB0901).

%%%%%%%%%%%%%%%%%%%%%%%%%%%%%%%%%%%%%%%%%%%%%%%%%%
\section*{Data Availability}

\bibliography{OHabs}{}
\bibliographystyle{aasjournal}



\end{document}